\documentclass[aps,prd,superscriptaddress,showpacs,floatfix,twocolumn]{revtex4-1}
\usepackage{graphicx}% Include figure files
\usepackage{amsmath}% for ``align''
\usepackage[caption=false]{subfig}
\usepackage{blindtext}

\usepackage[%
  colorlinks=true,
  urlcolor=blue,
  linkcolor=blue,
  citecolor=blue
]{hyperref}

\usepackage{color}

\usepackage[normalem]{ulem} % \sout{old text} for strikeout
\renewcommand\sout{\bgroup \color[rgb]{0.55,0.00,0.99} \ULdepth=-.5ex \ULset}

\usepackage{times}
\begin{document}

%\setpagewiselinenumbers 
%\modulolinenumbers[5]
%\linenumbers

\title{Fixed-target charmonium production and pion parton distributions}

\author{Wen-Chen Chang} 
\affiliation{Institute of Physics, Academia Sinica, Taipei 11529,
  Taiwan}

\author{Jen-Chieh Peng}
\affiliation{Department of Physics, University of Illinois at
  Urbana-Champaign, Urbana, Illinois 61801, USA}

\author{Stephane Platchkov}
\affiliation{IRFU, CEA, Universit\'{e} Paris-Saclay, 91191
  Gif-sur-Yvette, France}

\author{Takahiro Sawada}
\affiliation{Nambu Yoichiro Institute of Theoretical and Experimental
  Physics, Osaka Metropolitan University, Osaka 558-8585, Japan}

\date{\today}

\begin{abstract}

We investigate how charmonium hadroproduction at fixed-target energies
can be used to constrain the gluon distribution in pions. Using
nonrelativistic QCD (NRQCD) formulation, the $J/\psi$ and $\psi(2S)$
cross sections as a function of longitudinal momentum fraction $x_F$
from pions and protons colliding with light targets, as well as the
$\psi(2S)$ to $J/\psi$ cross section ratios, are included in the
analysis. The color-octet long-distance matrix elements are found to
have a pronounced dependence on the pion parton distribution functions
(PDFs). This study shows that the $x_F$ differential cross sections of
pion-induced charmonium production impose strong constraints on the
pion's quark and gluon PDFs. In particular, the pion PDFs with larger
gluon densities provide a significantly better description of the
data. It is also found that the production of the $\psi(2S)$ state is
associated with a larger quark-antiquark contribution, compared with
$J/\psi$.

\end{abstract}
% 12.38.Lg Other nonperturbative calculations  
% 14.20.Dh Protons and neutrons  
% 14.65.Bt Light quarks  
% 13.60.Hb Total and inclusive cross sections (including deep-inelastic processes)  
% \pacs{12.38.Lg,14.20.Dh,14.65.Bt,13.60.Hb}

% for revtex
\maketitle

%%%%%%%%%%%%%%%%%%%%%%
\section{Introduction}
\label{sec:introduction}
%%%%%%%%%%%%%%%%%%%%%%

The pion, as the lightest QCD bound state, plays an essential role in
the nucleon-nucleon interactions over nuclear-size
distances~\cite{Horn:2016rip}. Theoretically, its partonic structure
is easier to construct than that of the nucleon. Pion distribution
amplitudes and parton distribution functions (PDFs) have been
predicted by a number of recent calculations based on the chiral-quark
model~\cite{Nam:2012vm, Watanabe:2016lto, Watanabe:2017pvl},
Nambu-Jona-Lasinio model~\cite{Hutauruk:2016sug}, light-front
Hamiltonian~\cite{Lan:2019vui, Lan:2019rba, Lan:2021wok}, holographic
QCD~\cite{deTeramond:2018ecg, Watanabe:2019zny, Lan:2020fno}, maximum
entropy method~\cite{Han:2018wsw, Han:2020vjp}, Dyson-Schwinger
equations (DSE)~\cite{Chang:2014lva, Chang:2014gga, Chen:2016sno,
  Shi:2018mcb, Bednar:2018mtf, Ding:2019lwe, Cui:2020tdf,
  Freese:2021zne, Chang:2021utv, Cui:2021mom, Cui:2022bxn}, and
lattice QCD~\cite{Zhang:2018nsy, Sufian:2019bol, Izubuchi:2019lyk,
  Joo:2019bzr, Sufian:2020vzb, Chen:2019lcm, Gao:2020ito,
  Alexandrou:2020gxs, Alexandrou:2021mmi, Fan:2021bcr,
  Detmold:2021qln, Gao:2022iex, JAM:2022aix}. In contrast, the partonic structure
of pion is much less explored experimentally, due to the absence of a
pion target. The present knowledge on the pion PDFs comes primarily
from fixed-target pion-induced Drell-Yan (DY)
measurements~\cite{Chang:2013opa}. However, the DY data are mainly
sensitive to the valence-quark distributions, leaving the sea and
gluon distributions essentially unknown. The sea-quark contributions
can in principle be extracted by comparing measurements with the
positive and negative pion beams~\cite{Londergan:1995wp}, although the
existing measurements are scarce and of insufficient statistical
accuracy.

The gluon distribution in the pion can be accessed through processes
such as prompt-photon production~\cite{WA70:1987bai}, leading-neutron
deep-inelastic scattering (DIS)~\cite{Khoze:2006hw, McKenney:2015xis}
or heavy quarkonia production~\cite{Gluck:1977zm, Barger:1980mg}. Each
of these processes has its own advantages and limitations. With the
exception of Ref.~\cite{Owens:1984zj}, the pion-induced $J/\psi$ and
$\psi(2S)$ production data were not included in the global analysis,
possibly reflecting the concern that the production mechanism for
charmonium production was not well understood. Significant progress in
understanding the $J/\psi$ production mechanism has been made in
recent decades, and it is timely to investigate how the charmonium
production data can provide useful constraints on the pion PDFs.

The theoretical challenge in describing the charmonium production
comes from the treatment of the hadronization of $c \bar{c}$ pairs
into a charmonium bound state~\cite{Brambilla:2010cs,
  Lansberg:2019adr}. This nonperturbative process has been modeled in
several theoretical approaches including the color evaporation model
(CEM)~\cite{Einhorn:1975ua, Fritzsch:1977ay, Halzen:1977rs}, the
color-singlet model (CSM)~\cite{Chang:1979nn, Berger:1980ni,
  Baier:1983va}, and nonrelativistic QCD
(NRQCD)~\cite{Bodwin:1994jh}. The CEM, although successful for some
observables, fails to explain some others observables in charmonium
production~\cite{Bodwin:2005hm}. Within the more rigorous NRQCD
framework, the production of the heavy quark pair is treated
perturbatively, whereas its hadronization to a bound state is
described in terms of a set of long-distance matrix elements (LDMEs),
extracted phenomenologically from the data.

From the experimental perspective, charmonium production has one
important advantage: the cross sections are large, between one to two
orders of magnitude higher than the DY ones, depending on the
experimental conditions. A large number of fixed-target charmonium
production experiments have been performed in the past, including
experiments with pion beams~\cite{Schuler:1994hy, Vogt:1999cu}. These
data, collected mostly at CERN or at Fermilab, provide a wealth of
additional information on the pion structure, and are expected to shed
new light on its gluon distribution.

In this paper we investigate how charmonium production could help to
differentiate between the available pion PDFs by imposing further
constraints on the gluon distribution function~\cite{Chang:2020rdy,
  Hsieh:2021yzg}. In the fixed-target energy domain, charmonium
production is dominated by the quark-antiquark annihilation ($q
\bar{q}$) and gluon-gluon fusion ($GG$) partonic subprocesses. The
longitudinal momentum $x_F$-differential cross sections are sensitive
to the quark and gluon parton distributions of the colliding
hadrons. Since the nucleon PDFs are known with good accuracy, these
differential cross sections should provide additional constraints on
the pion's quark and gluon PDFs.

\begin{figure*}[!ht]
\centering
\includegraphics[width=1.8\columnwidth]{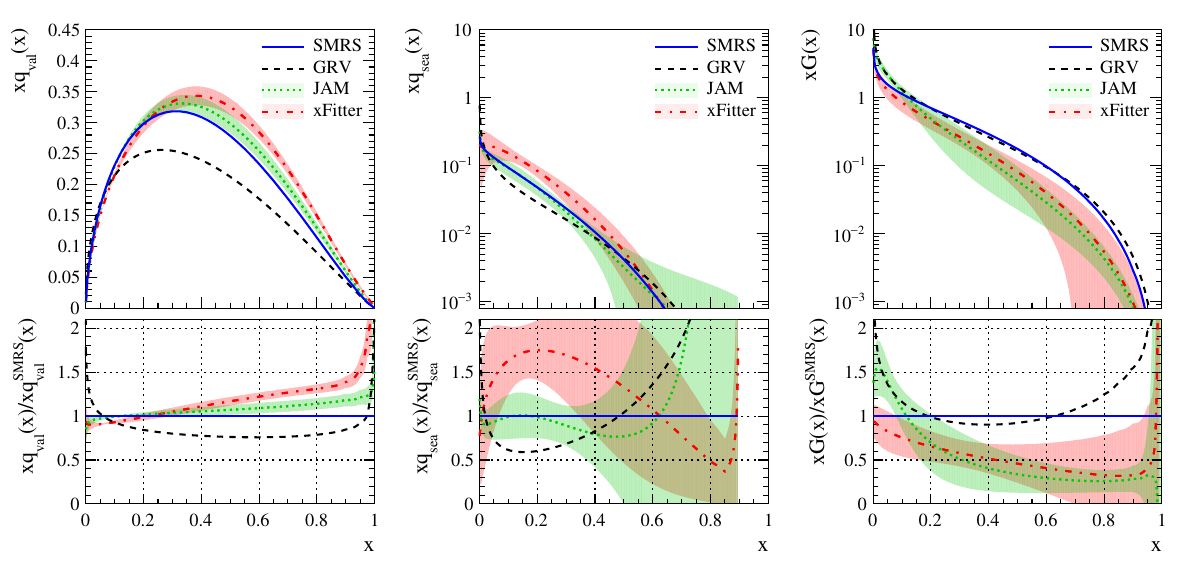}
\caption
[\protect{}] {Momentum density distributions $[xf(x)]$ of valence
  quarks, sea quarks and gluons of SMRS, GRV, xFitter and JAM pion
  PDFs and their ratios to the SMRS PDFs, at the scale of $J/\psi$
  mass ($Q^2$= 9.6 GeV$^2$). The uncertainty bands associated with JAM
  and xFitter PDFs are also shown.}
\label{fig_PDF}
\end{figure*}

To perform this study, we employ the NRQCD framework, along the lines
developed in Ref.~\cite{Beneke:1996tk}. Although limited to leading
order (LO), this approach provides an adequate description of the
fixed-target $J/\psi$ and $\psi(2S)$ production data and can be used
as a tool for accessing the pion PDFs. Our primary goal is to obtain a
good phenomenological description of both pion and proton-induced
data, and to explore the sensitivity of the results to the pion
quark-gluon structure. Assuming that the LDMEs are independent of the
beam species, the proton-induced cross sections are also included in
this analysis. Since the proton PDFs are well known, the
proton-induced data should help constraining the values of LDMEs
common to both the proton and pion data.

Results of an earlier study limited to total cross sections of
charmonium production were recently reported~\cite{Hsieh:2021yzg}. A
new set of color-octet LDMEs, leading to a good agreement between the
charmonium production data and the NRQCD fit was obtained. Here, we
extend the study by including the $x_F$-dependent cross sections for
$J/\psi$ and $\psi(2S)$ production as well as their ratios for both
pion and proton beams in the global fit. The distributions of the
differential $x_F$ cross sections are calculated by convolving the
partonic cross sections, the LDMEs of the various subprocesses and the
associated beam and target parton densities. An adequate NRQCD
description of such a large dataset should impose a strong constrain
on the pion PDFs. In order to minimize nuclear matter effects that are
not well understood, the present analysis is limited to data taken
with the lightest targets available: hydrogen, lithium and
beryllium. Data with heavier targets were considered only for the
$J/\psi$ to $\psi(2S)$ cross section ratios, assuming nuclear effects
are largely independent of the charmonium states.

This paper is organized as follows. In Sec.~\ref{sec:PDFs}, we
describes distinctive features of parton densities in four pion
PDFs. The NRQCD formalism used for this study is introduced in
Sec.~\ref{sec:NRQCD}. Section~\ref{sec:DATA} briefly describes the
$J/\psi$ and $\psi(2S)$ datasets used in the global fit. We present
the results of NRQCD calculations using various pion PDFs and the
comparison with the charmonium data in Sec.~\ref{sec:results}. Finally
we comment on the fit results in Sec.~\ref{sec:discussion} and
conclude in Sec.~\ref{sec:conclusion}

%%%%%%%%%%%%%%%%%%%%%%
\section{Pion PDFs}
\label{sec:PDFs}
%%%%%%%%%%%%%%%%%%%%%%

As mentioned before, pion-induced Drell-Yan data are used in all
global analyses for constraining the valence-quark distribution of the
pion PDFs. Without data from other processes, the sea and gluon
distributions can only be inferred through the momentum sum rule and
valence-quark sum rule. The two most recent global analyses dedicated
to the extraction of the pion PDFs are JAM~\cite{Barry:2018ort,
  Cao:2021aci, Barry:2021osv} and xFitter~\cite{Novikov:2020snp}. The
two groups consider the same DY data, but differ in the choice of the
additional processes. The xFitter group makes use of the pion-induced
prompt-photon production data, whereas the JAM collaboration includes
the leading-neutron DIS cross section measurements instead. The
Sutton-Martin-Roberts-Stirling (SMRS) global fit~\cite{Sutton:1991ay}
also incorporates the prompt-photon data, but instead of calculating
the fit uncertainties, it considers three different options for the
gluon and sea contents. Another widely used parametrization is the fit
of Gluck-Reya-Vogt (GRV)~\cite{Gluck:1991ey}, in which the gluon and
sea distributions are dynamically generated from the QCD evolution.

We utilize the LHAPDF framework~\cite{Whalley:2005nh, Buckley:2014ana}
to access these four pion PDFs for our study. The corresponding pion
PDF sets are ``SMRSPI.LHgrid'', ``GRVPI1'', ``JAM21PionPDFnlo'', and
``xFitterPI\_NLO\_EIG'', respectively. Out of the three possible
parametrizations for SMRS, we choose the one in which the sea quarks
carry 15\% of the pion momentum at $Q^2$= 4 GeV$^2$. Their valence,
sea and gluon momentum distributions $xf(x)$ at the scale of $J/\psi$
mass are compared in Fig.~\ref{fig_PDF}. Their ratios to SMRS are
shown in the bottom panel. Within the range of $x \sim$0.1--0.8, the
valence-quark distributions of SMRS, JAM and xFitter are close to each
other, whereas GRV is lower by up to 20\%--30\%. Not surprisingly, the
sea distribution is essentially unknown, as illustrated by the large
variations between the four PDFs. The gluon distributions also show
sizable differences; e.g., in the region of $x > 0.2$ the xFitter and
JAM distributions are smaller in comparison with SMRS and GRV, by up
to a factor of 2-3.

%%%%%%%%%%%%%%%%%%%%%%
\section{Heavy-Quark Pair Production and NRQCD Model}
\label{sec:NRQCD}
%%%%%%%%%%%%%%%%%%%%%%

Within the NRQCD theoretical framework, the heavy quarkonium
production is factorized into production of a heavy-quark pair ($Q
\bar{Q}$) at the parton level, and its subsequent hadronization into
quarkonium states.  The $Q \bar{Q}$ production cross section can be
calculated perturbatively~\cite{Nason:1987xz, Nason:1989zy,
  Mangano:1992kq}, whereas the hadronization probability of the $Q
\bar{Q}$ pair is encoded in the nonperturbative LDME parameters
$\langle \mathcal{O}_{n}^{H} [^{2S+1}L_{J}]\rangle$, depending on the
spin, orbital , and total angular momentum quantum numbers, $S$, $L$
and $J$, respectively, and on the color configuration ($n$). Parity,
charge conjugation and angular momentum conservation limit the allowed
quantum numbers to only a few. The LDMEs are assumed to be universal,
i.e., independent of the beam and target hadrons and of the energy
scale. The color singlet (CS) LDMEs are typically determined from
decay rate measurements using a potential model~\cite{Eichten:1995ch},
while the color octet (CO) LDMEs are obtained from a fit to the
experimental data.

In NRQCD, the differential cross section $d\sigma/dx_F$ for the
production of a charmonium state $H$ ($H$ = $J/\psi$, $\psi(2S)$, or
$\chi_{cJ}$) from the $hN$ collisions, where $h$ is the beam hadron
($h$ = $p$, $\bar{p}$, or $\pi$) and $N$ the target nucleon, is
expressed as~\cite{Vogt:1999dw}
\begin{align}
\label{eq:eq1}
\frac{d\sigma^{H}}{dx_F}=& \sum\limits_{i,j=q, \bar{q},
  G} \int_{0} ^{1} dx_{1} dx_{2} \delta(x_F - x_1 + x_2) \nonumber \\
 \times& f^{h}_{i}(x_1, \mu_{F}) f^{N}_{j}(x_2, \mu_{F}) \nonumber \\
\times& \hat{\sigma}[ij \rightarrow H](x_1 P_{h} , x_2 P_{N} , \mu_{F},
\mu_{R}, m_c), \\
\hat{\sigma}[ij \rightarrow H] =&  \sum\limits_{n} C^{ij}_{c \bar{c} [n]} (x_1 P_{h} , x_2 P_{N} , \mu_{F},
\mu_{R}, m_c) \nonumber \\
\times& \langle \mathcal{O}_{n}^{H}[^{2S+1}L_{J}] \rangle
\end{align}
where the indexes $i$ and $j$ run over the type of interacting partons
(gluons, quarks and antiquarks), and $C^{ij}_{c \bar{c} [n]}$ denotes
the hard-QCD production cross section for $c \bar c$ pair. The
parameter $m_c$ is the charm quark mass; $f^{h}$ and $f^{N}$ are the
incoming hadron and the target nucleon parton distribution functions,
evaluated at their respective Bjorken-$x$ values, $x_1$ and $x_2$. The
$\mu_F$ and $\mu_R$ are the factorization and renormalization
scales. The Feynman variable $x_F$ and the beam and target parton
momentum fractions $x_1$ and $x_2$ are:
\begin{align}
  x_F = \frac{2 p_L}{\sqrt{s}} \mbox{, }    x_{1,2} =  \frac{(x_F^2+4{M_{c \bar{c}}}^2/s)^{1/2} \pm x_F}{2}.
\end{align}
Here $M_{c \bar{c}}$ and $p_L$ are the mass and longitudinal momentum
of the $c \bar{c}$ pair in the center-of-mass frame. The total cross
sections are obtained by integrating over $x_F$.

In this study, we use the formula given in Ref.~\cite{Beneke:1996tk}
for computation of $J/\psi$, $\psi(2S)$, and $\chi_{cJ}$ production
via $GG$, $q \bar{q}$ and $qG$ subprocesses. The scattering subprocesses $q
\bar{q} \to Q \bar{Q}$ and $G G \to Q \bar{Q}$ at
$\mathcal{O}(\alpha_{s}^2)$ produce $Q \bar{Q}$ pairs in an $S$-wave CO
state or $P$-wave CS state. Table~\ref{tab:LDMEproc} summarizes the
relationships between the LDMEs and the scattering subprocesses for
$J/\psi$, $\psi(2S)$, $\chi_{c0}$, $\chi_{c1}$, and $\chi_{c2}$, up to
$\mathcal{O}(\alpha_{s}^3)$. For the $q \bar{q}$ subprocess, the $c
\bar{c}$ pairs are produced at $\mathcal{O}(\alpha_{s}^2)$ in color
octet states, which then hadronize into various charmonium states with
the LDMEs $\langle \mathcal{O}_{8}^{H}[^{3}S_1] \rangle$. For the $GG$
subprocess, both $J/\psi$, and $\psi(2S)$ can be produced from either the
CO $c \bar{c}$ at $\mathcal{O}(\alpha_{s}^2)$ or the CS $c \bar{c}$ at
$\mathcal{O}(\alpha_{s}^3)$. The CO $^{1}S_{0}$, $^{3}P_{0}$, and
$^{3}P_{2}$ are combined into a single LDME, $\Delta_8^{H}$, via the
relation: $\Delta_8^{H} = \langle \mathcal{O}_{8}^{H}[^{1}S_{0}]
\rangle + \frac{3}{m_c^2} \langle \mathcal{O}_{8}^{H}[^{3}P_{0}]
\rangle + \frac{4}{5m_c^2} \langle \mathcal{O}_{8}^{H}[^{3}P_{2}]
\rangle$.

%%%%%%%%%%%%%% LDME %%%%%%%%%%%%%%%%%%%%%%%%%
% exe sfitall#scandump_table idata=1 ipdf=1 ildme=1 ifit=0 isys=0
% exe sfitall#scandump_table idata=1 ipdf=1 ildme=2 ifit=0 isys=0
\begin{table}[!ht]   %\footnotesize
\setlength\tabcolsep{2pt}
\addtolength{\tabcolsep}{2pt}
\centering
%\begin{center}
\begin{tabular}{|c|c|c|c|}
%\hline
\hline
 $H$ & $q \bar{q}$ & $GG$ & $qG$ \\
\hline
$J/\psi$, $\psi(2S)$ & $\langle \mathcal{O}_{8}^{H}[^{3}S_{1}] \rangle$ ($\alpha_{s}^2$) & $ \Delta_8^{H}$ ($\alpha_{s}^2$) & \\
 & &  $\langle \mathcal{O}_{1}^{H}[^{3}S_{1}] \rangle$ ($\alpha_{s}^3$) & \\
\hline
$\chi_{c0}$ & $\langle \mathcal{O}_{8}^{H}[^{3}S_{1}] \rangle$ ($\alpha_{s}^2$) & $ \langle \mathcal{O}_{1}^{H}[^{3}P_{0}] \rangle$ ($\alpha_{s}^2$) & \\
\hline
$\chi_{c1}$ & $\langle \mathcal{O}_{8}^{H}[^{3}S_{1}] \rangle$ ($\alpha_{s}^2$) & $ \langle \mathcal{O}_{1}^{H}[^{3}P_{1}] \rangle$ ($\alpha_{s}^3$) & $ \langle \mathcal{O}_{1}^{H}[^{3}P_{1}] \rangle$ ($\alpha_{s}^3$)\\
\hline
$\chi_{c2}$ & $\langle \mathcal{O}_{8}^{H}[^{3}S_{1}] \rangle$ ($\alpha_{s}^2$) & $ \langle \mathcal{O}_{1}^{H}[^{3}P_{2}] \rangle$ ($\alpha_{s}^2$) & \\
\hline
\end{tabular}
%\end{center}
\caption {Relationship of LDMEs and the associated orders of
  $\alpha_s$ to the scattering subprocesses for various charmonium states
  in the NRQCD framework of Ref.~\cite{Beneke:1996tk}.  Here
  $\Delta_8^{H} = \langle \mathcal{O}_{8}^{H}[^{1}S_{0}] \rangle +
  \frac{3}{m_c^2} \langle \mathcal{O}_{8}^{H}[^{3}P_{0}] \rangle +
  \frac{4}{5m_c^2} \langle \mathcal{O}_{8}^{H}[^{3}P_{2}] \rangle $.}
\label{tab:LDMEproc}
\end{table}

The number of independent LDMEs is further reduced by applying the
spin symmetry relations~\cite{Beneke:1996tk, Maltoni:2006yp}:
\begin{align}
 \langle \mathcal{O}_{8}^{J/\psi, \psi(2S)} [^{3}P_{J}] \rangle &=
(2J+1) \langle \mathcal{O}_{8}^{J/\psi, \psi(2S)} [^{3}P_{0}] \rangle
\mbox{ for $J=2$} \nonumber \\
\langle \mathcal{O}_{8}^{\chi_{cJ}} [^{3}S_{1}] \rangle &= (2J+1)
\langle \mathcal{O}_{8}^{\chi_{c0}} [^{3}S_{1}] \rangle \mbox{ for
  $J=1, 2$} \nonumber \\
\langle \mathcal{O}_{1}^{\chi_{cJ}} [^{3}P_{J}] \rangle &= (2J+1)
\langle \mathcal{O}_{1}^{\chi_{c0}} [^{3}P_{0}] \rangle \mbox{ for
  $J=1, 2$}.
\end{align}

The LDMEs used in the present work exhibit sensitivity to different
elementary scattering subprocesses contributing to the charmonium
production. In the cases of $J/\psi$ and $\psi(2S)$ production, the CO
$\langle \mathcal{O}_{8}^{H}[^{3}S_{1}] \rangle$ LDME is related to
the $q\bar{q} \to Q \bar{Q}$ subprocess, while the $G G \to Q \bar{Q}$
subprocess is strongly dependent on the $\Delta_8^{H}$ term. More
details on the NRQCD framework used in this work can be found in
Refs.~\cite{Beneke:1996tk, Hsieh:2021yzg}. In the following study, the
CS $\langle \mathcal{O}_{1}^{H}[^{3}S_{1}] \rangle$ LDMEs for $J/\psi$
and $\psi(2S)$ and the CS $ \langle \mathcal{O}_{1}^{H}[^{3}P_{0}]
\rangle$ and CO $\langle \mathcal{O}_{8}^{H}[^{3}S_{1}] \rangle$ LDMEs
for $\chi_{c0}$ are fixed to be 1.16, 0.76, 0.044 and 0.0032,
respectively, which are the values used in Refs.~\cite{Beneke:1996tk,
  Hsieh:2021yzg}.

With the information of LDMEs, the direct production cross sections of
$J/\psi$, $\psi(2S)$ and three $\chi_{cJ}$ states as a function of
$x_F$ can be evaluated as shown in Eq.(\ref{eq:eq1}). The $J/\psi$
cross section is estimated taking into account the direct production
of $J/\psi$ and the feed-down from hadronic decays of $\psi(2S)$ and
radiative decays of three $\chi_{cJ}$ states as follows,
\begin{align}
\sigma_{J/\psi} & = \sigma_{J/\psi}^{direct} \nonumber \\
 & + Br(\psi(2S) \rightarrow J/\psi X) \sigma_{\psi(2S)} \nonumber \\
 & + \sum \limits_{J=0}^{2} Br( \chi_{cJ} \rightarrow J/\psi \gamma)\sigma_{\chi_{cJ}}
\end{align}
The various branching ratios $Br$ are taken from the PDG
2020~\cite{ParticleDataGroup:2020ssz}: $Br(\psi(2S) \rightarrow J/\psi
X) = 61.4\%$, $Br(\chi_{c0} \rightarrow J/\psi \gamma)= 1.4\%$,
$Br(\chi_{c1} \rightarrow J/\psi \gamma)= 34.3\%$, and $Br(\chi_{c2}
\rightarrow J/\psi \gamma)= 19.0\%$.

In the present analysis we use the convention of charm quark mass,
factorization and renormalization scales in Ref.~\cite{Beneke:1996tk}
for fixed-target hadroproduction of charmonium: $m_c=1.5$ GeV/$c^2$
and $\mu_F=\mu_R=2m_c$. The uncertainties associated with this choice
are evaluated by changing the reference scale from $m_c$ to
$3m_c$. The nucleon PDFs are taken from
CTEQ14nlo~\cite{Dulat:2015mca}. For the lithium, beryllium, silicon,
gold and tungsten targets, the nuclear EPPS16 PDFs
~\cite{Eskola:2016oht} are used.

%%%%%%%%%%%%%%%%%%%%%%
\section{OVERVIEW OF DATA USED}
\label{sec:DATA}
%%%%%%%%%%%%%%%%%%%%%% 

The present analysis is based on pion and proton-induced total and
differential cross sections for $J/\psi$ and $\psi(2S)$ production,
and on the differential $R_{\psi}(x_F) = \sigma_{\psi(2S)}(x_F) /
\sigma_{(J/\psi)}(x_F)$ ratios. The total cross sections for the
pion-induced data were taken from the compilations made in
Refs.~\cite{Schuler:1994hy} and \cite{Vogt:1999cu}. The proton-induced
total cross sections and ratios were taken from
Ref.~\cite{Maltoni:2006yp}. The proton-induced values for $R_{\psi}$
from HERA-B~\cite{HERA-B:2006bhy} and NA38~\cite{NA38:1994yau} and the
pion-induced ones from WA92~\cite{BEATRICE:1999mqh} and
WA39~\cite{jpsi_data19and20} were added to the selection. The
$x_F$-differential cross sections for pion-induced $J/\psi$
production~\cite{jpsi_data1, jpsi_data2and3, jpsi_data17and18and21,
  jpsi_data16, jpsi_data6and7and8, jpsi_data9and10} and $\psi(2S)$
production~\cite{jpsi_data1} were selected according to the targets
used: hydrogen, lithium and beryllium. Datasets with heavier targets
were not included. The same criterion was applied to the
proton-induced $J/\psi$ production~\cite{jpsi_data2and3,
  jpsi_data17and18and21}. The $R_{\psi}(x_F)$ ratios were taken from
Ref.~\cite{Heinrich:1991zm} for the pion-induced production and from
Refs.~\cite{HERA-B:2006bhy, NA50:2003fvu, E789:1995yhd, E771:1995ane}
for the proton-induced one. Assuming that nuclear effects are
identical for both charmonium states, no restriction on the target
employed was applied.

The datasets with $x_F$ dependent measurements are listed in
Table~\ref{tab:data}. In terms of pion-induced (proton-induced) data
sets, there are 8 (2) for $J/\psi$ production, 2 (0) for $\psi(2S)$
production and 1 (4) for $R_{\psi}(x_F)$. In total, there are 164 and
82 data points for the pion-induced and proton-induced data,
respectively. The beam momenta of the datasets cover the range of
39.5--515 GeV/$c$, corresponding to $\sqrt{s}$ values ranging from 8.6
to 31.1 GeV.

\begin{table*}[!ht]   %\footnotesize
\setlength\tabcolsep{3pt}
\addtolength{\tabcolsep}{3pt}
\centering
%\begin{center}
\begin{tabular}{|c|c|c|c|c|c|r|r|c|}
\hline
\hline
 Experiment & Beam & $P_{beam}$ (GeV/$c$) & Target & Data & $x_F$ & ndf & Norma.$^{a}$ & Ref. \\
\hline
\hline
FNAL E672, E706 & $\pi$ & 515  & Be & $\sigma^{J/\psi}$ & [0.11, 0.79] & 35 & 12.0 & \cite{jpsi_data1} \\
FNAL E705  & $\pi$ & 300  & Li & $\sigma^{J/\psi}$ & [-0.10, 0.45] & 12 & 9.5 & \cite{jpsi_data2and3} \\
CERN NA3$^{b}$ & $\pi$ & 280  & p  & $\sigma^{J/\psi}$ & [0.025, 0.825] & 17 & 13.0 & \cite{jpsi_data17and18and21} \\
CERN NA3$^{b}$  & $\pi$ & 200  & p  & $\sigma^{J/\psi}$ & [0.05, 0.75] & 8 & 13.0 & \cite{jpsi_data17and18and21} \\
CERN WA11$^{b}$ & $\pi$ & 190  & Be  & $\sigma^{J/\psi}$ & [-0.35, 0.75] & 12 & $^{c}$10.0 & \cite{jpsi_data16} \\
CERN NA3$^{b}$ & $\pi$ & 150  & p  & $\sigma^{J/\psi}$ & [0.025, 0.925] & 19 & 13.0  & \cite{jpsi_data17and18and21} \\
FNAL E537 & $\pi$ & 125  & Be  & $\sigma^{J/\psi}$ & [0.05, 0.95] & 10 & 6.0 & \cite{jpsi_data6and7and8} \\
CERN WA39$^{b}$ & $\pi$ & 39.5 & p  & $\sigma^{J/\psi}$ & [0.05, 0.85] & 9 & 15.0 & \cite{jpsi_data9and10} \\
FNAL E672, E706 & $\pi$ & 515  & Be & $\sigma^{\psi(2S)}$ & [0.17, 0.73] & 5 & 16.0 & \cite{jpsi_data1} \\
FNAL E615 & $\pi$ & 253  & W & $\sigma^{\psi(2S)}/\sigma^{J/\psi}$ & [0.275, 0.975] & 15 &  & \cite{Heinrich:1991zm} \\
HERA-B &  p & 920  & W & $\sigma^{\psi(2S)}/\sigma^{J/\psi}$ & [-0.3, 0.075] & 8 &  & \cite{HERA-B:2006bhy} \\
CERN NA50 &  p & 450  & W & $\sigma^{\psi(2S)}/\sigma^{J/\psi}$ & [-0.075, 0.075] & 4 &  & \cite{NA50:2003fvu} \\
FNAL E789 &  p & 800  & Au & $\sigma^{\psi(2S)}/\sigma^{J/\psi}$ & [0.00, 0.12] & 5 &  & \cite{E789:1995yhd} \\
FNAL E771 &  p & 800  & Si & $\sigma^{\psi(2S)}/\sigma^{J/\psi}$ & [0.00, 0.20] & 6 &  & \cite{E771:1995ane} \\
FNAL E705  & p & 300  & Li & $\sigma^{J/\psi}$ & [-0.10, 0.45] & 12 & 10.1 & \cite{jpsi_data2and3} \\
CERN NA3$^{b}$  & p & 200  & p  & $\sigma^{J/\psi}$ & [0.05, 0.75] & 8 & 13.0 & \cite{jpsi_data17and18and21} \\
\hline
\end{tabular}
%\end{center}
\caption {Differential cross sections datasets for charmonium
  production [$J/\psi$, $\psi(2S)$ and $R_{\psi}(x_F)$] used in the study,
  listed in order of decreasing beam momentum.\\ $^{a}$Percentage of
  uncertainty in the cross section normalization.\\ $^{b}$The
  numerical information was extracted from the published
  figures. \\ $^{c}$Information not available but an educated guess .}
\label{tab:data}
\end{table*}

%%%%%%%%%%%%%%%%%%%%%%
\section{Results of NRQCD calculations}
\label{sec:results}
%%%%%%%%%%%%%%%%%%%%%%

%%%%%%%%%%%%%%%%%%%%%%
\subsection{Reference NRQCD calculations}
%%%%%%%%%%%%%%%%%%%%%%

%%%%%%%%%%%%%%%%%%%%%%%%%%%%%%%%%%%%%%%%%%%%%%%%%%%%%%%%%%%%%%%%%%%%%%%%%%%
% exe sfitall2#scandump_table3_new2 for 1 (SMRS), 2 (GRV), 16 (JAM21) and 4 (xFitter)
% exe sfitall2#scandump_table3_new3 for 1 (SMRS), 2 (GRV), 10 (JAM21) and 4 (xFitter)
% exe sfitall2#scandump_table3_new4
% exe sfitall2#scandump_table3_new5

\begin{table*}[!ht]   %\footnotesize
\setlength\tabcolsep{3pt}
\addtolength{\tabcolsep}{3pt}
\centering
%\begin{center}
\begin{tabular}{|c|r|r||r|r||r|r||r|r|}
\hline
& \multicolumn{2}{c||}{SMRS} & \multicolumn{2}{c||}{GRV} & \multicolumn{2}{c||}{JAM} & \multicolumn{2}{c|}{xFitter} \\ 
\hline
 & REF & FIT & REF & FIT & REF & FIT & REF & FIT \\
\hline
 $\chi^2_{total}/\text{ndf}$  &   5.7 &   1.9 &   7.0 &   2.4 &  17.7 &   5.6 &  14.3 &   4.2 \\ 
 $\chi^2/\text{ndp}|^{\pi^-}_{x_F}$ &   5.3 &   1.8 &   7.6 &   2.4 &  25.5 &   5.9 &  19.5 &   4.5 \\ 
 $\chi^2/\text{ndp}|^{p}_{x_F}$ &  10.7 &   1.6 &  10.5 &   1.7 &  11.2 &   2.7 &  11.5 &   1.9 \\ 
 $\chi^2/\text{ndp}|^{\pi^-}_{\sqrt{s}}$ &   2.1 &   8.7 &   2.9 &   5.6 &   5.3 &  11.4 &   4.8 &   4.4 \\ 
 $\chi^2/\text{ndp}|^{p}_{\sqrt{s}}$ &   3.8 &   8.1 &   3.4 &   8.1 &   3.5 &   5.1 &   3.6 &   6.9 \\ 
 $\langle \mathcal{O}_{8}^{J/\psi}[^{3}S_{1}] \rangle$ & 0.0690 & 0.0259$\pm$0.0023 & 0.0950 & 0.0432$\pm$0.0038 & 0.0830 & 0.1192$\pm$0.0021 & 0.0740 & 0.0849$\pm$0.0041 \\ 
 $\Delta_8^{J/\psi}$ & 0.0250 & 0.0560$\pm$0.0016 & 0.0180 & 0.0521$\pm$0.0017 & 0.0200 & 0.0244$\pm$0.0016 & 0.0220 & 0.0393$\pm$0.0034 \\ 
 $\langle \mathcal{O}_{8}^{\psi(2S)}[^{3}S_{1}] \rangle$ & 0.0210 & 0.0132$\pm$0.0009 & 0.0260 & 0.0210$\pm$0.0013 & 0.0260 & 0.0237$\pm$0.0009 & 0.0230 & 0.0186$\pm$0.0012 \\ 
 $\Delta_8^{\psi(2S)}$ & 0.0017 & 0.0057$\pm$0.0003 & 0.0004 & 0.0042$\pm$0.0003 & 0.0004 & 0.0021$\pm$0.0003 & 0.0009 & 0.0040$\pm$0.0006 \\ 
\hline
\end{tabular}
%\end{center}
\caption{Results of the NRQCD calculation using the reference values
  of the LDMEs (columns labeled ``REF'') and of the fit of the LDMEs
  to the differential cross sections (columns ``FIT'').  The upper
  part of the table gives the values of the reduced
  $\chi^2/\text{ndf}$ of the entire dataset and the $\chi^2$ divided
  by the number of data point (ndp) for the pion-induced and
  proton-induced datasets separately. The subscript $x_F$ or $\sqrt
  s$ for $\chi^2/\text{ndp}$ refers to $x_F$-dependent data or $\sqrt
  s$-dependent $x_F$-integrated data. The lower part of the table
  displays the values of the reference and fitted LDMEs for SMRS, GRV,
  JAM and xFitter pion PDFs. All LDMEs are in units of $\rm{GeV}^3$.}
\label{tab:LDME_PDF}
\end{table*}

Before performing a fit to the data listed in Table~\ref{tab:data} to
obtain the best-fit LDMEs for the four pion PDFs, we first carry out
NRQCD calculations using the LDMEs found in a recent
study~\cite{Hsieh:2021yzg}, where only the pion and proton total cross
section data were fitted. The values of the LDMEs, obtained separately
for each pion PDF, are listed in Table~\ref{tab:LDME_PDF}. We then
compare the results of the NRQCD calculations for the $x_F$ dependent
charmonium production cross sections with the data listed in
Table~\ref{tab:data} and shown in Fig.~\ref{fig_jpsi_PDF1}. We call
these ``Reference NRQCD calculation" (REF), which provides the
reference information to be compared with that obtained later from a
fit to the $x_F$-dependent cross section data. Note that
Fig.~\ref{fig_jpsi_PDF1} is for the SMRS pion PDFs, and similar
figures for the other three pion PDFs can be found in the Supplemental
Material~\cite{Supplement}.

The total $\chi^2$/ndf, as well as the $\chi^2$/ndp (ndp denotes
``number of data points") for individual pion or proton datasets, are
listed in Table~\ref{tab:LDME_PDF} under the label ``REF".
Table~\ref{tab:LDME_PDF} shows that the reduced $\chi^2$/ndf for
``REF" are quite large, suggesting that the LDMEs deduced from the fit
to total cross section data are not optimal for describing the
$x_F$-dependent data. A further investigation shows that a significant
contribution to the overall $\chi^2$ comes from the absolute
normalization of the measured cross sections relative to the NRQCD
calculations. Despite the poor agreement between the data and the
calculation, it is interesting to note that calculations using the
SMRS and GRV pion PDFs are in a better agreement with the data than
the JAM and xFitter PDFs.

%%%%%%%%%%%%%%%%%%%%%%
\subsection{NRQCD fits}
%%%%%%%%%%%%%%%%%%%%%%

We now proceed to a refined determination of the color-octet $\langle
\mathcal{O}_{8}^{H}[^{3}S_{1}] \rangle$ and $ \Delta_8^{H}$ LDMEs for
$J/\psi$ and $\psi(2S)$ production by fitting the $x_F$ differential
cross sections and $R_{\psi}(x_F)$ ratios for proton and pion
beams. To avoid double counting, total cross sections data that result
from an integration over the associated differential cross sections
are not included in the fit. We note that the NRQCD calculations do
not require a normalization factor, as they predict absolute cross
sections. However, the experimental $x_F$-dependent $J/\psi$ and
$\psi(2S)$ cross sections are associated with experimental
normalization uncertainties $\delta_\sigma$, as quoted in
Table~\ref{tab:data}. An attempt to fit the data without taking into
account the normalization uncertainties only marginally reduces the
total $\chi^2/\text{ndf}$. In order to take into account these
uncertainties, a normalization parameter $F$ is added for each of the
$x_F$-differential datasets. Accordingly, a penalty term of
$((F-1)/\delta_\sigma)^2$ is included in the calculation of the
overall $\chi^2$. To avoid unrealistic values of $F$, we limit the
deviation of $F$ from 1.0 to be less than 2 $\delta_\sigma$. The
results of this approach are labeled as ``FIT'' below.

\begin{figure*}[!ht]
\centering \includegraphics[width=1.8\columnwidth]{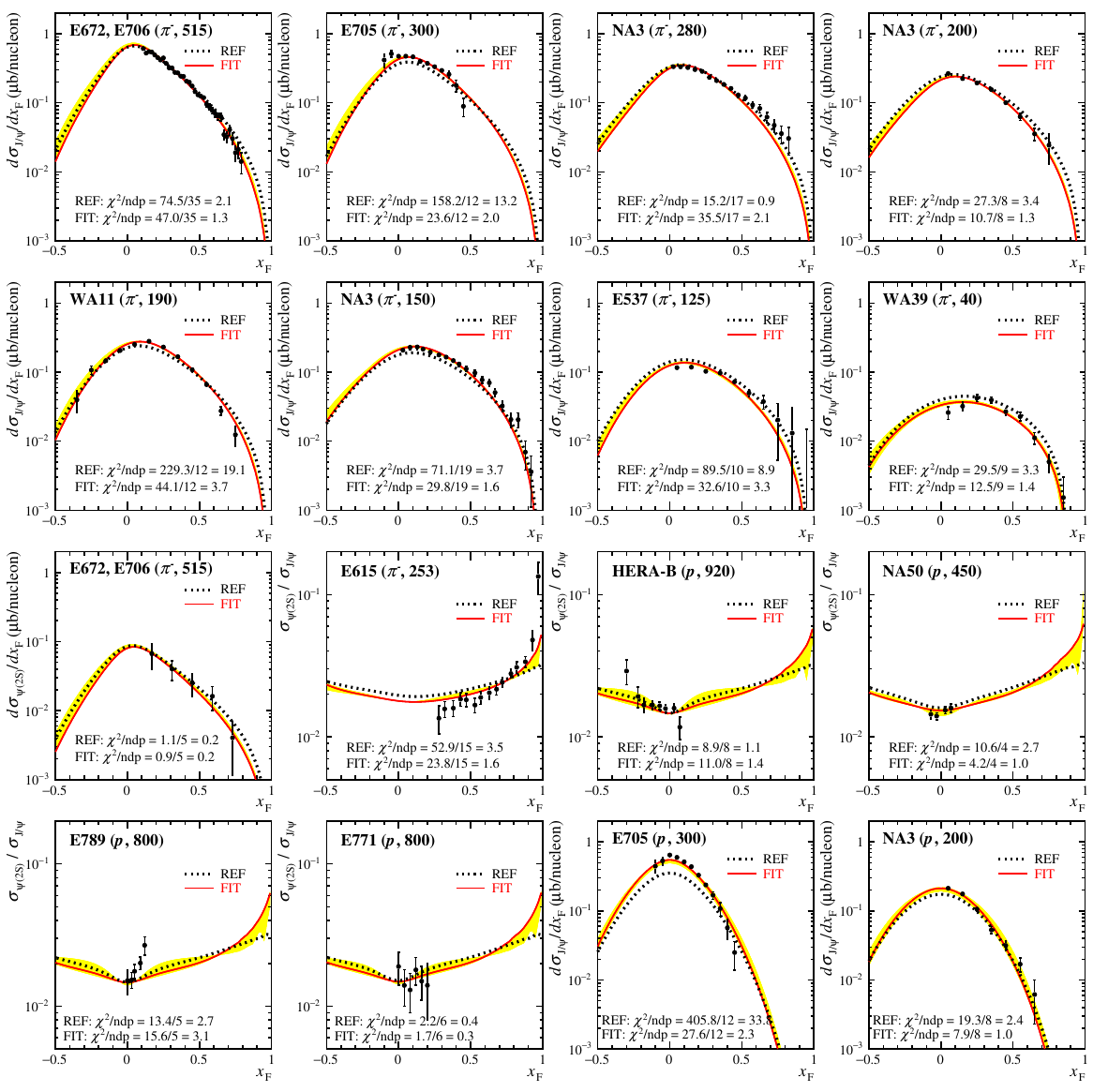}
\caption[\protect{}]{The $x_F$-dependent cross sections for $J/\psi$
  and $\psi(2S)$ production and $R_{\psi}(x_F)$ ratios in $\pi^- N$
  and $pN$ interactions, following the order given in
  Table.~\ref{tab:data}. The symbol and value in parenthesis denote
  the particle type and momentum of beam. The solid red and dotted
  black curves represent the NRQCD results of SMRS pion PDFs from the
  fit described in the text (``FIT'') and from the calculation using
  the LDMEs obtained in Ref.~\cite{Hsieh:2021yzg} (``REF''),
  respectively. The values of $\chi^2$ divided by the number of data
  point (ndp) for each dataset are also shown. The yellow bands
  represent the cross section uncertainties associated with the scale
  and charm quark mass systematic variations.}
\label{fig_jpsi_PDF1}
\end{figure*}

%%%%%%%%
% Differential xF plots
%%%%%%%%

Figure~\ref{fig_jpsi_PDF1} shows the new fit to the data for
$x_F$-differential data and ratios using the SMRS pion PDFs. The
newly-determined LDMEs parameters are shown in
Table~\ref{tab:LDME_PDF}. Except for the $J/\psi$ data of WA11, the
new NRQCD fit provides a reasonably good description of data for both
pion and proton beams. Table~\ref{tab:LDME_PDF} shows that for all
four pion PDFs and for nearly all datasets the individual
$\chi^2/\text{ndp}$ are significantly improved. The displayed yellow
uncertainty bands result from the scale and charm mass variations of
charm quark mass $m_c$ of 1.4 and 1.6 GeV/$c^2$ at $\mu_F = \mu_R = 2
m_c$, and $\mu_F = \mu_R =$ = 1 and 4 $m_c$ at $m_c$ = 1.5
GeV/$c^2$. The uncertainty is evaluated by the square root of
  the sum of squares of the cross section difference due to the
  individual variation. The corresponding LDMEs are obtained from a
new global fit for each configuration. The uncertainty bands are
relatively small and do not introduce an essential change in the
quality of data description. The systematic studies are further
discussed in Sec.~\ref{sec:systematic}. Similar figures for GRV, JAM
and xFitter are available in the Supplemental
Material~\cite{Supplement}.

Table~\ref{tab:LDME_PDF} also lists the $\chi^2$ values for both the
``REF'' and ``FIT'' calculations. The $\chi^2/\text{ndp}$ and the
fitted normalization factors for each dataset are summarized in
Table~\ref{tab:fit_xf}. The improved description of the differential
cross sections is also confirmed by the overall
$\chi^2_{total}/\text{ndf}$ and the $\chi^2/\text{ndp}$ values for
various datasets. The $\chi^2$/ndp of the pion-induced $x_F$ data
sets, $\chi^2/\text{ndp}|^\pi_{x_F}$, are 1.8, 2.4, 5.9 and 4.5 for
the SMRS, GRV, JAM and xFitter PDFs, respectively, an improvement of
about a factor of three over that of ``REF''. As expected, the
$\chi^2$/ndp of the proton-induced $x_F$ datasets,
$\chi^2/\text{ndp}|^p_{x_F}$, are of similar values, around 2.0 for
all four pion PDFs. In contrast, the $\chi^2/\text{ndp}$ of the
integrated cross sections ($\chi^2/\text{ndp}|^{\pi^-,p}_{\sqrt{s}}$ )
are now larger since these data are not included in the global fit.

Table~\ref{tab:LDME_PDF} also shows the newly fitted LDMEs. In
comparison with the ``REF'' calculation, the ``FIT'' results give
smaller $\langle \mathcal{O}_{8}^{H}[^{3}S_{1}] \rangle$ values for
both SMRS and GRV PDFs, while the corresponding $\Delta_8^{H}$ LDMEs
are slightly larger.  For the JAM and xFitter PDFs the ``REF'' and
``FIT'' LDMEs remain consistent within their uncertainties. The
$\chi^2/\text{ndp}|^{p}_{x_F, \sqrt{s}}$ have a mild dependence on the
pion PDFs, only through the correlation of LDMEs and PDFs in the
global fit.

In NRQCD, the relative weighting between $q \bar{q}$ and $GG$
subprocesses is set by a convolution of the pQCD partonic cross
sections, the associated parton densities, and the LDMEs. The $F$
factor does not modify the shape of $d\sigma/d x_F$. Therefore,
adequate shapes of $d\sigma/d x_F$ distributions of individual $GG$
and $q \bar{q}$ contributions from NRQCD calculations are required to
achieve a reasonable description of the data, particularly for
$x_F>0.5$. Since the partonic cross sections and the nucleon PDFs
involved in the calculations of the cross sections remain the same,
the variation of the results originates from the difference in the
pion PDFs and the LDMEs.

%%%%%%%%%%%%%%%%%%%%%%%%%%%%%%%%%%%%%%%%%%%%%%%%%%%%%%%%%%%%%%%%%%%%%%%%%%%
% exe sfitall2#scandump_table5 ifit=5
\begin{table*}[!ht]   %\footnotesize
\setlength\tabcolsep{3pt}
\addtolength{\tabcolsep}{3pt}
\centering
\begin{tabular}{|c|c|c||c|c||c|c||c|c|}
\hline
Data  & \multicolumn{2}{|c||}{SMRS} & \multicolumn{2}{|c||}{GRV} & \multicolumn{2}{|c||}{JAM} & \multicolumn{2}{|c|}{xFitter} \\
\hline
Exp & $\chi^2$/\text{ndp} & $F$  & $\chi^2$/\text{ndp} & $F$ & $\chi^2$/\text{ndp} & $F$ & $\chi^2$/\text{ndp} & $F$ \\
\hline
 E672, E706 ($\sigma^{J/\psi}$) &   1.3 &  0.80 $\pm$  0.01 &   2.6 &  0.79 $\pm$  0.01 &   6.1 &  1.14 $\pm$  0.01 &   4.2 &  1.08 $\pm$  0.02 \\ 
 E705 ($\sigma^{J/\psi}$) &   2.0 &  0.98 $\pm$  0.02 &   1.7 &  0.96 $\pm$  0.02 &   4.1 &  1.19 $\pm$  0.01 &   2.6 &  1.18 $\pm$  0.01 \\ 
 NA3 ($\sigma^{J/\psi}$) &   2.1 &  0.86 $\pm$  0.02 &   2.3 &  0.87 $\pm$  0.02 &   2.7 &  1.00 $\pm$  0.02 &   2.9 &  1.01 $\pm$  0.02 \\ 
 NA3 ($\sigma^{J/\psi}$) &   1.3 &  0.87 $\pm$  0.02 &   0.9 &  0.89 $\pm$  0.02 &   1.8 &  0.92 $\pm$  0.02 &   1.5 &  0.95 $\pm$  0.02 \\ 
 WA11 ($\sigma^{J/\psi}$) &   3.7 &  1.02 $\pm$  0.02 &   8.5 &  1.02 $\pm$  0.02 &  29.9 &  1.09 $\pm$  0.01 &  22.0 &  1.12 $\pm$  0.02 \\ 
 NA3 ($\sigma^{J/\psi}$) &   1.6 &  1.24 $\pm$  0.03 &   1.3 &  1.23 $\pm$  0.03 &   1.5 &  1.10 $\pm$  0.02 &   1.6 &  1.18 $\pm$  0.03 \\ 
 E537 ($\sigma^{J/\psi}$) &   3.3 &  0.88 $\pm$  0.00 &   1.6 &  0.88 $\pm$  0.01 &   2.6 &  0.88 $\pm$  0.00 &   2.1 &  0.88 $\pm$  0.01 \\ 
 WA39 ($\sigma^{J/\psi}$) &   1.4 &  1.30 $\pm$  0.04 &   1.4 &  1.18 $\pm$  0.07 &   2.9 &  0.70 $\pm$  0.00 &   1.3 &  0.70 $\pm$  0.05 \\ 
 E672, E706 ($\sigma^{\psi(2S)}$) &   0.2 &  0.80 $\pm$  0.01 &   0.2 &  0.79 $\pm$  0.01 &   0.3 &  1.14 $\pm$  0.01 &   0.2 &  1.08 $\pm$  0.02 \\ 
 E615 ($\sigma^{\psi(2S)}/\sigma^{J/\psi}$) &   1.6 & 1 $\pm$ 0 &   1.7 & 1 $\pm$ 0 &   5.0 & 1 $\pm$ 0 &   4.3 & 1 $\pm$ 0 \\ 
 HERA-B ($\sigma^{\psi(2S)}/\sigma^{J/\psi}$) &   1.4 & 1 $\pm$ 0 &   1.5 & 1 $\pm$ 0 &   1.2 & 1 $\pm$ 0 &   1.2 & 1 $\pm$ 0 \\ 
 NA50 ($\sigma^{\psi(2S)}/\sigma^{J/\psi}$) &   1.0 & 1 $\pm$ 0 &   1.6 & 1 $\pm$ 0 &   1.3 & 1 $\pm$ 0 &   1.1 & 1 $\pm$ 0 \\ 
 E789 ($\sigma^{\psi(2S)}/\sigma^{J/\psi}$) &   3.1 & 1 $\pm$ 0 &   3.3 & 1 $\pm$ 0 &   2.8 & 1 $\pm$ 0 &   2.9 & 1 $\pm$ 0 \\ 
 E771 ($\sigma^{\psi(2S)}/\sigma^{J/\psi}$) &   0.3 & 1 $\pm$ 0 &   0.3 & 1 $\pm$ 0 &   0.3 & 1 $\pm$ 0 &   0.3 & 1 $\pm$ 0 \\ 
 E705 ($\sigma^{J/\psi}$) &   2.3 &  1.20 $\pm$  0.00 &   2.2 &  1.20 $\pm$  0.00 &   5.7 &  1.20 $\pm$  0.00 &   3.1 &  1.20 $\pm$  0.00 \\ 
 NA3 ($\sigma^{J/\psi}$) &   1.0 &  1.00 $\pm$  0.01 &   1.2 &  1.00 $\pm$  0.01 &   1.9 &  1.00 $\pm$  0.01 &   1.6 &  1.00 $\pm$  0.01 \\ 
\hline
\end{tabular}
\caption {Results of the NRQCD fits. The columns display the
  $\chi^2$/ndp values and fitted normalization factors $F$ for each of
  the selected datasets and for SMRS, GRV, JAM and xFitter pion
  PDFs.}
\label{tab:fit_xf}
\end{table*}

\subsection{Differential cross sections for $J/\psi$}

%%%% Selected datasets of d\sigma/dxF %%%%%%

\begin{figure}[!ht]
\centering
\includegraphics[width=1.0\columnwidth]{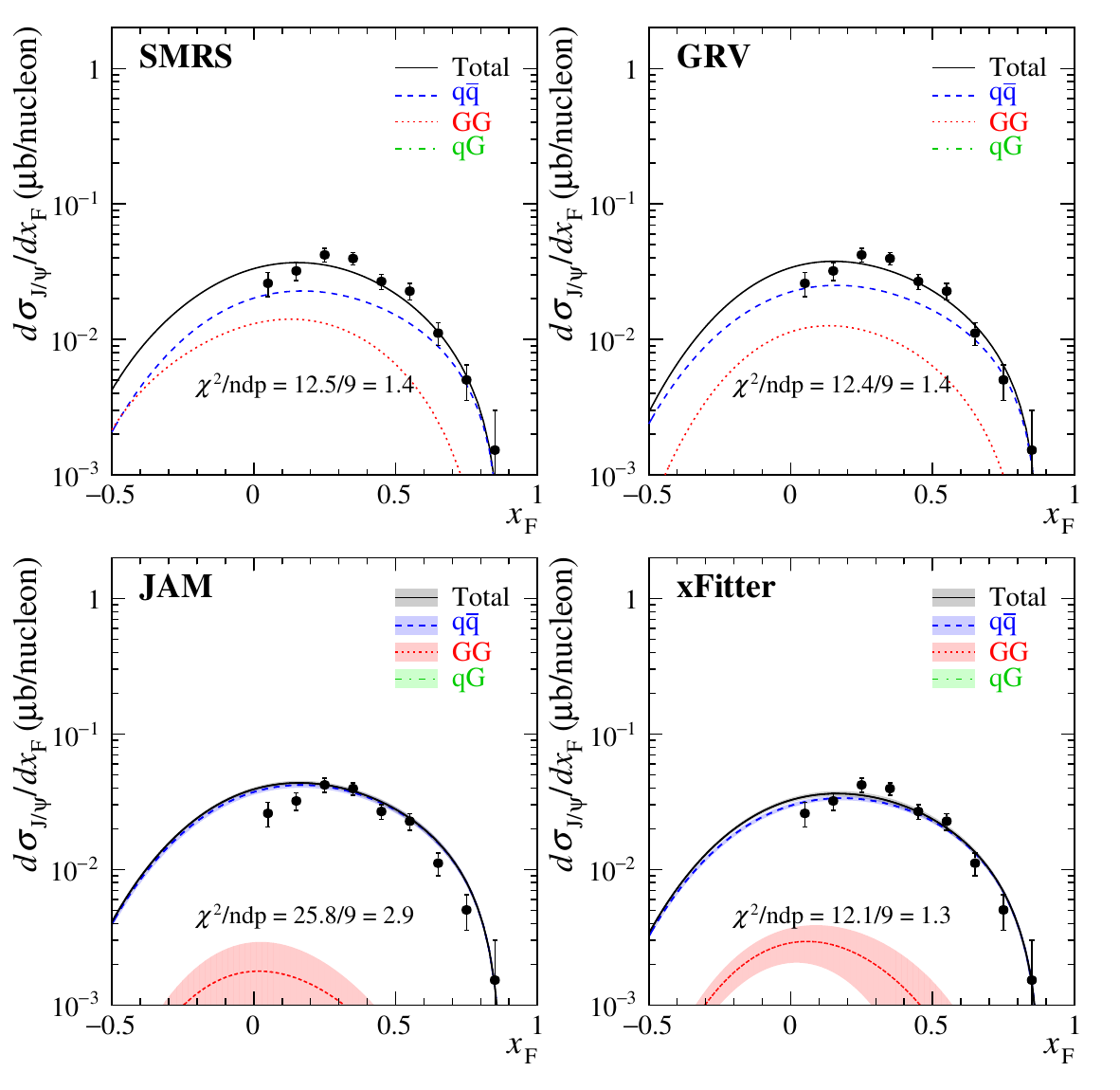}
\caption[\protect{}]{Differential cross sections for $J/\psi$
  production with a 39.5-GeV/$c$ $\pi^-$
  beam~\cite{jpsi_data9and10}. The data are compared to the NRQCD fit
  results for the SMRS, GRV, xFitter, and JAM PDFs. The total cross
  sections and $q \bar{q}$, $GG$, and $qG$ contributions are denoted
  as solid black, dashed blue, dotted red, and dot-dashed green lines,
  respectively. The uncertainty bands associated with JAM and xFitter
  PDFs are also shown.}
\label{fig_jpsi_data08}
\end{figure}

\begin{figure}[!ht]
\centering \includegraphics[width=1.0\columnwidth]{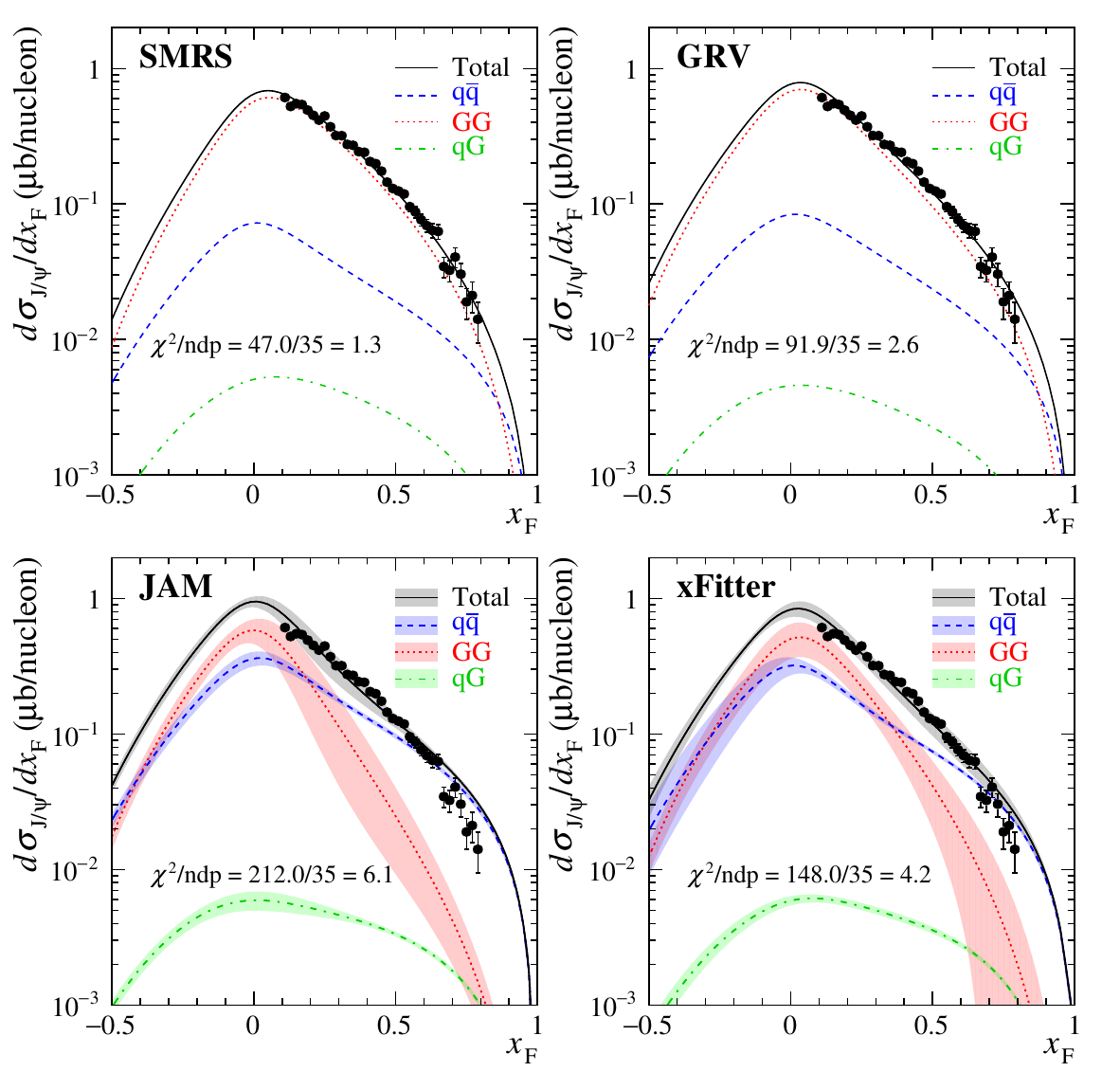}
\caption[\protect{}]{Same as Fig.~\ref{fig_jpsi_data01} for $J/\psi$
  production data with a a 515~GeV/$c$ $\pi^-$
  beam~\cite{jpsi_data1}.}
\label{fig_jpsi_data01}
\end{figure}

A comparison of the $J/\psi$ production data and the NRQCD
calculations in terms of the subprocess contributions has been made
for all of the datasets included in the fit. Irrespective of the pion
PDFs, the relative weighting of $q \bar{q}$ and $GG$ shows a strong
energy dependence. At the lowest energy, the $q \bar{q}$ term provides
the major contribution to the cross section, similar to the DY
production, while the $GG$ contribution is dominant at the highest
beam energies. A global analysis of charmonium datasets with a wide
range of beam energy could simultaneously constrain both pion's
valence quark and gluon distributions. It is instructive to compare
the results obtained with each of the four pion PDFs. This comparison
is illustrated in Figs.~\ref{fig_jpsi_data08} and
~\ref{fig_jpsi_data01} for the data with pion beam momenta of 39.5
GeV/$c$~\cite{jpsi_data9and10} and 515 GeV/$c$~\cite{jpsi_data1}. The
$\chi^2$/ndf values are displayed in the plots.

At the lowest beam momentum of 39.5 GeV/$c$
(Fig.~\ref{fig_jpsi_data08}), the $q \bar{q}$ subprocess provides the
largest contribution to the cross section over the whole $x_F$
region. The $GG$ contribution is much reduced, so that the shape of
the $x_F$ distribution is essentially determined by the shape of the
$q \bar{q}$ contribution. Since the pion valence-quark distribution is
well determined from the DY data, good $\chi^2/\text{ndf}$ values are
obtained for the four PDFs. Nevertheless, the agreement with the data
is less satisfactory for JAM. Figure ~\ref{fig_jpsi_data08} also
suggests that future $J/\Psi$ data at negative $x_F$ with low beam
energies could further constrain the pion valence-quark distribution
at lower $x$.

At the highest beam momentum of 515 GeV/$c$, where the $GG$
contribution becomes dominant, Fig.~\ref{fig_jpsi_data01} shows that
SMRS and GRV are favored over JAM and xFitter. The fraction of the
$GG$ component is maximized around $x_F = 0$, corresponding to the
gluon distribution $G_{\pi}(x)$ around $x \sim$0.1--0.2. As a result
of the rapid drop of the $G_{\pi}(x)$ toward $x=1$, the $GG$
contribution quickly decreases at large $x_F$. In contrast, the $q
\bar{q}$ contribution has a slower fall-off toward high $x_F$ because
of a relatively strong pion valence antiquark density, in comparison
with the gluon one, at large $x$. Consequently, the $q \bar{q}$
contribution has a broader $x_F$ distribution than that of the $GG$
contribution and the relative importance of $q \bar{q}$ rises at the
large $x_F$ region. The ratio of $q \bar{q}$ to $GG$ shows a strong
$x_F$ dependence, making the $x_F$-differential cross sections at high
energies particularly sensitive to the shape of pion $G_{\pi}(x)$.

Similar conclusions can be drawn for the intermediate energies used in
this analysis. The corresponding figures are available in the
Supplemental Material~\cite{Supplement}. As a general observation, the
$q\bar{q}$ and $GG$ contributions have quite similar strengths for the
fits with SMRS and GRV, whereas the $q\bar{q}$ contribution is the
dominant component for the fits with JAM and xFitter. In terms of
$\chi^2$/ndf, the data show a slight preference for GRV and SMRS.

%%%%%%%%%%%%%%%%%%%%%%%%%%%

\subsection{Differential cross sections for $\psi(2S)$ and the $R_{\psi}(x_F)$ ratios}

%%%%%%
%$\psi(2S)$
%%%%%%

\begin{figure}[!ht]
\includegraphics[width=1.0\columnwidth]{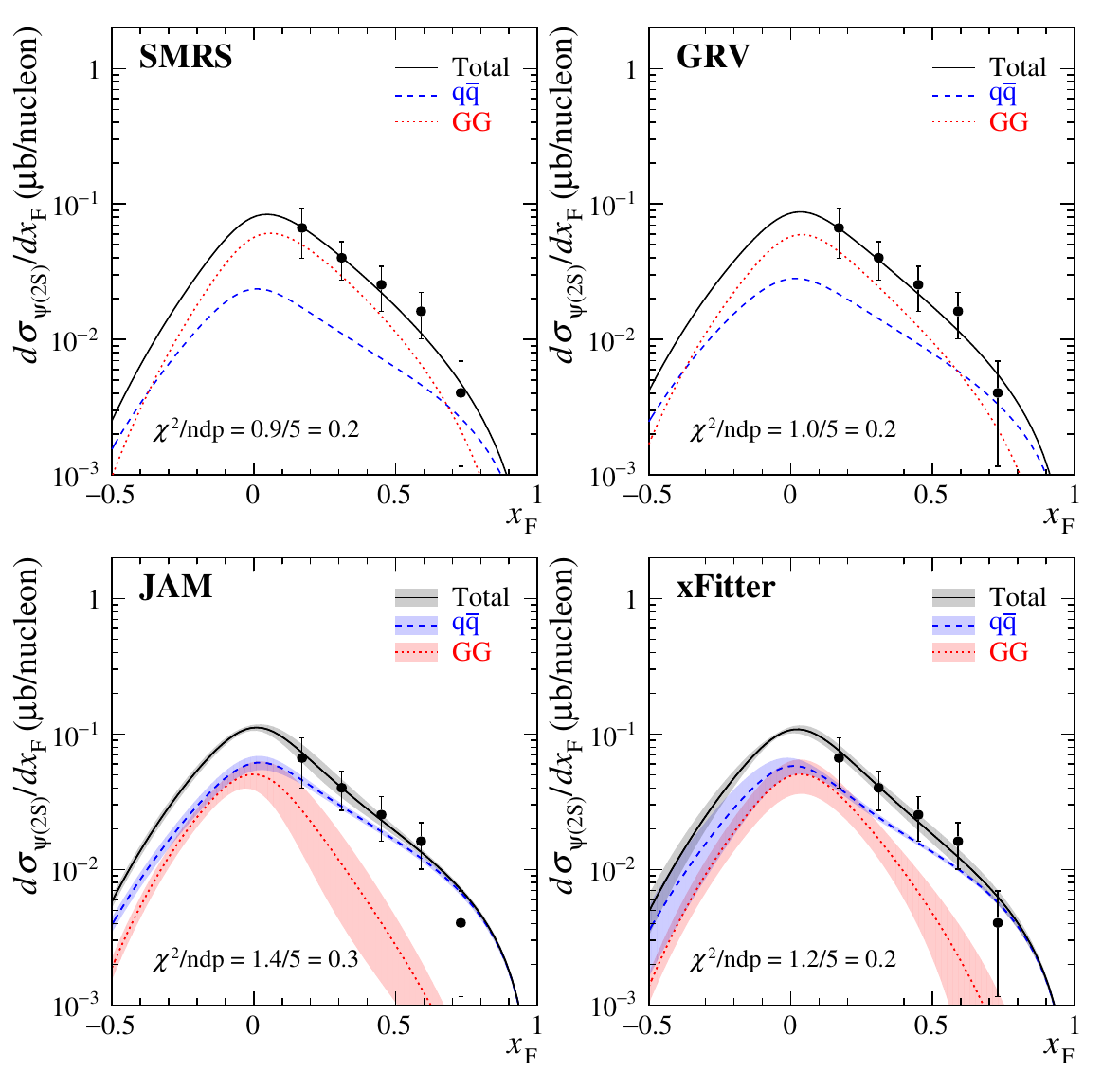}
\caption[\protect{}]{Same as Fig.~\ref{fig_jpsi_data01} for $\psi(2S)$
  production with a 515-GeV/$c$ $\pi^-$ beam~\cite{jpsi_data1}.}
\label{fig_jpsi_data09}
\end{figure}

Additional information on the charmonium production mechanism can be
obtained by comparing the production of the two charmonium states,
$J/\psi$ and $\psi(2S)$. The strengths of their $q\bar{q}$ and $GG$
subprocesses are controlled by the associated LDMEs. In comparison
with the $J/\psi$, the smaller cross section for the $\psi(2S)$
production implies also smaller LDMEs. The fitted LDMEs are indeed
smaller, but interestingly, not in the same proportion. As shown in
Table~\ref{tab:LDME_PDF}, the values of the $\langle
\mathcal{O}_{8}^{\psi(2S)}[^{3}S_{1}] \rangle$ LDMEs for $\psi(2S)$
are smaller than that for $J/\psi$ by about a factor of two. In
contrast, the $\Delta_8^{\psi(2S)}$ values for $\psi(2S)$ are an order
of magnitude smaller. This is illustrated in
Fig.~\ref{fig_jpsi_data09} for the E672/E706 $\psi(2S)$ data taken at
515 GeV/$c$~\cite{jpsi_data1}. In comparison with the production of
$J/\psi$ at the same energy (Fig.~\ref{fig_jpsi_data01}), the
$q\bar{q}$ contribution is greatly enhanced in $\psi(2S)$
production. Figure~\ref{fig_jpsi_data09} shows that this observation
is valid for all pion PDFs, and the $q\bar{q}$ term is even dominant
for JAM and xFitter. For the fit with the SMRS pion PDFs around
$x_F=0$, the $q\bar{q}$ component accounts for about 15\% of the
direct part (feed-down excluded) of the $J/\psi$ cross section.  Its
fraction rises to nearly 30\% for $\psi(2S)$. Obviously, the increase
of the $q\bar{q}$ term is compensated by a decrease of the $GG$
term. This significant difference between the two charmonium states
can only be partially explained by the larger $\psi(2S)$ mass. Its
full understanding would require further investigations.

\begin{figure}[!ht]
\includegraphics[width=1.0\columnwidth]{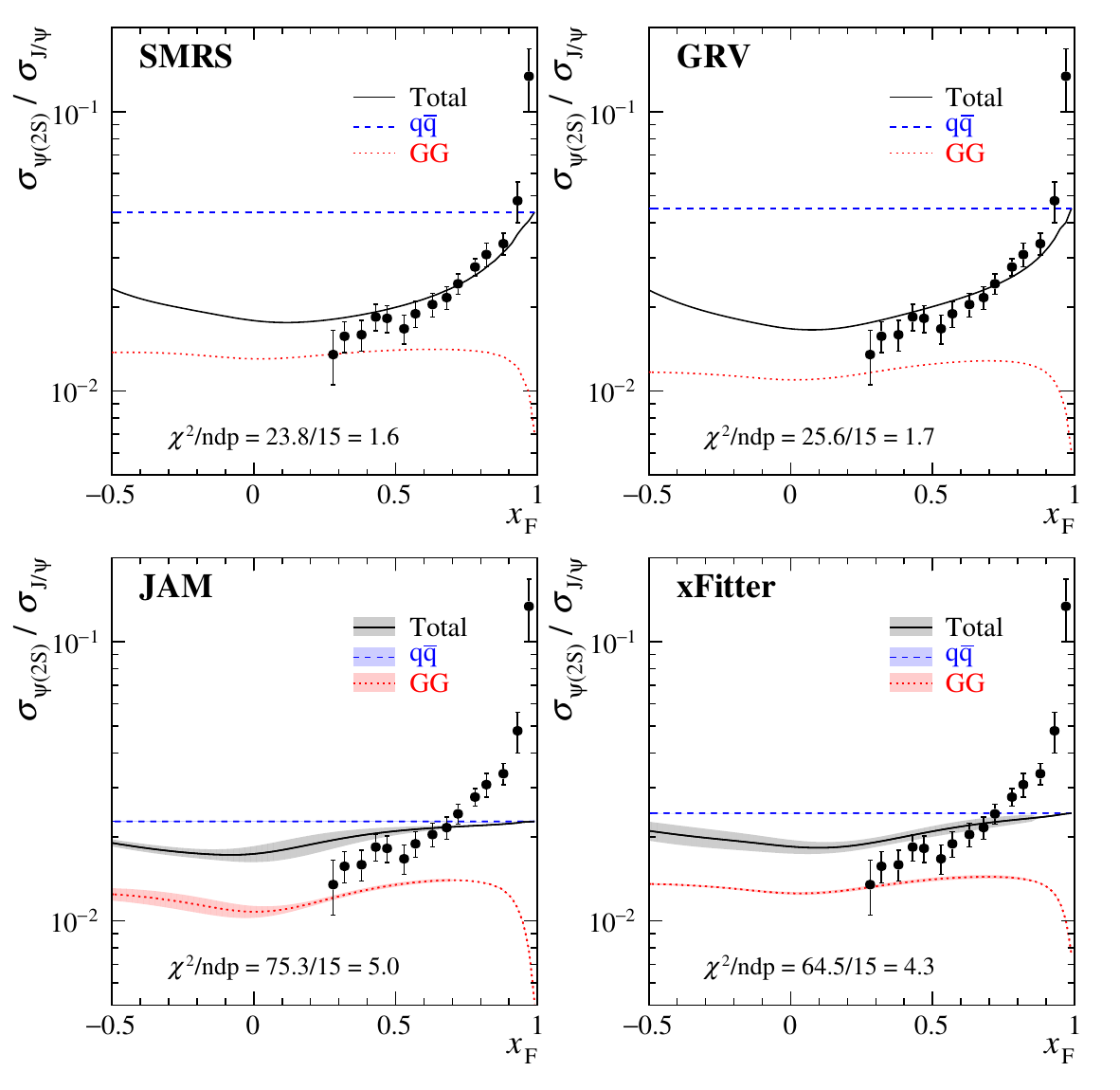}
\caption[\protect{}]{The $\psi(2S)$ to $J/\psi$ cross section ratios
  $R_{\psi}(x_F)$ for $J/\psi$ and $\psi(2S)$ production with a
  252-GeV/$c$ $\pi^-$ beam~\cite{Heinrich:1991zm}. The data are
  compared to the NRQCD fit results for the SMRS, GRV, xFitter, and
  JAM PDFs. The ratios of total cross sections and individual $R^{q
    \bar q}_\psi (x_F)$ and $R^{GG}_\psi (x_F)$ contributions are
  denoted as solid black, dashed blue, and dotted red lines,
  respectively.}
\label{fig_jpsi_data10}
\end{figure}

The observations above are consistent with the measurements of the
$\psi(2S)$ to $J/\psi$ ratios, $R_{\psi}(x_F)$. The largest statistics
on $R_{\psi}(x_F)$ have been collected by the E615 experiment for an
incident pion momentum of 252 GeV/$c$~\cite{Heinrich:1991zm}. The data
are compared to the NRQCD fits with each of the four pion PDFs in
Fig.~\ref{fig_jpsi_data10}. The $R_{\psi}(x_F)$ shows a strong $x_F$
dependence and this suggests that the relative weights of the
individual subprocesses $q \bar{q}$ and $GG$ components in $J/\psi$
and $\psi(2S)$ production are distinctly different. We note that the
CEM models predicts an $x_F$-independent
$R_{\psi}(x_F)$~\cite{HERA-B:2006bhy}, since the fractions of $q
\bar{q}$ and $GG$ components are identical for each charmonium
state. In NRQCD, an $x_F$-dependent $R_{\psi}(x_F)$ is possible due to
the different LDMEs associated with the $q \bar{q}$ and $GG$ channels
in producing $J/\psi$ and $\psi(2S)$. The pronounced $x_F$ dependence
of $R_\psi(x_F)$ in Fig.~\ref{fig_jpsi_data10} clearly disfavors the
CEM model.

As shown in Figs.~\ref{fig_jpsi_data01} and~\ref{fig_jpsi_data09}, the
$q \bar q$ subprocess gives a significantly broader $x_F$ distribution
than the $GG$ subprocess. This is caused by the slower fall-off of the
valence-quark distribution than the gluon distribution toward
$x=1$. Therefore, the pronounced rise in the $R_\psi(x_F)$ data at
forward $x_F$, shown in Fig.~\ref{fig_jpsi_data10}, clearly indicates
that the $ q \bar q$ subprocess is more important for the $\psi(2S)$
production than for the $J/\psi$ production.

It is also instructive to examine the $x_F$ dependence of $R_\psi
(x_F)$ from the $q \bar q$ and $GG$ subprocesses separately. In
Fig.~\ref{fig_jpsi_data10}, the dashed blue and dotted red curves
correspond, respectively, to
\begin{equation}
R^{q \bar q}_\psi (x_F) \equiv \frac {\sigma^{q \bar q}_{\psi(2S)} (x_F)}
{\sigma^{q \bar q}_{J/\psi} (x_F)};~~
R^{GG}_\psi (x_F) \equiv \frac {\sigma^{GG}_{\psi(2S)} (x_F)}
{\sigma^{GG}_{J/\psi} (x_F)},
\end{equation}
where the superscripts $q \bar q$ and $GG$ denote the two
subprocesses. Neglecting the tiny contribution from the $q G$
subprocess, one can then obtain
\begin{align}
R_\psi(x_F) \equiv & \frac {\sigma_{\psi(2S)} (x_F)} {\sigma_{J/\psi} (x_F)} \\ \nonumber
= & [A(x_F) R^{q \bar q}_\psi (x_F) + B(x_F) R^{GG}_\psi (x_F)],
\end{align}
where
\begin{equation}
A(x_F)=\frac{\sigma^{q \bar q}_{J/\psi}(x_F)}{\sigma_{J/\psi}(x_F)} \mbox{~~and~~} 
B(x_F)=\frac{\sigma^{GG}_{J/\psi}(x_F)}{\sigma_{J/\psi}(x_F)} \\ \nonumber
\end{equation}
have the property $0 \leq A(x_F) \leq 1$ and $0 \leq B(x_F) \leq
1$. It follows that $R_\psi(x_F)$ must be bounded by
$R^{GG}_\psi(x_F)$ and $R^{q \bar q}_\psi(x_F)$ in
Fig.~\ref{fig_jpsi_data10}. As shown in Fig.~\ref{fig_jpsi_data10},
the $R_\psi(x_F)$ data largely fall within these two bounds for
calculations with the SMRS and GRV PDFs, while a large fraction of the
data are outside of these bounds for the calculations using the JAM
and xFitter PDFs. The striking contrast between the SMRS/GRV and the
JAM/xFitter PDFs in their ability to describe the $R_\psi (x_F)$ data
in Fig.~\ref{fig_jpsi_data10} illustrates the advantages of the
$R_\psi (x_F)$ data in constraining the pion PDFs. We also note that
none of the pion PDFs can explain the sharp rise of the
$R_{\psi}(x_F)$ data beyond $x_F = 0.8$. This incompatibility at large
$x_F$ could be due to either higher-twist
effects~\cite{Heinrich:1991zm} or higher-order QCD processes that are
beyond the present leading-order NRQCD analysis.

%%%%%%%%%%%
%\subsection{Decomposition of CS and CO components}

\begin{figure}[!ht]
\centering
\includegraphics[width=1.0\columnwidth]{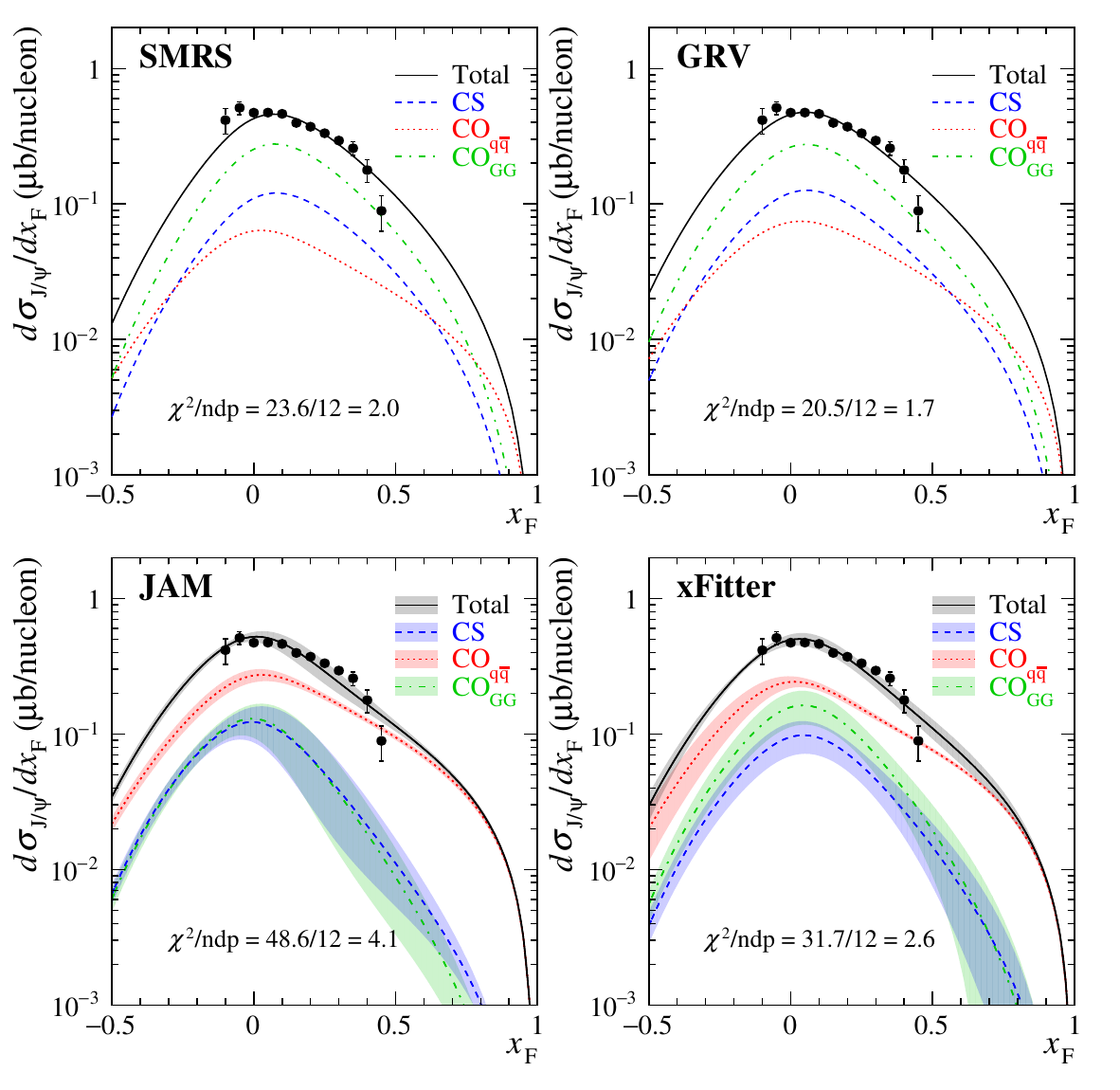}
\caption[\protect{}]{Differential cross sections for $J/\psi$
  production with a 300~GeV/$c$ $\pi^-$
  beam~\cite{jpsi_data2and3}. The data are compared to the fit results
  with SMRS, GRV, xFitter, and JAM PDFs. The total cross section
    and its decomposition into contributions from CS, CO $q \bar q$
    and CO $GG$ subprocesses are denoted as solid black, dashed blue,
    dotted red and dot-dashed green lines, respectively.}
\label{fig_jpsi_data02_2}
\end{figure}

Our analysis also shows that fixed-target charmonium production data
are particularly sensitive to the color octet contribution to the
cross section. This is illustrated in Fig.~\ref{fig_jpsi_data02_2}
which displays the decomposition of the $J/\psi$ $x_F$-dependent cross
sections from the E705 experiment~\cite{jpsi_data2and3} into color
octet and color singlet contributions. The CO contribution plays a
dominant role in the $J/\psi$ production across the entire $x_F$
range, and this observation is valid for any of the four pion
PDFs. Further information can be obtained by separating the CO
contribution into $GG$ and $q\bar q$ components. Only the CO $GG$
component, controlled by the $\Delta_8^{H}$ LDME, is displayed. For
the SMRS and GRV pion PDFs it provides the largest part of the CO
contribution. In contrast, its relative magnitude is significantly
reduced for the JAM and xFitter PDFs, an observation that is in line
with their smaller gluon distributions.

%The enhancement of importance of CO-$q \bar{q}$ contributions toward
%large $x_F$ is commonly observed for all four pion PDFs.

\subsection{Integrated cross sections}

%%%%%%%%%%%%%%%%%%%%%%%%%%%%%%%%%%%%%%%%%%%%%%%%%%%%%%%%%%%%%%%%%%%%%%

\begin{figure*}[!ht]
\centering
\includegraphics[width=1.5\columnwidth]{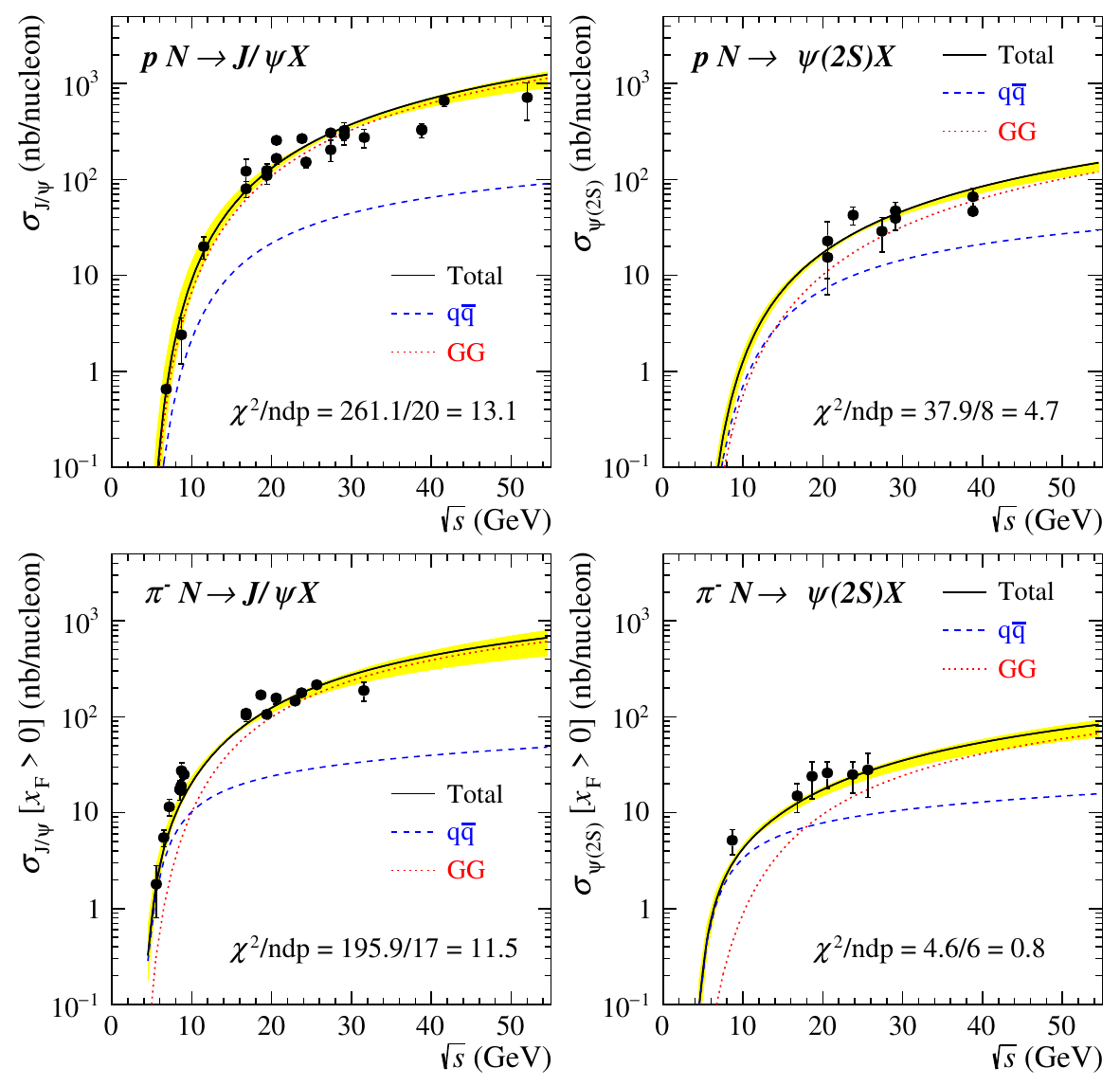}
\caption[\protect{}] {Integrated charmonium cross sections in $pN$ and
  $\pi^- N$ collisions. The data for $J/\psi$ and $\psi(2S)$
  production are compared to the results of NRQCD calculations with
  the SMRS pion PDFs and the ``FIT'' LDMEs in
  Table~\ref{tab:LDME_PDF}. The total cross section and its $q
  \bar{q}$ and $GG$ contributions are denoted as solid black, dashed
  blue and dotted red lines, respectively. The yellow bands represent
  the cross section uncertainties associated with the scale and charm
  quark mass systematic variations.}
\label{fig_sdep_SMRS}
\end{figure*}

Because of the presence of valence antiquarks in the pion, the $q
\bar{q}$ and $GG$ subprocesses to the $J/\psi$ production with proton
and pion beams have different contributions to the integrated cross
sections. In the production with a proton beam the $GG$ contribution
is dominant across all center-of-mass energies $\sqrt{s}$ except near
threshold. With pion beams the $q \bar{q}$ contribution is
significantly enhanced. It dominates at low energies, with $GG$
contribution gradually becoming important as $\sqrt{s}$ increases.

Our analysis of the differential cross sections shows that the
relative contributions of the $q \bar{q}$ and $GG$ subprocesses in the
production of $J/\psi$ and $\psi(2S)$ differ considerably. The same
conclusion can be drawn from the integrated cross
sections. Figure~\ref{fig_sdep_SMRS} shows the comparison of data and
NRQCD calculations for the $J/\psi$ and $\psi(2S)$ production cross
sections in $pN$ and $\pi^-N$ collisions with the SMRS pion PDFs and
the ``FIT'' LDMEs in Table~\ref{tab:LDME_PDF}. The fractions of $q
\bar{q}$ and $GG$ contributions as a function of $\sqrt{s}$ vary
considerably, reflecting the differences of the corresponding gluon
and quark parton distributions between the pion PDFs. For SMRS, whose
gluon strength at large $x$ is relatively strong, the $GG$
contribution starts to dominate the cross section beyond $\sqrt{s}=$18
and 10 GeV for the production of $\psi(2S)$ and $J/\psi$,
respectively, while the transition happens at larger $\sqrt{s}$ for
the results with JAM, in consequence of a relatively weak gluon
strength. The uncertainty bands estimated in the same fashion as in
Fig.~\ref{fig_jpsi_PDF1} are displayed. The plots for GRV, JAM and
xFitter pion PDFs are provided in the Supplemental
Material~\cite{Supplement}. All these observations confirm our
previous conclusion: the $q \bar{q}$ contribution plays a much more
important role in the $\psi(2S)$ production, compared to $J/\psi$.

\subsection{Systematic studies}
\label{sec:systematic}

\begin{figure*}[!ht]
\includegraphics[width=2.0\columnwidth]{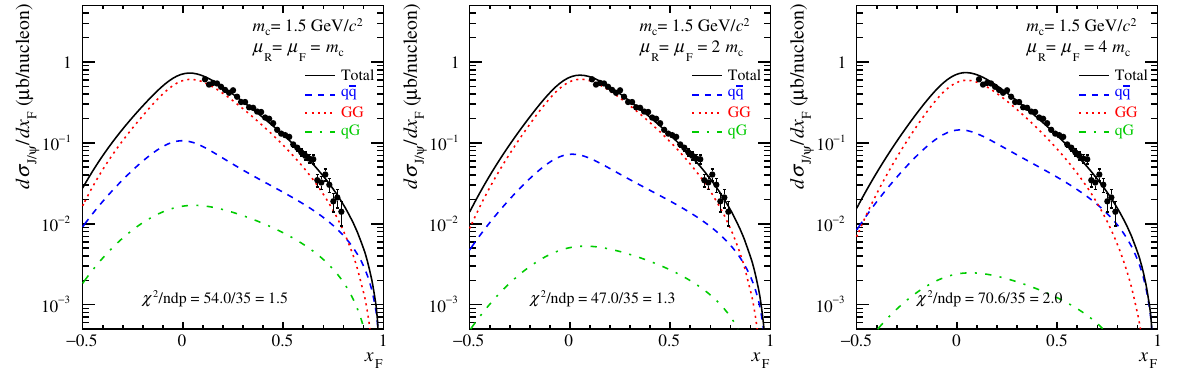}
\caption[\protect{}]{The NRQCD results with variation of charm quark
  mass $m_c$ and renormalization scale $\mu_R$, compared with the
  $d\sigma/dx_F$ data of $J/\psi$ production off the beryllium target
  with a 515-GeV/$c$ $\pi^-$ beam from the E672/E706
  experiment~\cite{jpsi_data1}. The pion PDFs used for the calculation
  is SMRS. The total cross sections and $q \bar{q}$, $GG$, and $qG$
  contributions are denoted as solid black, dashed blue, dotted red
  and dot-dashed green lines, respectively. The charm quark mass
  $m_c$, factorization scale $\mu_F$, and renormalization scale
  $\mu_R$ used for the NRQCD calculation as well as the fit
  $\chi^2$/ndf are displayed in each plot.}
\label{fig_sys_SMRS}
\end{figure*}

So far only the uncertainties associated with the parametrizations of
JAM and xFitter PDFs have been taken into account. Our results are
also sensitive to the NRQCD input parameters and to the choice of the
nuclear PDFs. We have checked that fits performed with the
nCTEQ15~\cite{Kovarik:2015cma} parametrization instead of EPPS16
result in negligible differences. Fits with the
factorization/renormalization scale parameter $\mu_R$ set to 1, 2, and
4 $m_c$, with $m_c$ = 1.4, 1.5, and 1.6~GeV/$c^2$, have also been
made. The values of the total $\chi^2$/ndf do not vary much: they
remain nearly unchanged between $\mu_R = m_c$ and $\mu_R = 4 m_c$ at
$m_c$ = 1.5~GeV/$c^2$. The effect on the values of the LDMEs is more
important. For both $J/\psi$ and $\psi(2S)$ the fitted LDMEs increase
by nearly a factor of four when $\mu_R$ increases from $\mu_R=m_c$ to
$\mu_R=4 m_c$. Nevertheless, the shape and the magnitude of the final
cross section remain nearly unchanged, as illustrated in
Fig.~\ref{fig_sys_SMRS} for the fit with the SMRS PDFs. The relative
contributions of the $q \bar{q}$ and $GG$ subprocesses for the three
values of $\mu_R$ are only slightly modified. The charm quark mass
correlates with the LDMEs in the partonic cross
sections. Consequently, the variation of $m_c$ around its nominal
value affects the values of the best-fit LDMEs and the overall quality
of fits remains stable. The systematic studies with GRV, JAM and
xFitter pion PDFs lead to results fully consistent with these
conclusions. The corresponding figures and tables are available in the
Supplemental Material~\cite{Supplement}.

\begin{figure}[!ht]
\centering
\subfloat[]
{\includegraphics[width=1.0\columnwidth]{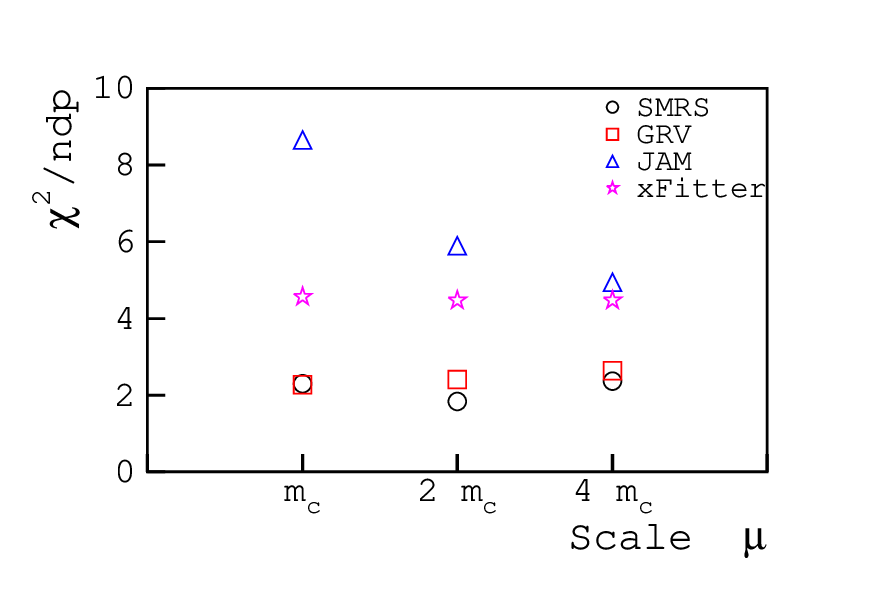}}
\hfill
%\quad
\vspace{-1cm}
\subfloat[]
{\includegraphics[width=1.0\columnwidth]{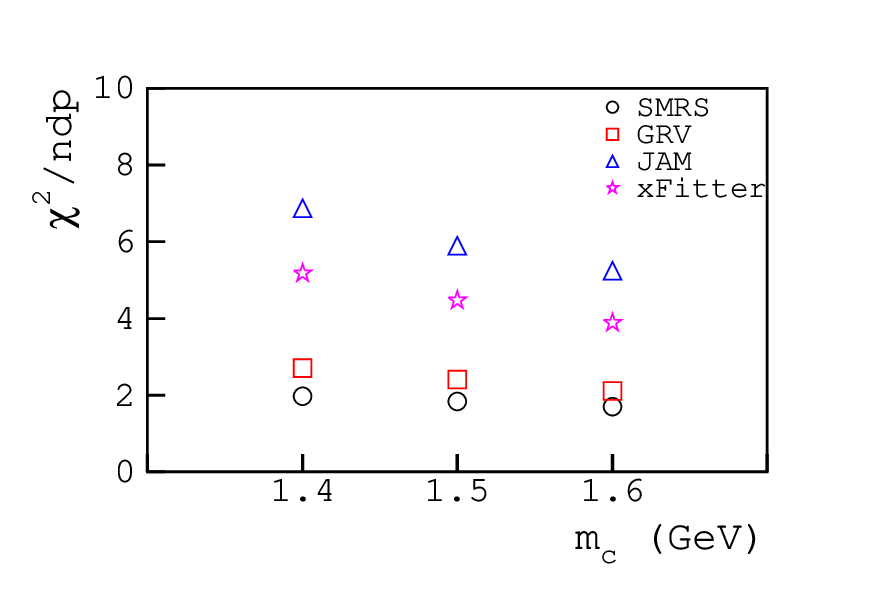}}
\caption[\protect{}] {The $\chi^2$ divided by the number of data point
  (ndp) of the pion-induced $x_F$-dependent data for four pion PDFs
  versus: (a) the scale parameter $\mu_R$ (b) and charm mass $m_c$.}
\label{fig_syserr}
\end{figure}

The overall $\chi^2$/ndf for the pion-induced J/$\psi$ and $\psi(2S)$
$x_F$-dependent data versus different choices of scale and $m_c$ for
four pion PDFs are shown in Fig.~\ref{fig_syserr}. The $\chi^2$/ndf
values of SMRS and GRV remain consistently better than those of JAM
and xFitter. The systematic variation of the scale and mass parameters
do not change the preference of the data for GRV and SMRS.

In addition, the theoretical uncertainties corresponding to the
variations of $m_c$ from 1.4 to 1.6 GeV at $\mu_R$ = 2 $m_c$, and
those of $\mu$ from $m_c$ to 4 $m_c$ at $m_c$ = 1.5 GeV/$c^2$ with the
fixed LDMEs labeled as ``FIT'' in Table~\ref{tab:LDME_PDF} for the
total and differential $x_F$ cross sections are displayed as yellow
bands in Figs.~22 and ~23, respectively, in the Supplemental
Material~\cite{Supplement}. Compared to Figs.~\ref{fig_jpsi_PDF1} and
~\ref{fig_sdep_SMRS}, the uncertainty bands in these two additional
figures are significantly larger, with the overall $\chi^2/\text{ndp}$
rising by a factor of 20 to 50. We note however that the increase in
$\chi^2$ is primarily due to the changes in the overall normalization,
common to all pion PDFs, while the shapes of the $x_F$ dependence are
largely preserved. This suggests that the ability to discriminate
various pion PDFs, based on their predicted shapes of the xF
distributions, is insensitive to the choice of $m_c$ and $\mu$.

%%%%%%%%%%%%%%%%%%%%%%%%%%%%%%%%%%%%%%%%%%%%%%%%%%%%%%%%%%%%%%%%%%%%%
% Discussion %
\section{Discussion}
\label{sec:discussion}

Our analysis shows that the $x_F$-dependent proton and pion-induced
$J/\psi$ and $\psi(2S)$ production data can be simultaneously
described within the NRQCD framework. The results exhibit a strong
dependence on the pion PDFs and particularly on the gluon
distribution. The conclusions drawn here fully corroborate the results
obtained previously~\cite{Chang:2020rdy} using the more
phenomenological color evaporation model. The similarity between the
results of the two studies indicate that our main findings are quite
independent of the charmonium production models. 

We note that our analysis is performed in leading order only and in
the region of small $p_T$, in which a proof of factorization is still
lacking. Our work is based on the assumption adopted in
Refs.~\cite{Beneke:1996tk} and ~\cite{Maltoni:2006yp} that NRQCD can
lead to a satisfactory description of proton-induced charmonium
production at fixed-target energies. In order to evaluate the
theoretical uncertainties associated with these limitations, we also
investigated the sensitivity of the results to the NRQCD input
parameters. Varying the scale and the charm mass parameters within the
commonly accepted ranges leads to the error bands shown in
Figs.~\ref{fig_jpsi_PDF1} and~\ref{fig_sdep_SMRS}. The calculations
with each of the four pion PDFs are all modified consistently,
preserving the dependence already observed for the best fits.

The values of the color-octet LDMEs, resulting from the fits to the
data may contain model uncertainties, although they provide a good
description of the data. The formalism used is limited to LO and is
able to determine individually the CO $\langle
\mathcal{O}_{8}^{H}[^{3}S_{1}] \rangle$ LDMEs for $J/\psi$ and
$\psi(2S)$ only. The $\Delta_8^{J/\psi}$ and $\Delta_8^{\psi(2S)}$
terms combine each three additional color octet LDMEs. Furthermore,
most of the data included in the analysis have transverse momenta
$p_T$ smaller than 3~GeV/$c$. This is in sharp contrast with most of
the available LDMEs that result from fits at much larger energies and
for transverse momenta $p_T$ larger than
5~GeV/$c$~\cite{Lansberg:2006dh} and often even larger than
10~GeV/$c$~\cite{Bodwin:2014gia}. Assuming the approximate
universality of the LDMEs, a comparison with the published values
remains qualitative and can be solely used as an indirect criterion
for the significance of our results~\cite{Rothstein:1996vh}.

For the fits on the $J/\psi$ data sample, the $\langle
\mathcal{O}_{8}^{J/\psi}[^{3}S_{1}] \rangle$ values obtained,
e.g. $(2.59\pm 0.23)\times10^{-2}$ GeV$^{3}$ for the SMRS pion PDFs,
are nearly an order of magnitude larger than some of the published
LDMEs \cite{Butenschoen:2010rq, Chao:2012iv}.  Yet, they are only a
factor of 2.5 larger than the values of $(1.0 \pm 0.3)\times 10^{-2}$
GeV$^{3}$ reported in Ref.~\cite{Zhang:2014ybe} derived from data on
$\eta_c$ production using spin symmetry relations and $(1.1 \pm
1.0)\times 10^{-2}$ GeV$^{3}$ obtained in Ref.~\cite{Bodwin:2014gia}
from fits to Tevatron and LHC data. For the $\psi(2S)$, the fitted
$\langle \mathcal{O}_{8}^{\psi(2S)}[^{3}S_{1}]\rangle$ LDME with the
SMRS pion PDFs has a value of $(1.32 \pm 0.90)\times10^{-2}$
GeV$^{3}$, about a factor of four larger than the values quoted in
Refs.~\cite{Ma:2010jj,Gong:2012ug} and more recently in
Ref.~\cite{Butenschoen:2022orc}. A value with a different sign has
also been reported~\cite{Bodwin:2015iua}.  The comparison of our
$\Delta_8^{J/\psi}$ and $\Delta_8^{\psi(2S)}$ LDMEs with the
individual CO $\langle\mathcal{O}_{8}^{J/\psi}[^{1}S_{0}] \rangle$
values is only indicative. The $\Delta_8^{J/\psi}$ value is compatible
with the values derived in Refs.~\cite{Butenschoen:2010rq,
  Chao:2012iv,Zhang:2014ybe,Bodwin:2014gia}.  The
$\Delta_8^{\psi(2S)}$ LDME is also inside the range defined by the
values quoted in Refs.~\cite{Ma:2010jj, Gong:2012ug,
  Butenschoen:2022orc}. Within the systematic uncertainties associated
with the fits and given the assumptions made, the comparison can be
considered satisfactory, providing an indirect support for the present
analysis.

Our analysis is performed using a leading-order NRQCD framework
only. The results obtained may vary if a more advanced NRQCD formalism
with higher order terms is applied. In addition, for most of the
fixed-target data considered here, the mean transverse momenta are
smaller than the $J/\psi$ mass. Inclusion of higher-order corrections
could therefore provide a better description, but probably would not
change the general conclusions. The analysis has been also limited to
data taken with only light targets. A large amount of data of
$x_F$-differential cross sections with heavier targets have been
collected in the past. These data could be included in a more complete
analysis if the energy loss effects~\cite{Arleo:2018zjw} responsible
for the suppression of the charmonium cross section in hadron-nucleus
collisions are reliably accounted for.

%%%%%%%%%%%%%%%%%%%%%%%%%%%%%%%%%%%%%%%%%%%%%%%%%%%%%%%%%%%%%%%%%%%%%
% Conclusion %
\section{Conclusion}
\label{sec:conclusion}

We have analyzed fixed-target experimental cross sections for $J/\psi$
and $\psi(2S)$ production using the NRQCD framework. To minimize
nuclear matter effects, only data on hydrogen, lithium and beryllium
targets were selected. Heavier targets were only considered for the
data on the $J/\psi$ to $\psi(2S)$ ratios. Assuming the universality
of the NRQCD approach, both pion and proton-induced datasets were
included in the analysis. Fits to the individual $x_F$-differential
cross sections and their ratios have been made, using four different
pion PDF parametrizations. The proton data, although not directly
sensitive to the pion PDFs, enrich the selection and contribute to the
stability of the final results.

A simultaneous fit to all pion and proton datasets has been
achieved. The results of these common fits show that the relative
fractions of the $q\bar{q}$ and $GG$ contributions to the cross
sections strongly depend on the beam particle, on its incoming energy
and on the $x_F$ region considered. A strong dependence on the pion
PDF parametrization used is observed and particularly on the magnitude
of the pion gluon distribution. The results indicate a clear
preference for parametrizations with larger gluon distributions at
relatively large $x$. Good agreement with the data is obtained with
the SMRS and GRV PDFs. The fits with the recent JAM and xFitter
parametrizations turn out to show much larger deviations for most of
the datasets. 

The comparison between the results for $J/\psi$ and $\psi(2S)$
production leads to an important new observation: the strengths of the
$q\bar{q}$ and $GG$ contributions to these two charmonium states are --
unexpectedly -- quite different. The $q\bar{q}$ component of the
$\psi(2S)$ cross section is, proportionally, few times larger than the
$q\bar{q}$ component of the $J/\psi$ cross section. This interesting
feature is confirmed for both differential and integrated cross
sections and for both pion and proton beams. The production of
$\psi(2S)$ appears to be more sensitive to the pion's valence quark
distribution than that of $J/\psi$. This observation could be relevant
for a better understanding of the charmonium production mechanism.

In the kinematical domain of the available fixed-target data --
relatively small center-of-mass energy and therefore small transverse
momenta -- the theoretical uncertainties could be substantial. A proof
of factorization is still lacking and additional higher-order
corrections may play a role. Conversely, the conclusions drawn rely on
a simultaneous study of the pion and proton-induced cross sections and
ratios, both $x_F$-differential and integrated, for all of which the
agreement achieved is quite good. The conclusions are also fully
supported by the results from our previous study done with the color
evaporation model~\cite{Chang:2020rdy}. While further theoretical
efforts are required to better understand the reaction mechanism for
quarkonium production, the inclusion of the charmonium data in a new
global analysis to extract the pion PDFs would be very
informative~\cite{Bourrely:2022mjf}.

New results of Drell-Yan as well as $J/\psi$ measurements in $\pi A$
reactions will be available from the CERN
COMPASS~\cite{COMPASS:2017jbv} and AMBER~\cite{Adams:2018pwt}
experiments in the near future. These data will be important in
providing better knowledge of the pion PDFs. For the longer-term
electron-ion collider projects in U.S. and China, the pion as well
kaon structures are planned to be explored using the tagged DIS
process~\cite{Aguilar:2019teb, Anderle:2021wcy, Chavez:2021koz}.

\section*{Acknowledgments}
\label{sec:acknowledgments}

This work was supported in part by the U.S. National Science
Foundation and National Science and Technology Council of Taiwan (R.O.C.).

%%%%%%%%%%%%%%%%%%%%%%%%%%%%%%%%%%%%%%%%%%%%%%%%%%%%%%%%%%%%%%%%%%%%%%%
%bibtex
\bibliography{ref}
%%%%%%%%%%%%%%%%%%%%%%%%%%%%%%%%%%%%%%%%%%%%%%%%%%%%%%%%%%%%%%%%%%%%%%%

% this is the right order!
\setcounter{figure}{0}
\onecolumngrid
\newpage
\section*{Supplemental materials}

\begin{figure*}[!ht]
\centering
\includegraphics[width=1.0\columnwidth]{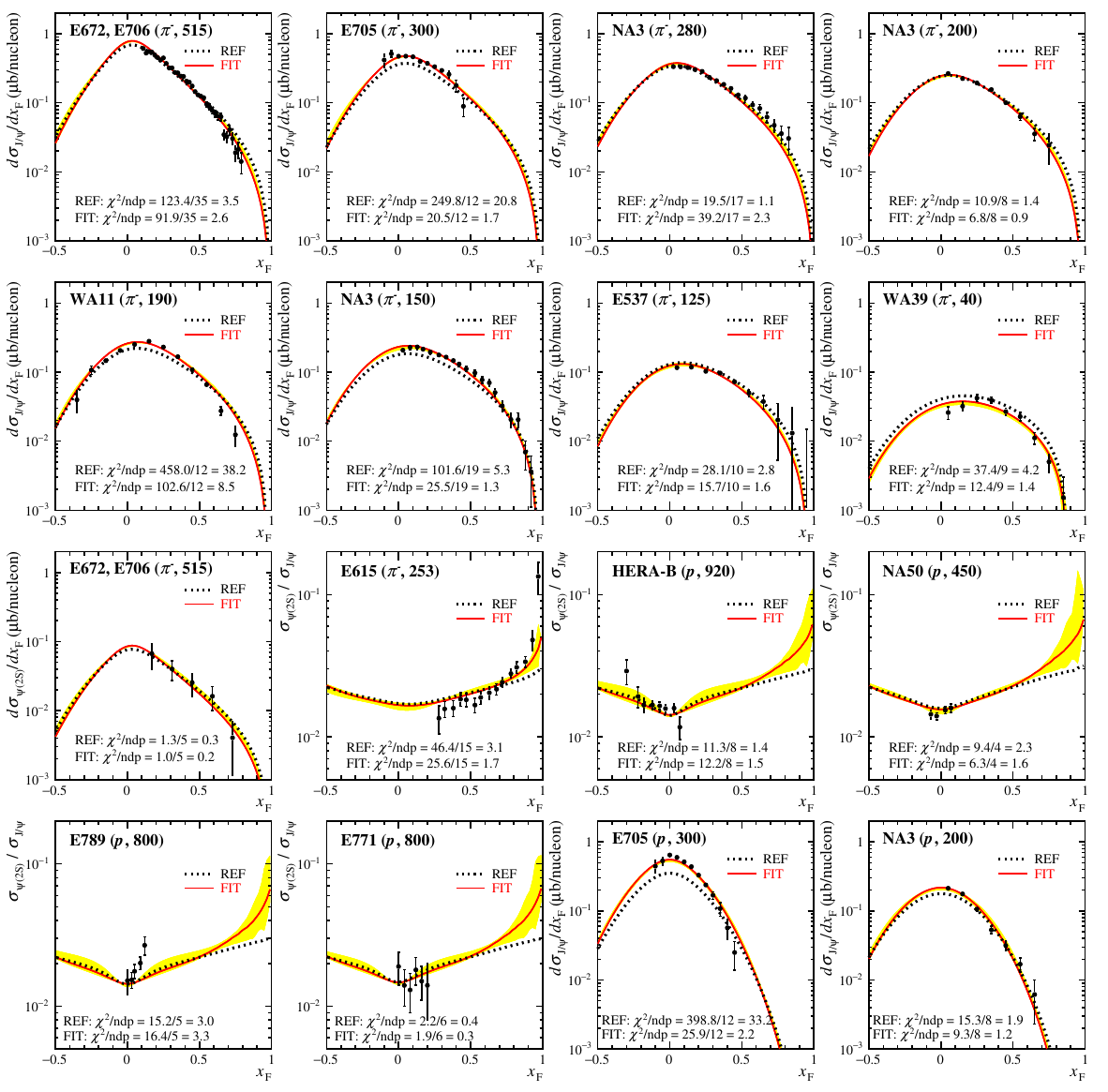}
\caption[\protect{}]{The $x_F$-dependent cross sections for $J/\psi$
  and $\psi(2S)$ production and $R_{\psi}(x_F)$ ratios in $\pi^- N$
  and $pN$ interactions, following the order given in
  Table.~II. The symbol and value in parenthesis denote
  the particle type and momentum of beam. The solid red and dotted
  black curves represent the NRQCD results of GRV pion PDFs from the
  fit described in the text (``FIT'') and from the calculation using
  the LDMEs obtained in Ref.~\cite{Hsieh:2021yzg} (``REF''),
  respectively. The values of $\chi^2$ divided by the number of data
  point (ndp) for each dataset are also shown.}
\label{fig_jpsi_PDF2}
\end{figure*}

\begin{figure*}[!ht]
\centering
\includegraphics[width=1.0\columnwidth]{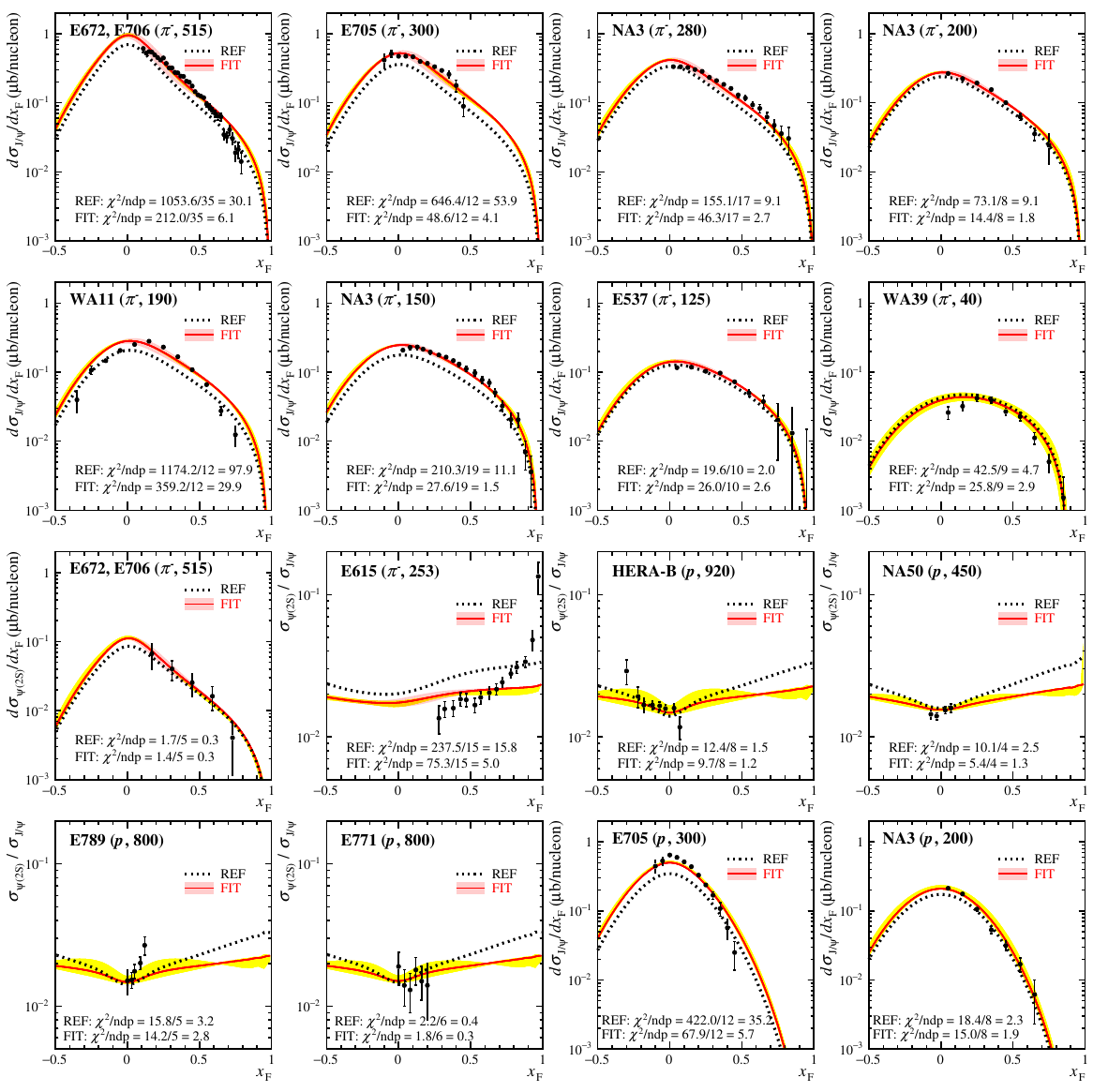}
\caption[\protect{}]{The $x_F$-dependent cross sections for $J/\psi$
  and $\psi(2S)$ production and $R_{\psi}(x_F)$ ratios in $\pi^- N$
  and $pN$ interactions, following the order given in Table.~II. The
  symbol and value in parenthesis denote the particle type and
  momentum of beam. The solid red and dotted black curves represent
  the NRQCD results of JAM pion PDFs from the fit described in the
  text (``FIT'') and from the calculation using the LDMEs obtained in
  Ref.~\cite{Hsieh:2021yzg} (``REF''), respectively. The values of
  $\chi^2$ divided by the number of data point (ndp) for each dataset
  are also shown. The yellow bands represent the cross section
  uncertainties associated with the scale and charm quark mass
  systematic variations.}
\label{fig_jpsi_PDF3}
\end{figure*}

\begin{figure*}[!ht]
\centering
\includegraphics[width=1.0\columnwidth]{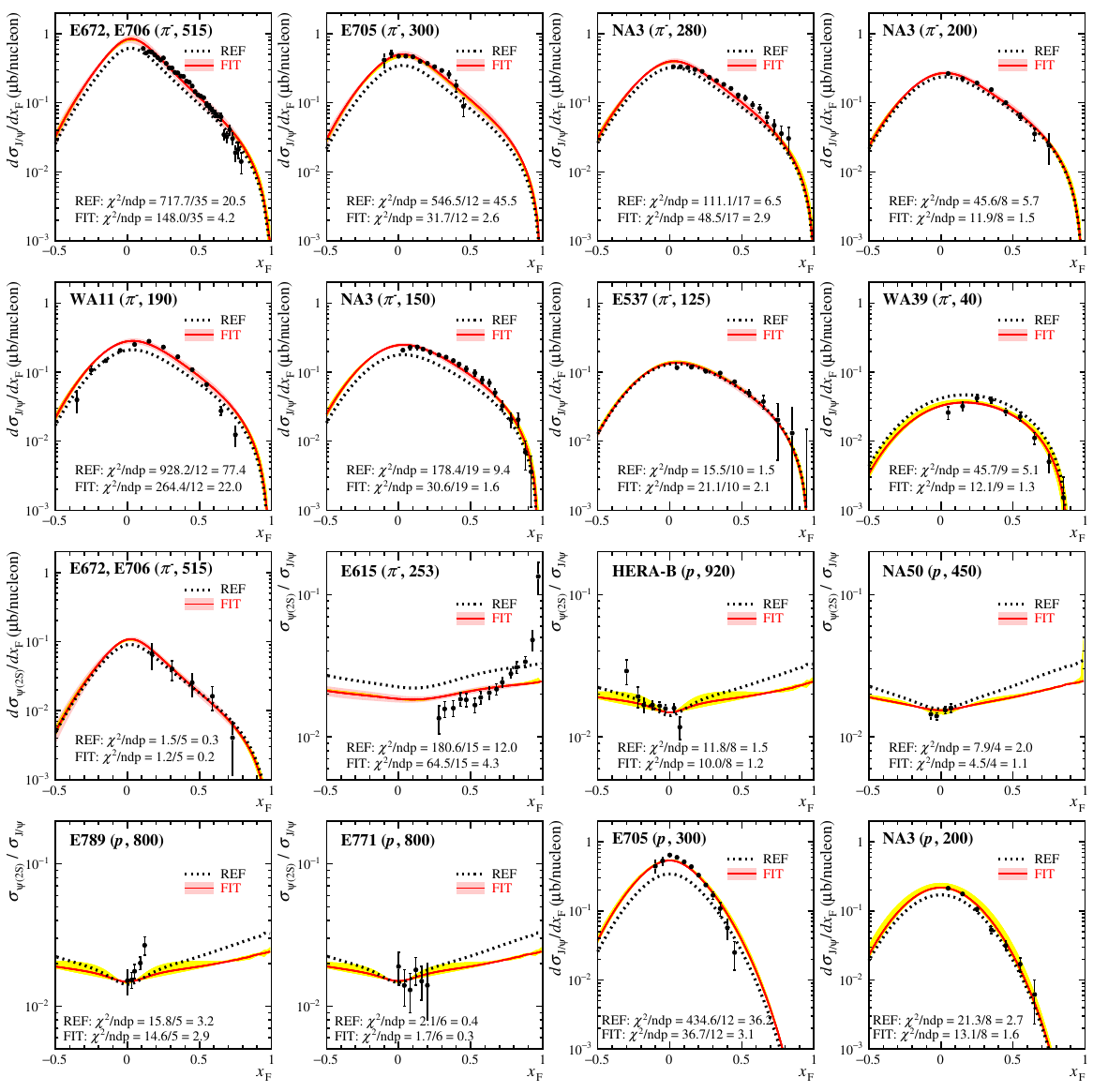}
\caption[\protect{}]{The $x_F$-dependent cross sections for $J/\psi$
  and $\psi(2S)$ production and $R_{\psi}(x_F)$ ratios in $\pi^- N$
  and $pN$ interactions, following the order given in Table.~II. The
  symbol and value in parenthesis denote the particle type and
  momentum of beam. The solid red and dotted black curves represent
  the NRQCD results of xFitter pion PDFs from the fit described in the
  text (``FIT'') and from the calculation using the LDMEs obtained in
  Ref.~\cite{Hsieh:2021yzg} (``REF''), respectively. The values of
  $\chi^2$ divided by the number of data point (ndp) for each dataset
  are also shown. The yellow bands represent the cross section
  uncertainties associated with the scale and charm quark mass
  systematic variations.}
\label{fig_jpsi_PDF4}
\end{figure*}

%%%%%%%%%%%%%%%%%%%%%%%%%%%%%%%%%%%%%%%%%%%%%%%%%%%%%%%%%%%%%%%%%%%%%%%%%%

\begin{figure}[!ht]
\centering
\includegraphics[width=0.8\columnwidth]{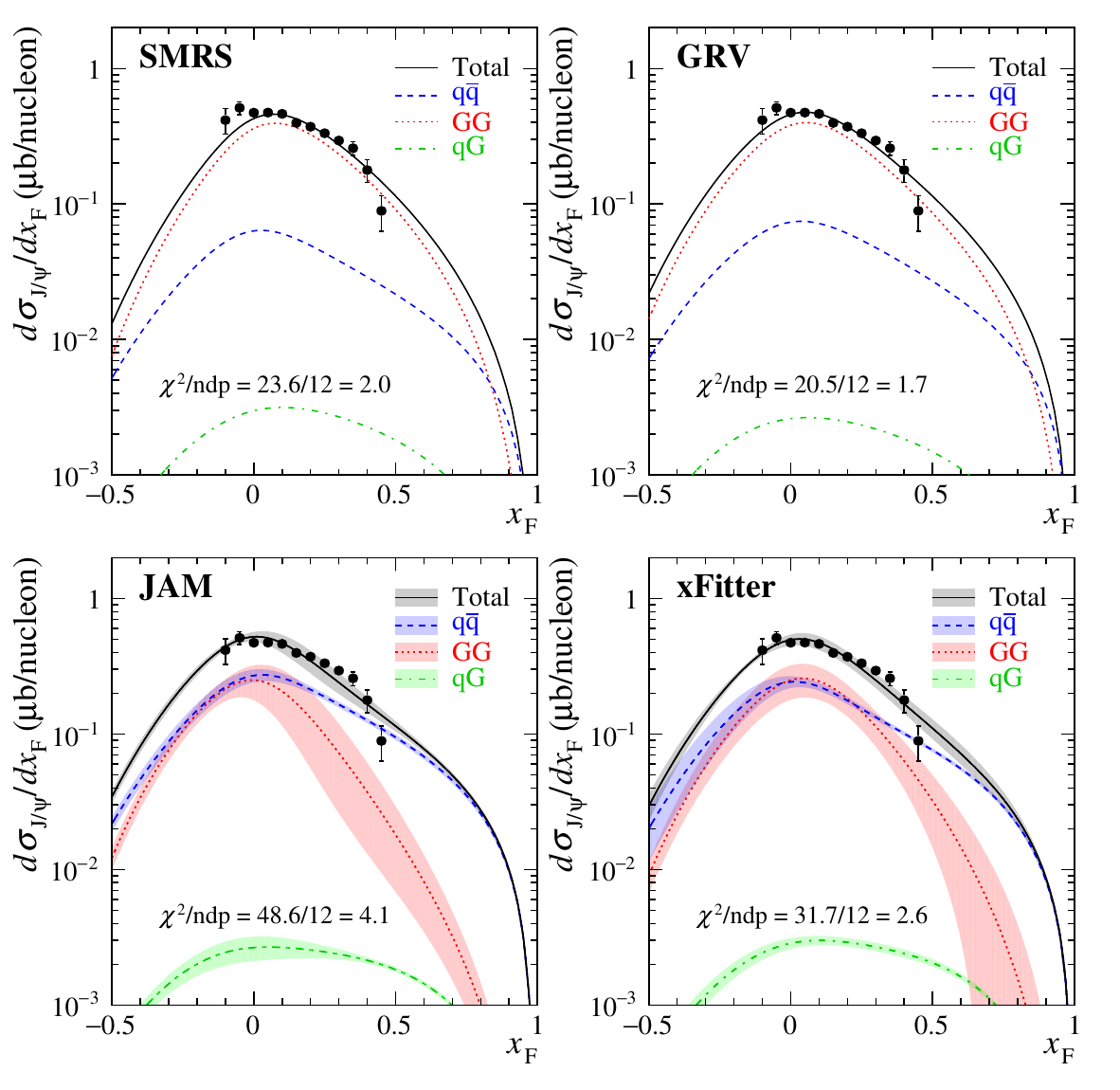}
\caption[\protect{}]{Differential cross sections for $J/\psi$
  production with a 300-GeV/$c$ $\pi^-$ beam from the E705
  experiment~\cite{jpsi_data2and3}. The data are compared to the NRQCD
  fit results for the SMRS, GRV, xFitter, and JAM PDFs. The total
  cross sections and $q \bar{q}$, $GG$, and $qG$ contributions are
  denoted as solid black, dashed blue, dotted red, and dot-dashed
  green lines, respectively. The uncertainty bands associated with JAM
  and xFitter PDFs are also shown.}
\label{fig_jpsi_data02}
\end{figure}

\begin{figure}[!ht]
\centering
\includegraphics[width=0.8\columnwidth]{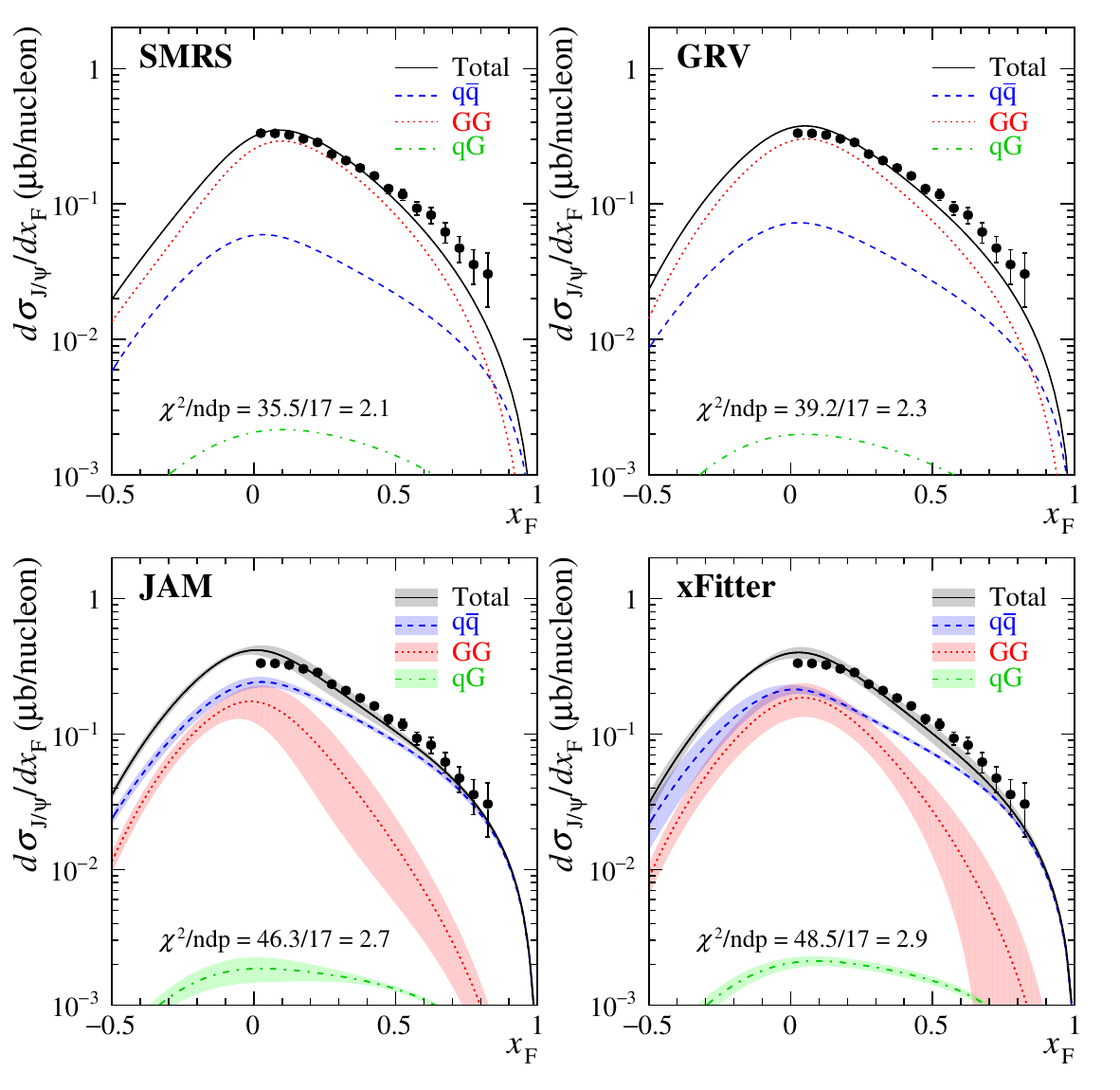}
\caption[\protect{}]{Differential cross sections for $J/\psi$
  production with a 280-GeV/$c$ $\pi^-$ beam from the NA3
  experiment~\cite{jpsi_data17and18and21}. The data are compared to
  the NRQCD fit results for the SMRS, GRV, xFitter, and JAM PDFs. The
  total cross sections and $q \bar{q}$, $GG$, and $qG$ contributions
  are denoted as solid black, dashed blue, dotted red, and dot-dashed
  green lines, respectively. The uncertainty bands associated with JAM
  and xFitter PDFs are also shown.}
\label{fig_jpsi_data03}
\end{figure}

%\blindtext

\begin{figure}[!ht]
\centering
\includegraphics[width=0.8\columnwidth]{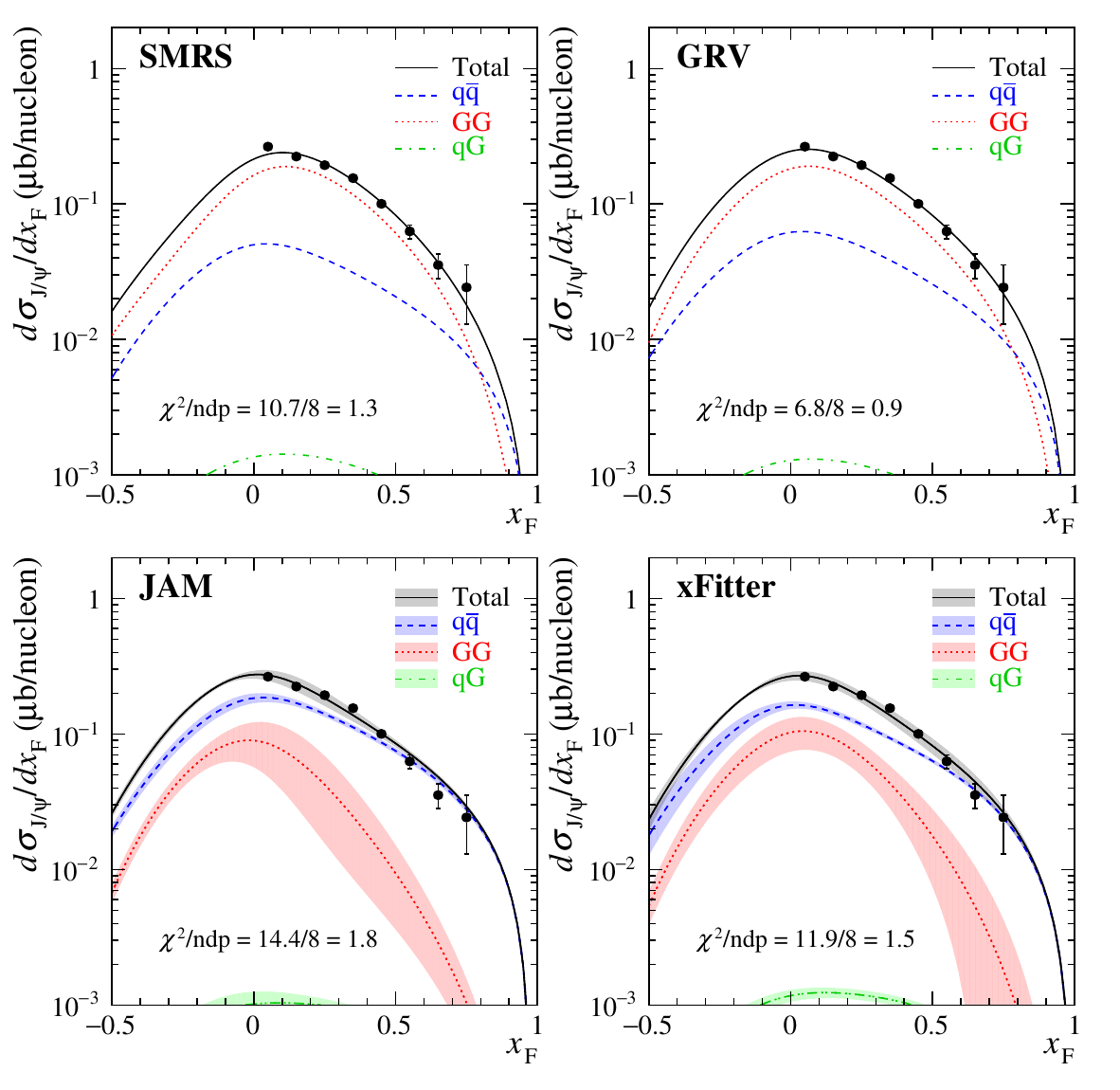}
\caption[\protect{}]{Differential cross sections for $J/\psi$
  production with a 200-GeV/$c$ $\pi^-$ beam from the NA3
  experiment~\cite{jpsi_data17and18and21}. The data are compared to
  the NRQCD fit results for the SMRS, GRV, xFitter, and JAM PDFs. The
  total cross sections and $q \bar{q}$, $GG$, and $qG$ contributions
  are denoted as solid black, dashed blue, dotted red, and dot-dashed
  green lines, respectively. The uncertainty bands associated with JAM
  and xFitter PDFs are also shown.}
\label{fig_jpsi_data04}
\end{figure}

\begin{figure}[!ht]
\centering
\includegraphics[width=0.8\columnwidth]{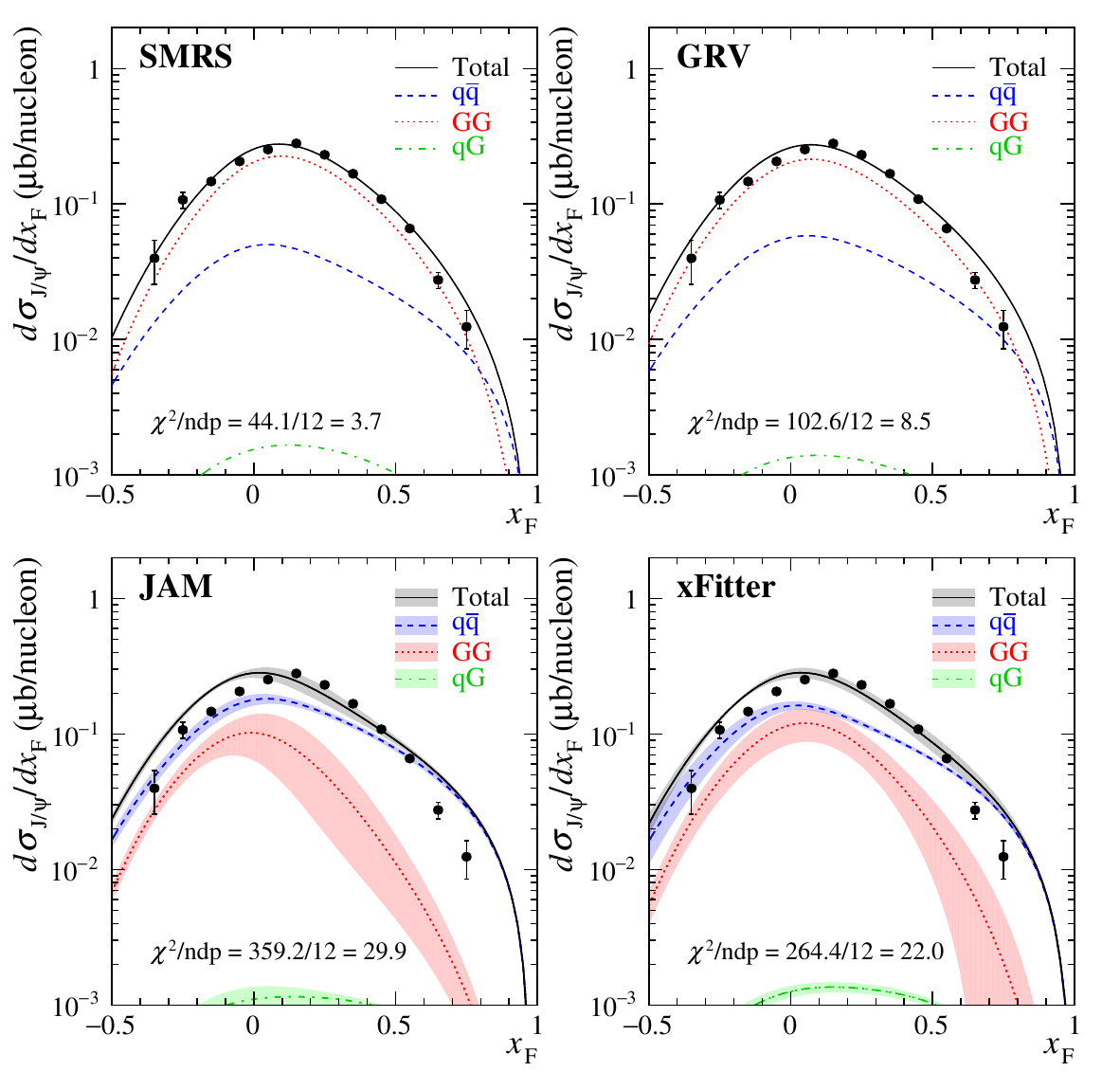}
\caption[\protect{}]{Differential cross sections for $J/\psi$
  production with a 190-GeV/$c$ $\pi^-$ beam from the WA11
  experiment~\cite{jpsi_data16}. The data are compared to the NRQCD
  fit results for the SMRS, GRV, xFitter, and JAM PDFs. The total
  cross sections and $q \bar{q}$, $GG$, and $qG$ contributions are
  denoted as solid black, dashed blue, dotted red, and dot-dashed
  green lines, respectively. The uncertainty bands associated with JAM
  and xFitter PDFs are also shown.}
\label{fig_jpsi_data05}
\end{figure}

\begin{figure}[!ht]
\centering
\includegraphics[width=0.8\columnwidth]{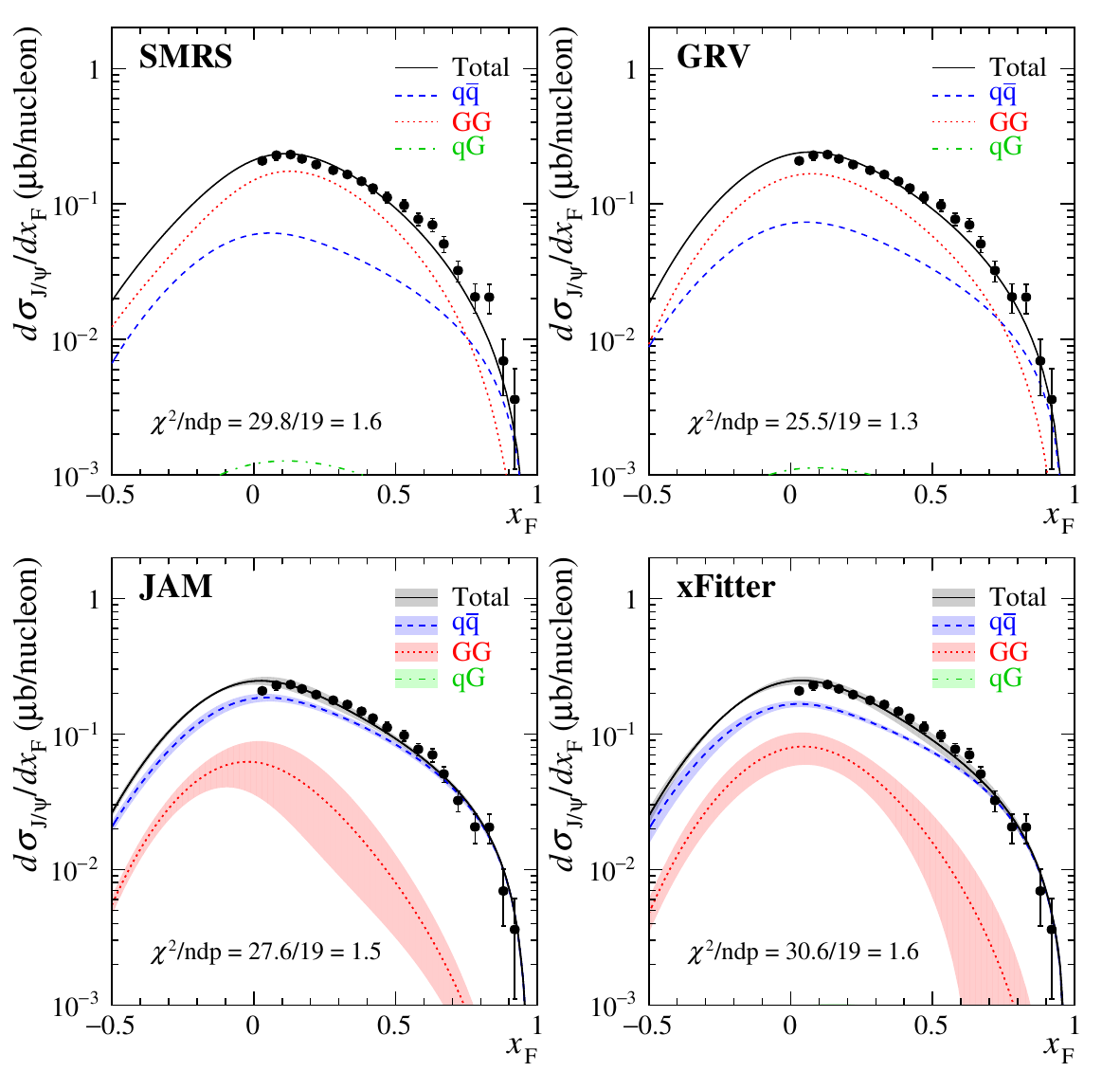}
\caption[\protect{}]{Differential cross sections for $J/\psi$
  production with a 150-GeV/$c$ $\pi^-$ beam from the NA3
  experiment~\cite{jpsi_data17and18and21}. The data are compared to
  the NRQCD fit results for the SMRS, GRV, xFitter, and JAM PDFs. The
  total cross sections and $q \bar{q}$, $GG$, and $qG$ contributions
  are denoted as solid black, dashed blue, dotted red, and dot-dashed
  green lines, respectively. The uncertainty bands associated with JAM
  and xFitter PDFs are also shown.}
\label{fig_jpsi_data06}
\end{figure}

\begin{figure}[!ht]
\centering 
\includegraphics[width=0.8\columnwidth]{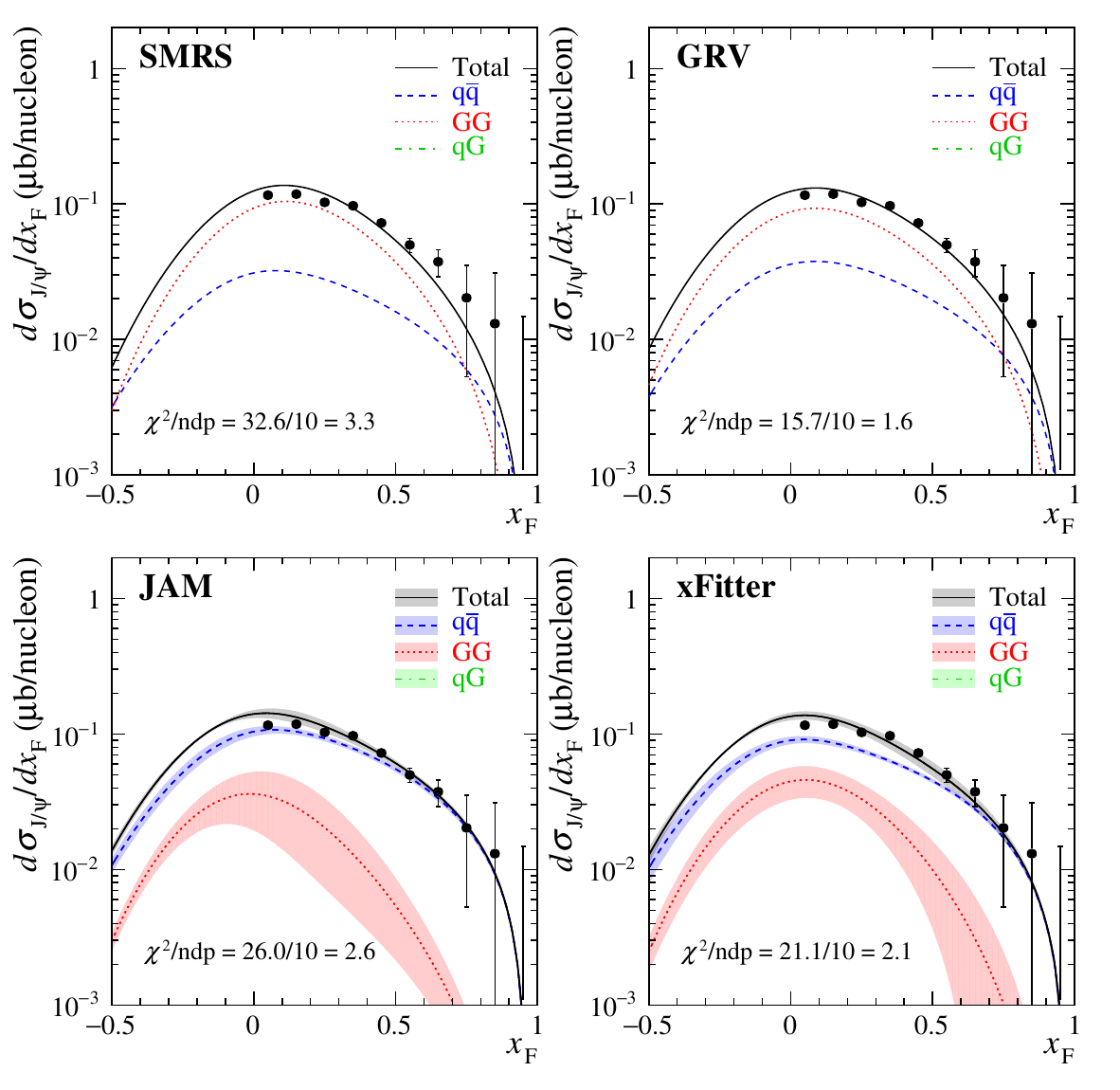}
\caption[\protect{}]{Differential cross sections for $J/\psi$
  production with a 125-GeV/$c$ $\pi^-$ beam from the E537
  experiment~\cite{jpsi_data6and7and8}. The data are compared to the
  NRQCD fit results for the SMRS, GRV, xFitter, and JAM PDFs. The
  total cross sections and $q \bar{q}$, $GG$, and $qG$ contributions
  are denoted as solid black, dashed blue, dotted red, and dot-dashed
  green lines, respectively. The uncertainty bands associated with JAM
  and xFitter PDFs are also shown.}
\label{fig_jpsi_data07}
\end{figure}

\begin{figure}[!ht]
\centering 
\includegraphics[width=0.8\columnwidth]{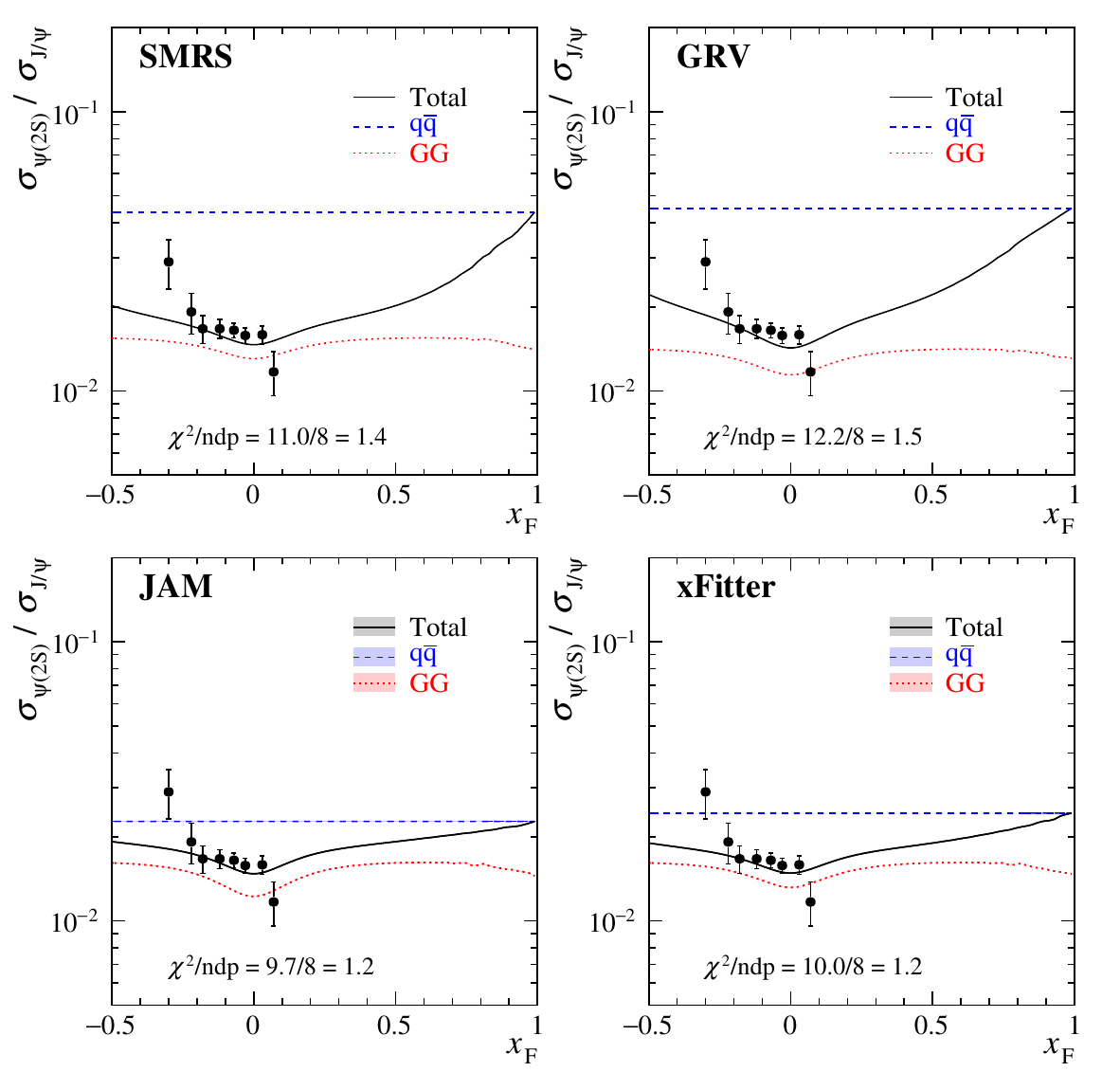}
\caption[\protect{}]{The $\psi(2S)$ to $J/\psi$ cross section ratios
  $R_{\psi}(x_F)$ for $J/\psi$ and $\psi(2S)$ production with a
  920-GeV/$c$ proton beam from the HERA-B
  experiment~\cite{HERA-B:2006bhy}. The data are compared to the NRQCD
  fit results for the SMRS, GRV, xFitter, and JAM PDFs. The ratios of
  total cross sections and individual $R^{q \bar q}_\psi (x_F)$ and
  $R^{GG}_\psi (x_F)$ contributions are denoted as solid black, dashed
  blue, and dotted red lines, respectively.}
\label{fig_jpsi_data11}
\end{figure}

\begin{figure}[!ht]
\centering 
\includegraphics[width=0.8\columnwidth]{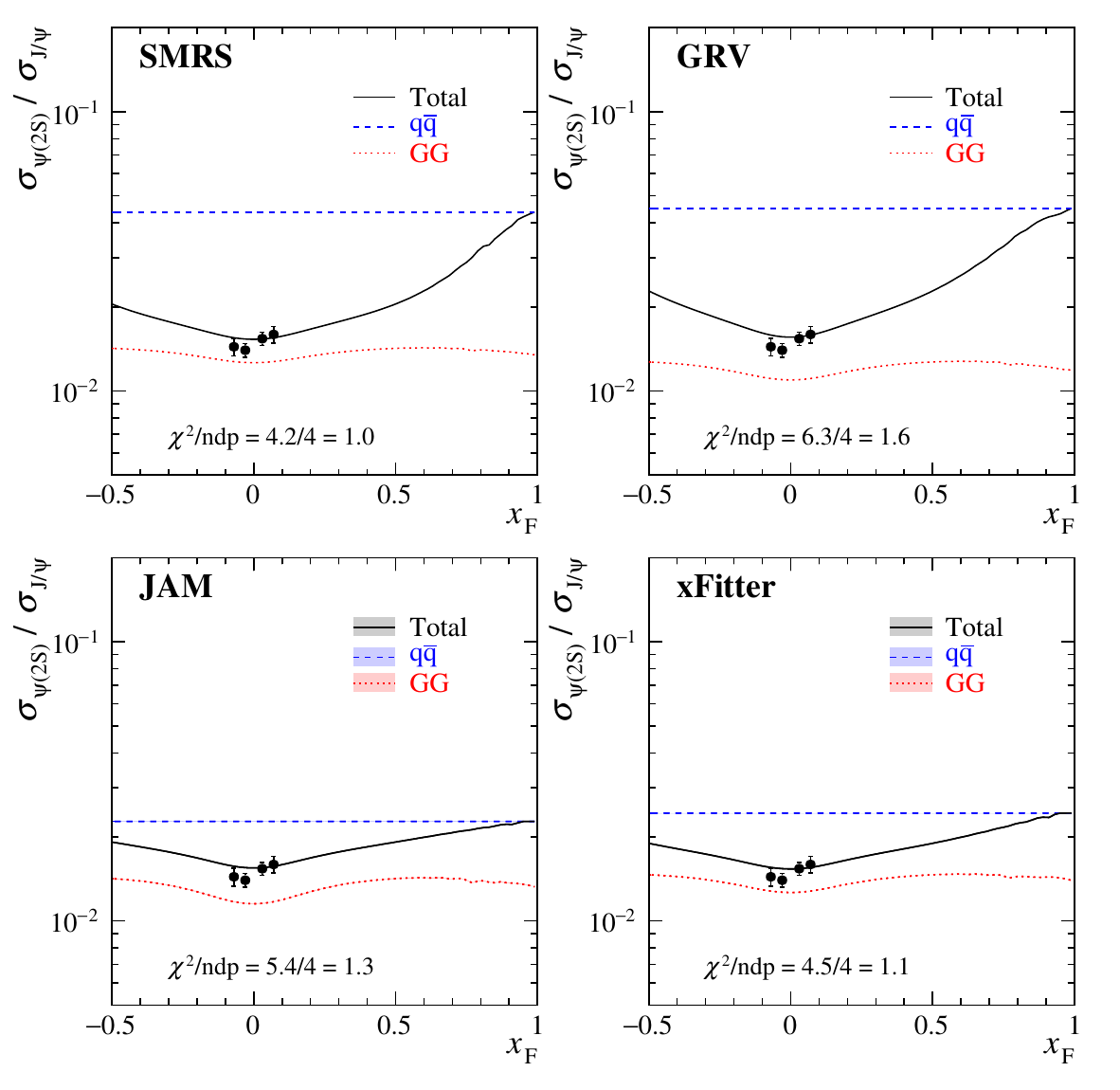}
\caption[\protect{}]{The $\psi(2S)$ to $J/\psi$ cross section ratios
  $R_{\psi}(x_F)$ for $J/\psi$ and $\psi(2S)$ production with a
  450-GeV/$c$ proton beam from the NA50
  experiment~\cite{NA50:2003fvu}. The data are compared to the NRQCD
  fit results for the SMRS, GRV, xFitter, and JAM PDFs. The ratios of
  total cross sections and individual $R^{q \bar q}_\psi (x_F)$ and
  $R^{GG}_\psi (x_F)$ contributions are denoted as solid black, dashed
  blue, and dotted red lines, respectively.}
\label{fig_jpsi_data12}
\end{figure}

\begin{figure}[!ht]
\centering 
\includegraphics[width=0.8\columnwidth]{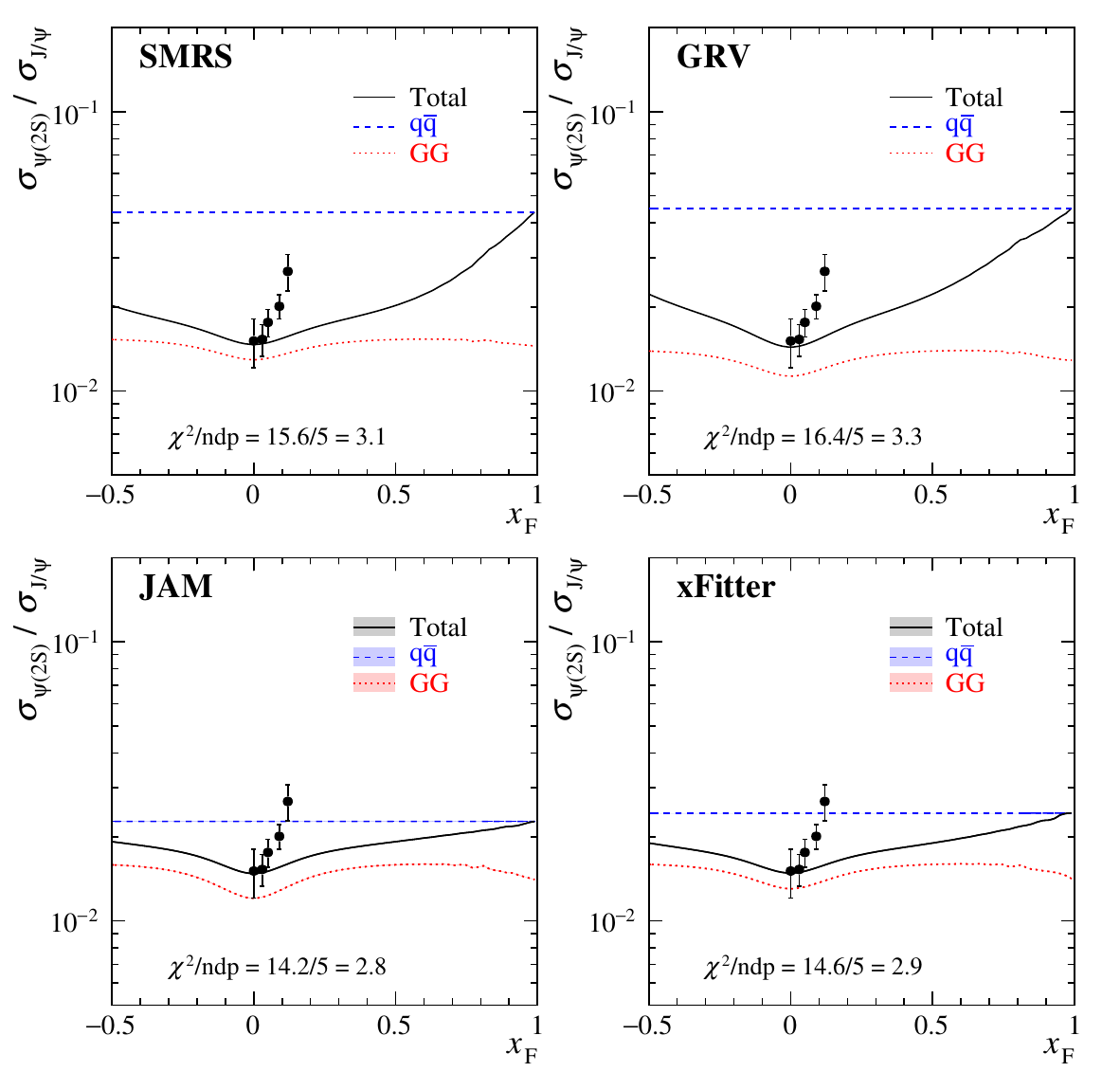}
\caption[\protect{}]{The $\psi(2S)$ to $J/\psi$ cross section ratios
  $R_{\psi}(x_F)$ for $J/\psi$ and $\psi(2S)$ production with a
  450-GeV/$c$ proton beam from the NA50
  experiment~\cite{NA50:2003fvu}. The data are compared to the NRQCD
  fit results for the SMRS, GRV, xFitter, and JAM PDFs. The ratios of
  total cross sections and individual $R^{q \bar q}_\psi (x_F)$ and
  $R^{GG}_\psi (x_F)$ contributions are denoted as solid black, dashed
  blue, and dotted red lines, respectively.}
\label{fig_jpsi_data13}
\end{figure}

\begin{figure}[!ht]
\centering 
\includegraphics[width=0.8\columnwidth]{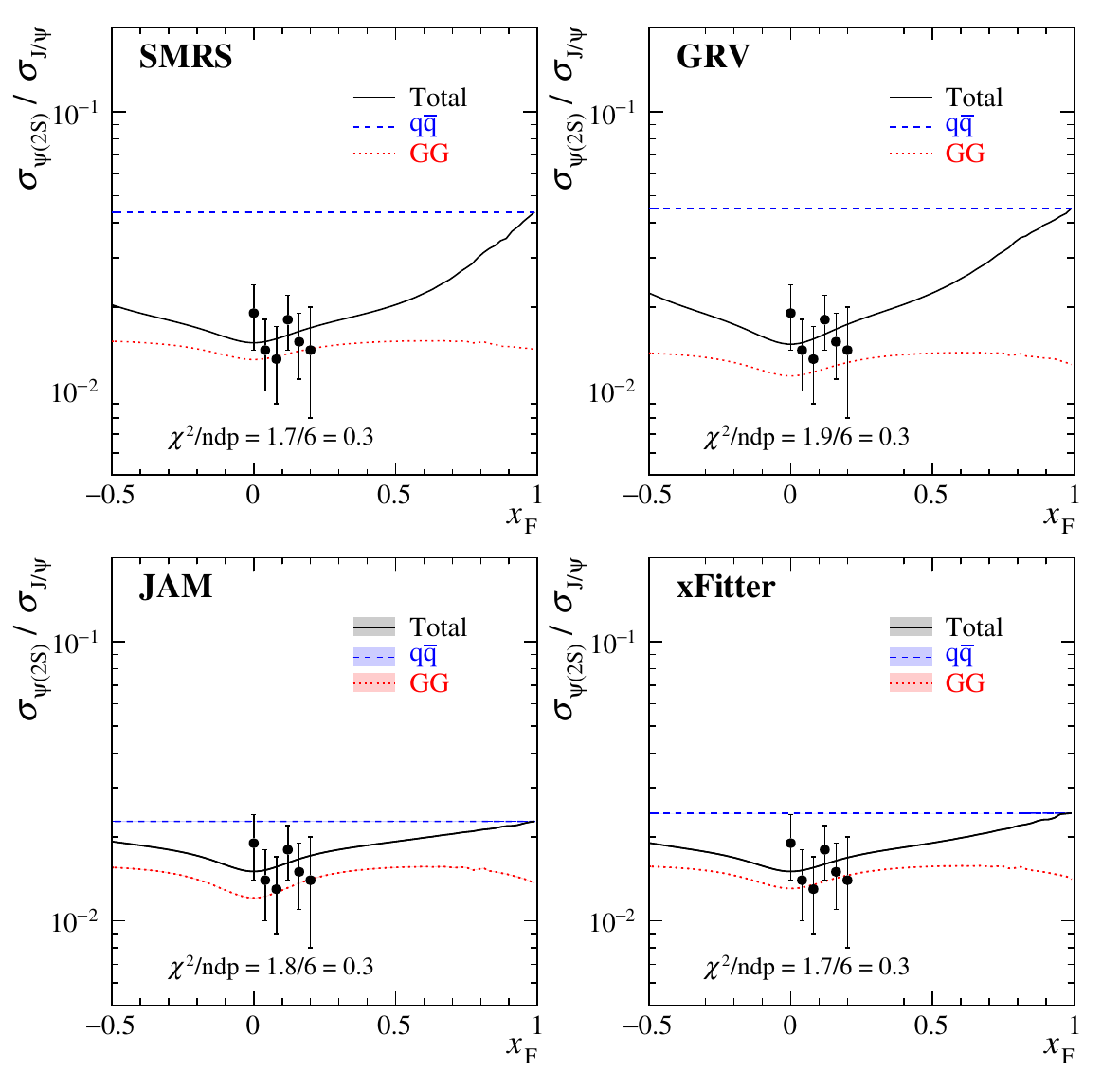}
\caption[\protect{}]{The $\psi(2S)$ to $J/\psi$ cross section ratios
  $R_{\psi}(x_F)$ for $J/\psi$ and $\psi(2S)$ production with a
  800-GeV/$c$ proton beam from the E771
  experiment~\cite{E771:1995ane}. The ratios of total cross sections
  and individual $R^{q \bar q}_\psi (x_F)$ and $R^{GG}_\psi (x_F)$
  contributions are denoted as solid black, dashed blue, and dotted
  red lines, respectively.}
\label{fig_jpsi_data14}
\end{figure}

\begin{figure}[!ht]
\centering 
\includegraphics[width=0.8\columnwidth]{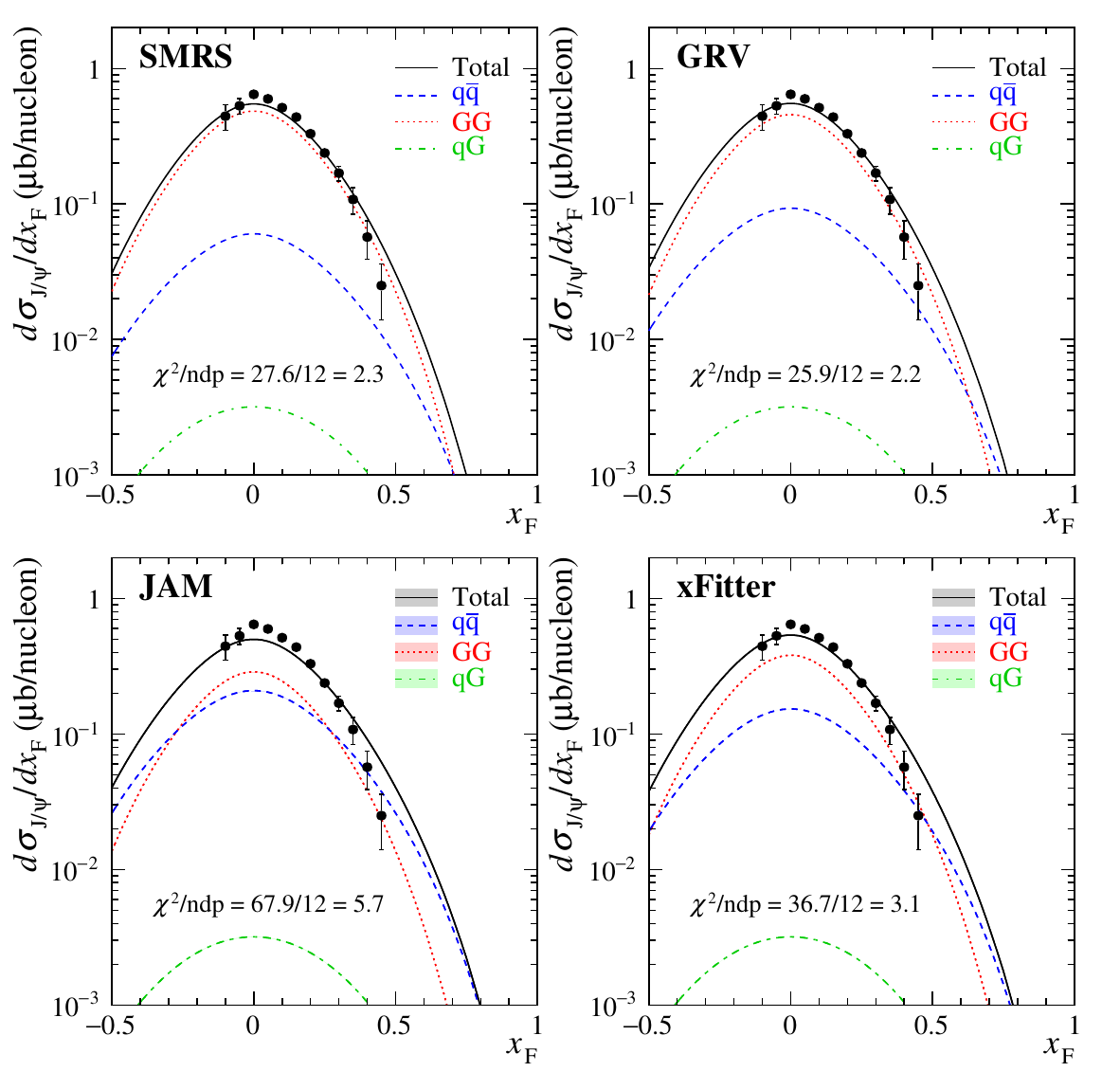}
\caption[\protect{}]{Differential cross sections for $J/\psi$
  production with a 300-GeV/$c$ proton beam from the E705
  experiment~\cite{jpsi_data2and3}. The data are compared to the NRQCD
  fit results of LDMEs for the SMRS, GRV, xFitter, and JAM PDFs. The
  total cross sections and $q \bar{q}$, $GG$, and $qG$ contributions
  are denoted as solid black, dashed blue, dotted red, and dot-dashed
  green lines, respectively.}
\label{fig_jpsi_data15}
\end{figure}

\begin{figure}[!ht]
\centering 
\includegraphics[width=0.8\columnwidth]{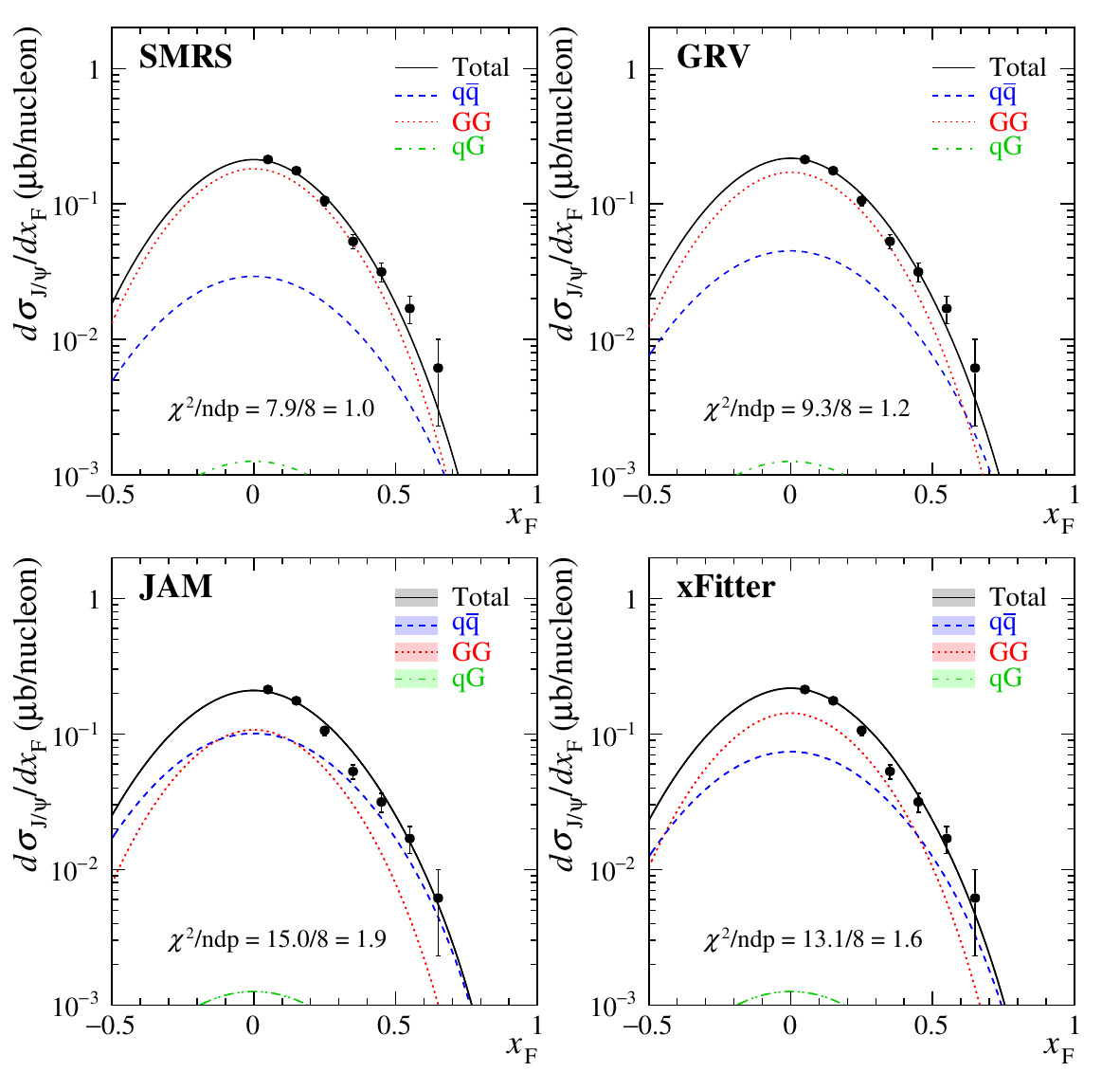}
\caption[\protect{}]{Differential cross sections for $J/\psi$
  production with a 200-GeV/$c$ proton beam from the NA3
  experiment~\cite{jpsi_data17and18and21}. The data are compared to
  the NRQCD fit results of LDMEs for the SMRS, GRV, xFitter, and JAM
  PDFs. The total cross sections and $q \bar{q}$, $GG$, and $qG$
  contributions are denoted as solid black, dashed blue, dotted red,
  and dot-dashed green lines, respectively.}
\label{fig_jpsi_data16}
\end{figure}

%%%%%%%%%%%%%%%%%%%%%%%%%%%%%%%%%%%%%%%%%%%%%%%%%%%%%%%%%%%%%%%%%%%%%%%%%%

\begin{figure*}[!ht]
\centering 
\includegraphics[width=0.8\columnwidth]{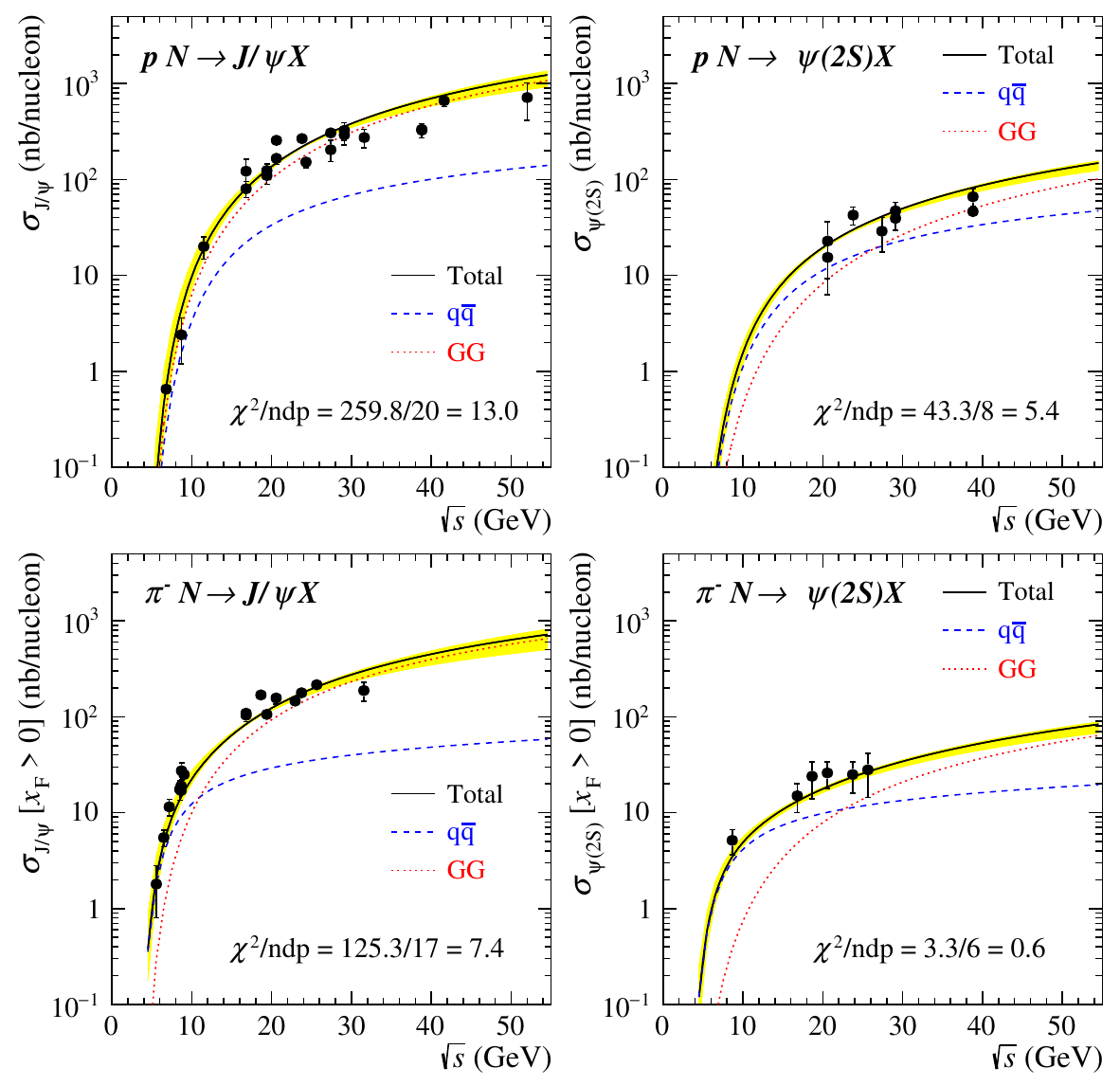}
\caption[\protect{}]{Integrated charmonium cross sections in $pN$ and
  $\pi^- N$ collisions. The data for $J/\psi$ and $\psi(2S)$
  production are compared to the fit made using the GRV pion PDFs. The
  total cross section and its $q \bar{q}$ and $GG$ contributions are
  denoted as solid black, dashed blue and dotted red lines,
  respectively. The yellow bands represent the cross section
  uncertainties associated with the scale and charm quark mass
  systematic variations.}
\label{fig_sdep_GRV}
\end{figure*}

\begin{figure*}[!ht]
\centering
\includegraphics[width=0.8\columnwidth]{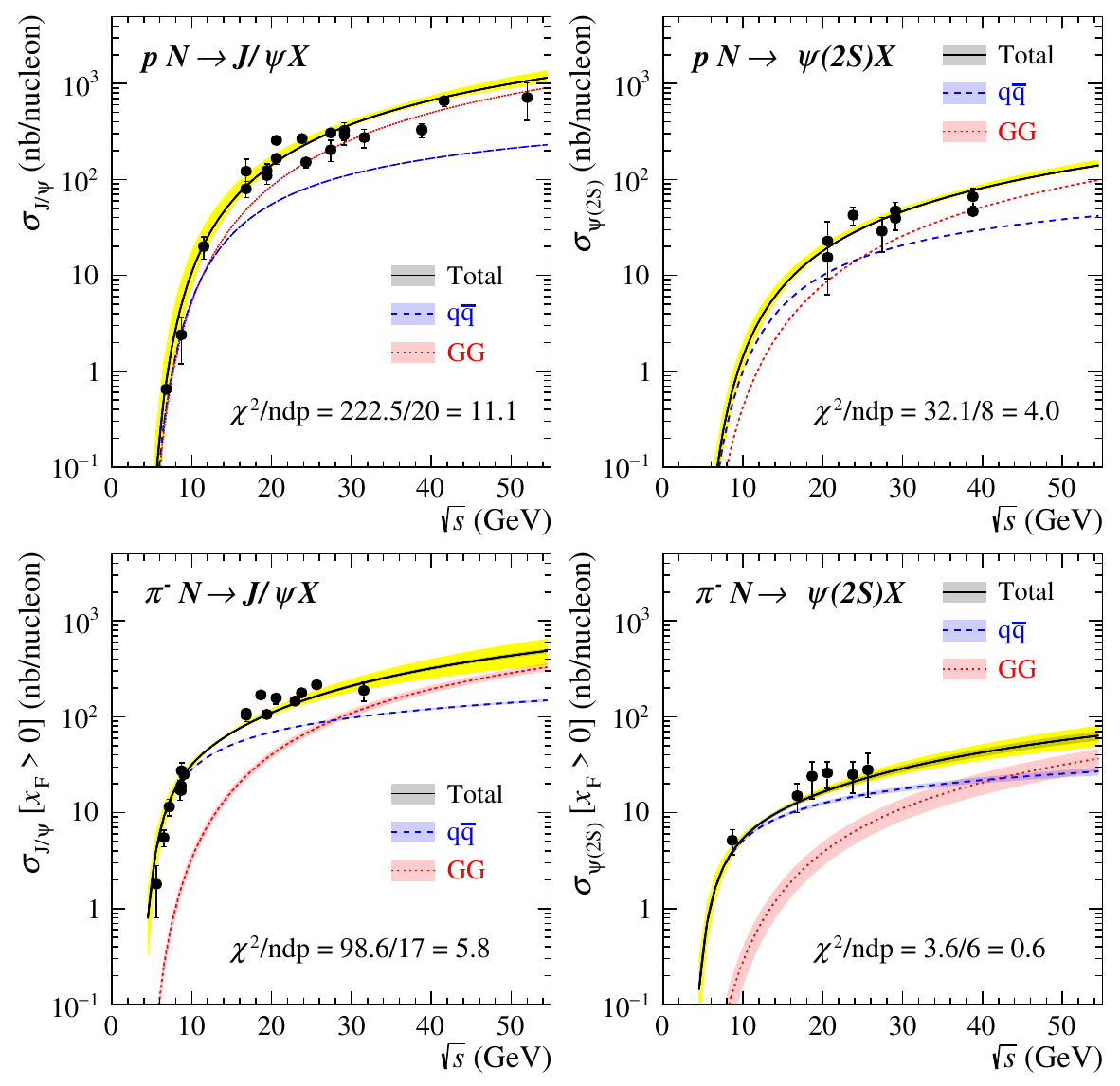}
\caption[\protect{}]{Integrated charmonium cross sections in $pN$ and
  $\pi^- N$ collisions. The data for $J/\psi$ and $\psi(2S)$
  production are compared to the fit made using the JAM pion PDFs. The
  total cross section and its $q \bar{q}$ and $GG$ contributions are
  denoted as solid black, dashed blue and dotted red lines,
  respectively. The uncertainty bands associated with JAM PDFs are
  also shown. The yellow bands represent the cross section
  uncertainties associated with the scale and charm quark mass
  systematic variations.}
\label{fig_sdep_JAM}
\end{figure*}

\begin{figure*}[!ht]
\centering
\includegraphics[width=0.8\columnwidth]{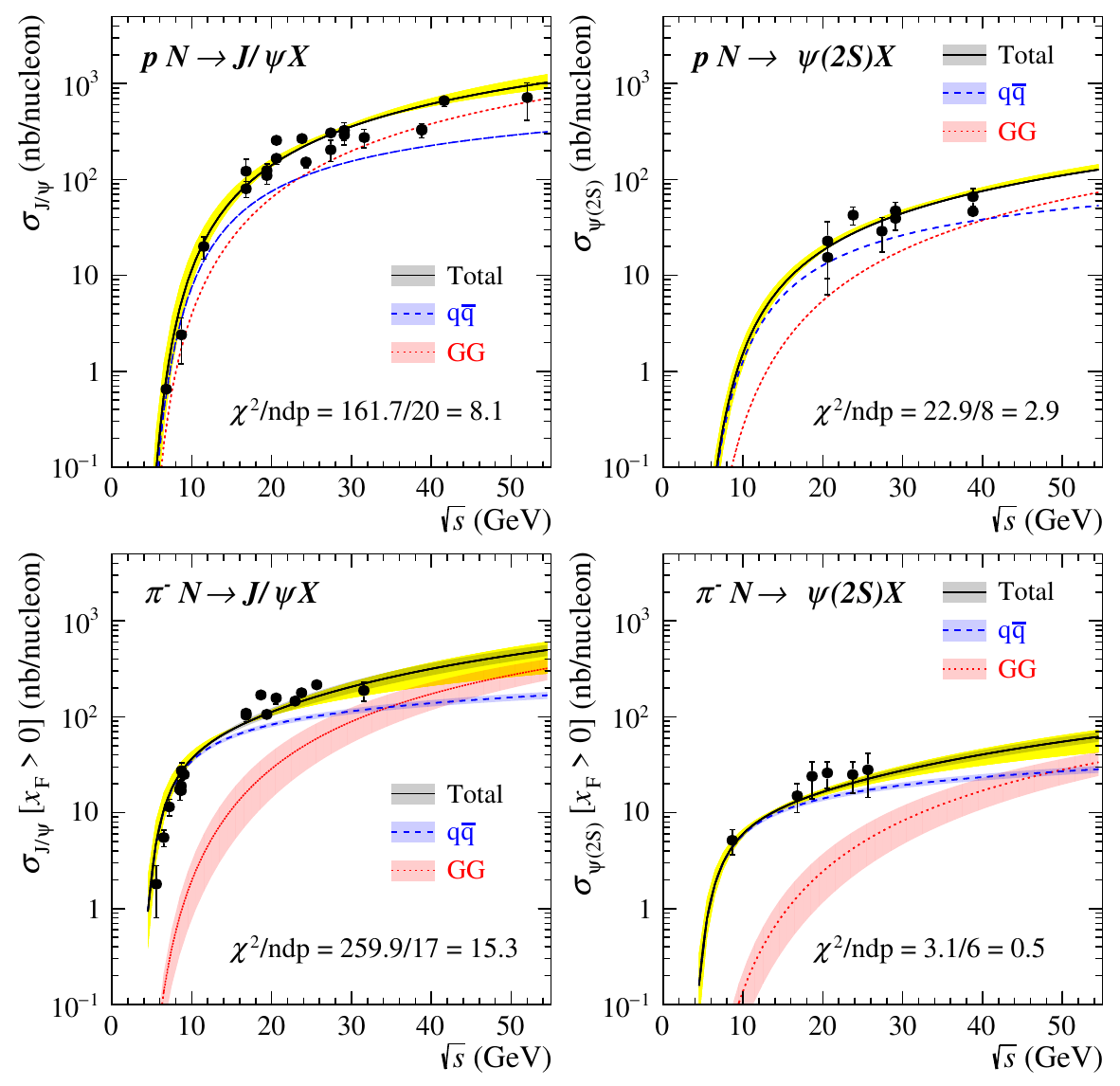}
\caption[\protect{}]{Integrated charmonium cross sections in $pN$ and
  $\pi^- N$ collisions. The data for $J/\psi$ and $\psi(2S)$
  production are compared to the fit made using the xFitter pion
  PDFs. The total cross section and its $q \bar{q}$ and $GG$
  contributions are denoted as solid black, dashed blue and dotted red
  lines, respectively. The uncertainty bands associated with xFitter
  PDFs are also shown. The yellow bands represent the cross section
  uncertainties associated with the scale and charm quark mass
  systematic variations.}
\label{fig_sdep_xFitter}
\end{figure*}

%%%%%%%%%%%%%%%%%%%%%%%%%%%%%%%%%%%%%%%%%%%%%%%%%%%%%%%%%%%%%%%%%%%%%%%%%%

\begin{figure*}[!ht]
\includegraphics[width=1.0\columnwidth]{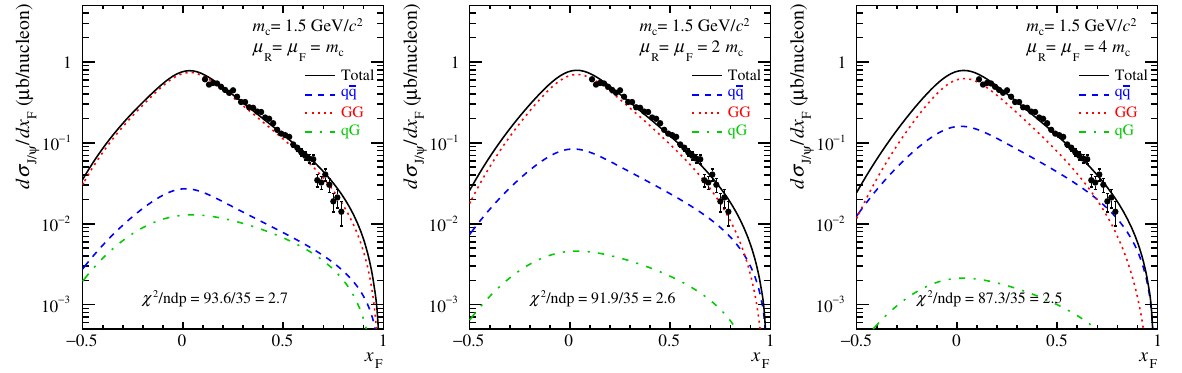}
\caption[\protect{}]{The NRQCD results with variation of charm quark
  mass $m_c$ and renormalization scale $\mu_R$, compared with the
  $d\sigma/dx_F$ data of $J/\psi$ production off the beryllium target
  with a 515-GeV/$c$ $\pi^-$ beam from the E672/E706
  experiment~\cite{jpsi_data1}. The pion PDFs used for the calculation
  is GRV. The total cross sections and $q \bar{q}$, $GG$, and $qG$
  contributions are denoted as solid black, dashed blue, dotted red
  and dot-dashed green lines, respectively. The charm quark mass
  $m_c$, factorization scale $\mu_F$, and renormalization scale
  $\mu_R$ used for the NRQCD calculation as well as the fit
  $\chi^2$/ndf are displayed in each plot.}
\label{fig_sys_GRV}
\end{figure*}

\begin{figure*}[!ht]
\includegraphics[width=1.0\columnwidth]{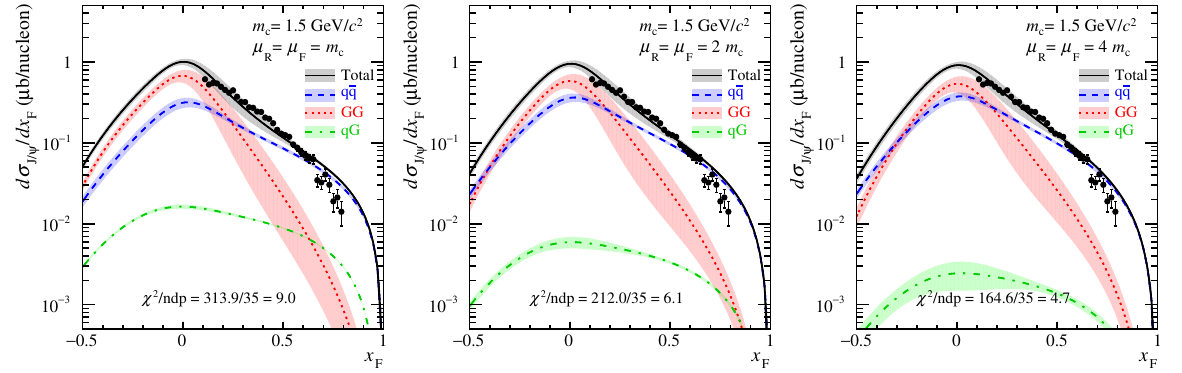}
\caption[\protect{}]{Same as Fig.~19 but with the input of JAM pion
  PDFs.}
\label{fig_sys_JAM}
\end{figure*}

\begin{figure*}[!ht]
\includegraphics[width=1.0\columnwidth]{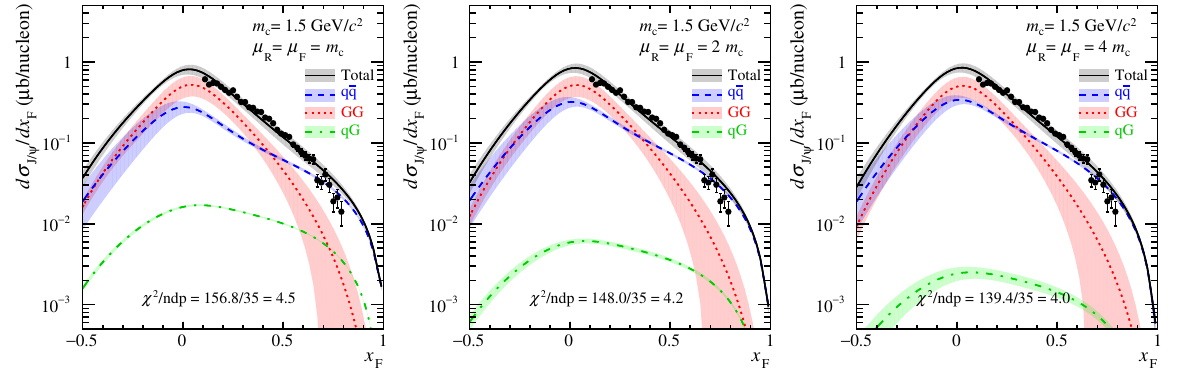}
\caption[\protect{}]{Same as Fig.~19 but with the input of xFitter pion
  PDFs.}
\label{fig_sys_xFitter}
\end{figure*}

%%%%%%%%%%%%%%%%%%%%%%%%%%%%%%%%%%%%%%%%%%%%%%%%%%%%%%%%%%%%%%%%%%%%%%%%%%%
% exe sfitall2#scandump_table2_new3 1
% ipdf for 1 (SMRS), 2 (GRV), 10 (JAM21) and 4 (xFitter)
\begin{table*}[!ht]   %\footnotesize
\setlength\tabcolsep{3pt}
\addtolength{\tabcolsep}{3pt}
\centering
%\begin{center}
\begin{tabular}{|c|r|r|r||r|r|r|}
\hline
 & \multicolumn{6}{|c|}{SMRS} \\
\hline
$m_c$ (GeV/$c^2$) & \multicolumn{3}{c||}{1.5} & 1.4 & 1.5 & 1.6 \\
\hline
$\mu/m_c$ & 1 & 2 & 4 & \multicolumn{3}{c|}{2} \\
\hline
 $\chi^2_{total}/ndf$  &   2.5 &   1.9 &   2.7 &   2.0 &   1.9 &   2.0 \\ 
 $\chi^2/ndp|^{\pi^-}_{x_F}$ &   2.3 &   1.8 &   2.4 &   2.0 &   1.8 &   1.7 \\ 
 $\chi^2/ndp|^{p}_{x_F}$ &   2.3 &   1.6 &   3.0 &   1.4 &   1.6 &   2.1 \\ 
 $\chi^2/ndp|^{\pi^-}_{\sqrt{s}}$ &   4.6 &   8.7 &   4.3 &   6.7 &   8.7 &  10.7 \\ 
 $\chi^2/ndp|^{p}_{\sqrt{s}}$ &   5.0 &   8.1 &   8.7 &   6.6 &   8.1 &   9.4 \\ 
 $\langle \mathcal{O}_{8}^{J/\psi}[^{3}S_{1}] \rangle$ & 1.6E-02 & 2.6E-02 & 9.7E-02 & 1.2E-02 & 2.6E-02 & 5.2E-02 \\ 
  & $\pm$1.6E-03 & $\pm$2.3E-03 & $\pm$5.7E-03 & $\pm$1.6E-03 & $\pm$2.3E-03 & $\pm$4.6E-03 \\ 
 $\Delta_8^{J/\psi}$ & 1.3E-02 & 5.6E-02 & 9.3E-02 & 2.0E-02 & 5.6E-02 & 1.1E-01 \\ 
 //' & $\pm$7.9E-04 & $\pm$1.6E-03 & $\pm$2.4E-03 & $\pm$9.8E-04 & $\pm$1.6E-03 & $\pm$2.3E-03 \\ '
 $\langle \mathcal{O}_{8}^{\psi(2S)}[^{3}S_{1}] \rangle$ & 7.7E-03 & 1.3E-02 & 2.9E-02 & 8.0E-03 & 1.3E-02 & 2.2E-02 \\ 
  & $\pm$4.0E-04 & $\pm$8.6E-04 & $\pm$1.4E-03 & $\pm$4.8E-04 & $\pm$8.6E-04 & $\pm$1.5E-03 \\ 
 $\Delta_8^{\psi(2S)}$ & 2.5E-04 & 5.7E-03 & 9.1E-03 & 1.7E-03 & 5.7E-03 & 1.1E-02 \\ 
  & $\pm$1.9E-04 & $\pm$2.9E-04 & $\pm$5.4E-04 & $\pm$1.5E-04 & $\pm$2.9E-04 & $\pm$5.8E-04 \\ 
\hline
\end{tabular}
\caption {The reduced $\chi^2/\text{ndf}$ of values for the whole data
  sets and the $\chi^2$ divided by the number of data point (ndf) for
  the pion-induced and proton-induced datasets with the systematic
  variation of charm quark mass $m_c$ of 1.4, 1.5 and 1.6 GeV/$c^2$,
  and $\mu = \mu_R = \mu_F$ at 1.0, 2.0, and 4.0 $m_c$ in NRQCD
  calculations and the corresponding input or best-fit LDMEs for SMRS
  pion PDFs. All LDMEs are in units of $\rm{GeV}^3$.}
\label{tab:SYS_SMRS}
\end{table*}

%%%%%%%%%%%%%%%%%%%%%%%%%%%%%%%%%%%%%%%%%%%%%%%%%%%%%%%%%%%%%%%%%%%%%%%%%%%
% exe sfitall2#scandump_table2_new3 2
% ipdf for 1 (SMRS), 2 (GRV), 10 (JAM21) and 4 (xFitter)
\begin{table*}[!ht]   %\footnotesize
\setlength\tabcolsep{3pt}
\addtolength{\tabcolsep}{3pt}
\centering
%\begin{center}
\begin{tabular}{|c|r|r|r||r|r|r|}
\hline
 & \multicolumn{6}{|c|}{GRV} \\
\hline
$m_c$ (GeV/$c^2$) & \multicolumn{3}{c||}{1.5} & 1.4 & 1.5 & 1.6 \\
\hline
$\mu/m_c$ & 1 & 2 & 4 & \multicolumn{3}{c|}{2} \\
\hline
 $\chi^2_{total}/ndf$  &   2.4 &   2.4 &   2.7 &   2.7 &   2.4 &   2.3 \\ 
 $\chi^2/ndp|^{\pi^-}_{x_F}$ &   2.3 &   2.4 &   2.6 &   2.7 &   2.4 &   2.1 \\ 
 $\chi^2/ndp|^{p}_{x_F}$ &   2.0 &   1.7 &   2.1 &   1.9 &   1.7 &   2.2 \\ 
 $\chi^2/ndp|^{\pi^-}_{\sqrt{s}}$ &   8.4 &   5.6 &   2.8 &   2.1 &   5.6 &   8.7 \\ 
 $\chi^2/ndp|^{p}_{\sqrt{s}}$ &   5.5 &   8.1 &   9.8 &   5.8 &   8.1 &   9.4 \\ 
 $\langle \mathcal{O}_{8}^{J/\psi}[^{3}S_{1}] \rangle$ & 1.5E-04 & 4.3E-02 & 1.4E-01 & 3.5E-02 & 4.3E-02 & 7.3E-02 \\ 
  & $\pm$1.3E-04 & $\pm$3.8E-03 & $\pm$8.4E-03 & $\pm$5.5E-05 & $\pm$3.8E-03 & $\pm$3.9E-03 \\ 
 $\Delta_8^{J/\psi}$ & 1.9E-02 & 5.2E-02 & 8.8E-02 & 1.3E-02 & 5.2E-02 & 1.1E-01 \\ 
 //' & $\pm$1.4E-04 & $\pm$1.7E-03 & $\pm$2.8E-03 & $\pm$2.2E-05 & $\pm$1.7E-03 & $\pm$1.7E-03 \\ '
 $\langle \mathcal{O}_{8}^{\psi(2S)}[^{3}S_{1}] \rangle$ & 8.4E-03 & 2.1E-02 & 4.2E-02 & 1.5E-02 & 2.1E-02 & 3.3E-02 \\ 
  & $\pm$5.8E-04 & $\pm$1.3E-03 & $\pm$3.1E-03 & $\pm$3.2E-05 & $\pm$1.3E-03 & $\pm$1.5E-03 \\ 
 $\Delta_8^{\psi(2S)}$ & 5.2E-04 & 4.2E-03 & 7.2E-03 & 1.0E-04 & 4.2E-03 & 9.3E-03 \\ 
  & $\pm$2.5E-04 & $\pm$2.9E-04 & $\pm$8.3E-04 & $\pm$7.8E-05 & $\pm$2.9E-04 & $\pm$3.7E-04 \\ 
\hline
\end{tabular}
\caption {The reduced $\chi^2/\text{ndf}$ of values for the whole data
  sets and the $\chi^2$ divided by the number of data point (ndf) for
  the pion-induced and proton-induced datasets with the systematic
  variation of charm quark mass $m_c$ of 1.4, 1.5 and 1.6 GeV/$c^2$,
  and $\mu = \mu_R = \mu_F$ at 1.0, 2.0, and 4.0 $m_c$ in NRQCD
  calculations and the corresponding input or best-fit LDMEs for GRV
  pion PDFs. All LDMEs are in units of $\rm{GeV}^3$.}
\label{tab:SYS_GRV}
\end{table*}

%%%%%%%%%%%%%%%%%%%%%%%%%%%%%%%%%%%%%%%%%%%%%%%%%%%%%%%%%%%%%%%%%%%%%%%%%%%
% exe sfitall2#scandump_table2_new3 10
% ipdf for 1 (SMRS), 2 (GRV), 10 (JAM21) and 4 (xFitter)
\begin{table*}[!ht]   %\footnotesize
\setlength\tabcolsep{3pt}
\addtolength{\tabcolsep}{3pt}
\centering
%\begin{center}
\begin{tabular}{|c|r|r|r||r|r|r|}
\hline
 & \multicolumn{6}{|c|}{JAM} \\
\hline
$m_c$ (GeV/$c^2$) & \multicolumn{3}{c||}{1.5} & 1.4 & 1.5 & 1.6 \\
\hline
$\mu/m_c$ & 1 & 2 & 4 & \multicolumn{3}{c|}{2} \\
\hline
 $\chi^2_{total}/ndf$  &   8.2 &   5.6 &   4.7 &   6.3 &   5.6 &   5.0 \\ 
 $\chi^2/ndp|^{\pi^-}_{x_F}$ &   8.7 &   5.9 &   4.9 &   6.9 &   5.9 &   5.2 \\ 
 $\chi^2/ndp|^{p}_{x_F}$ &   4.0 &   2.7 &   2.5 &   2.5 &   2.7 &   2.7 \\ 
 $\chi^2/ndp|^{\pi^-}_{\sqrt{s}}$ &  31.3 &  11.4 &   4.9 &  31.4 &  11.4 &   9.4 \\ 
 $\chi^2/ndp|^{p}_{\sqrt{s}}$ &   6.8 &   5.1 &   7.5 &   5.4 &   5.1 &   7.7 \\ 
 $\langle \mathcal{O}_{8}^{J/\psi}[^{3}S_{1}] \rangle$ & 6.0E-02 & 1.2E-01 & 2.1E-01 & 7.2E-02 & 1.2E-01 & 1.9E-01 \\ 
  & $\pm$1.2E-03 & $\pm$2.1E-03 & $\pm$4.0E-03 & $\pm$1.7E-03 & $\pm$2.1E-03 & $\pm$4.9E-01 \\ 
 $\Delta_8^{J/\psi}$ & 3.5E-03 & 2.4E-02 & 6.2E-02 & 1.9E-03 & 2.4E-02 & 6.8E-02 \\ 
 //' & $\pm$6.9E-04 & $\pm$1.6E-03 & $\pm$2.6E-03 & $\pm$8.0E-04 & $\pm$1.6E-03 & $\pm$3.3E+00 \\ '
 $\langle \mathcal{O}_{8}^{\psi(2S)}[^{3}S_{1}] \rangle$ & 1.1E-02 & 2.4E-02 & 4.0E-02 & 1.4E-02 & 2.4E-02 & 3.7E-02 \\ 
  & $\pm$3.6E-04 & $\pm$8.5E-04 & $\pm$1.5E-03 & $\pm$4.2E-04 & $\pm$8.5E-04 & $\pm$4.4E-01 \\ 
 $\Delta_8^{\psi(2S)}$ & 1.9E-09 & 2.1E-03 & 5.9E-03 & 3.1E-08 & 2.1E-03 & 6.7E-03 \\ 
  & $\pm$1.1E-05 & $\pm$3.2E-04 & $\pm$5.4E-04 & $\pm$2.9E-05 & $\pm$3.2E-04 & $\pm$4.3E-01 \\ 
\hline
\end{tabular}
\caption {The reduced $\chi^2/\text{ndf}$ of values for the whole data
  sets and the $\chi^2$ divided by the number of data point (ndf) for
  the pion-induced and proton-induced datasets with the systematic
  variation of charm quark mass $m_c$ of 1.4, 1.5 and 1.6 GeV/$c^2$,
  and $\mu = \mu_R = \mu_F$ at 1.0, 2.0, and 4.0 $m_c$ in NRQCD
  calculations and the corresponding input or best-fit LDMEs for JAM
  pion PDFs. All LDMEs are in units of $\rm{GeV}^3$.}
\label{tab:SYS_JAM}
\end{table*}

%%%%%%%%%%%%%%%%%%%%%%%%%%%%%%%%%%%%%%%%%%%%%%%%%%%%%%%%%%%%%%%%%%%%%%%%%%%
% exe sfitall2#scandump_table2_new3 4
% ipdf for 1 (SMRS), 2 (GRV), 10 (JAM21) and 4 (xFitter)
\begin{table*}[!ht]   %\footnotesize
\setlength\tabcolsep{3pt}
\addtolength{\tabcolsep}{3pt}
\centering
%\begin{center}
\begin{tabular}{|c|r|r|r||r|r|r|}
\hline
 & \multicolumn{6}{|c|}{xFitter} \\
\hline
$m_c$ (GeV/$c^2$) & \multicolumn{3}{c||}{1.5} & 1.4 & 1.5 & 1.6 \\
\hline
$\mu/m_c$ & 1 & 2 & 4 & \multicolumn{3}{c|}{2} \\
\hline
 $\chi^2_{total}/ndf$  &   4.8 &   4.2 &   4.2 &   4.7 &   4.2 &   3.7 \\ 
 $\chi^2/ndp|^{\pi^-}_{x_F}$ &   4.6 &   4.5 &   4.5 &   5.2 &   4.5 &   3.9 \\ 
 $\chi^2/ndp|^{p}_{x_F}$ &   3.9 &   1.9 &   1.9 &   1.8 &   1.9 &   2.0 \\ 
 $\chi^2/ndp|^{\pi^-}_{\sqrt{s}}$ &   9.7 &   4.4 &   2.5 &   9.3 &   4.4 &   2.1 \\ 
 $\chi^2/ndp|^{p}_{\sqrt{s}}$ &  11.0 &   6.9 &   9.5 &   6.2 &   6.9 &   9.7 \\ 
 $\langle \mathcal{O}_{8}^{J/\psi}[^{3}S_{1}] \rangle$ & 3.6E-02 & 8.5E-02 & 1.6E-01 & 5.1E-02 & 8.5E-02 & 1.3E-01 \\ 
  & $\pm$9.9E-04 & $\pm$4.1E-03 & $\pm$4.7E-03 & $\pm$1.0E-03 & $\pm$4.1E-03 & $\pm$6.0E-03 \\ 
 $\Delta_8^{J/\psi}$ & 1.7E-02 & 3.9E-02 & 8.3E-02 & 9.6E-03 & 3.9E-02 & 9.4E-02 \\ 
 //' & $\pm$9.5E-04 & $\pm$3.4E-03 & $\pm$4.1E-03 & $\pm$6.3E-04 & $\pm$3.4E-03 & $\pm$5.2E-03 \\ '
 $\langle \mathcal{O}_{8}^{\psi(2S)}[^{3}S_{1}] \rangle$ & 9.3E-03 & 1.9E-02 & 3.4E-02 & 1.2E-02 & 1.9E-02 & 2.9E-02 \\ 
  & $\pm$3.6E-04 & $\pm$1.2E-03 & $\pm$1.4E-03 & $\pm$4.5E-04 & $\pm$1.2E-03 & $\pm$1.6E-03 \\ 
 $\Delta_8^{\psi(2S)}$ & 9.1E-04 & 4.0E-03 & 8.7E-03 & 6.2E-04 & 4.0E-03 & 1.0E-02 \\ 
  & $\pm$1.8E-04 & $\pm$6.4E-04 & $\pm$7.0E-04 & $\pm$1.5E-04 & $\pm$6.4E-04 & $\pm$9.5E-04 \\ 
\hline
\end{tabular}
\caption {The reduced $\chi^2/\text{ndf}$ of values for the whole data
  sets and the $\chi^2$ divided by the number of data point (ndf) for
  the pion-induced and proton-induced datasets with the systematic
  variation of charm quark mass $m_c$ of 1.4, 1.5 and 1.6 GeV/$c^2$,
  and $\mu = \mu_R = \mu_F$ at 1.0, 2.0, and 4.0 $m_c$ in NRQCD
  calculations and the corresponding input or best-fit LDMEs for xFitter
  pion PDFs. All LDMEs are in units of $\rm{GeV}^3$.}
\label{tab:SYS_xFitter}
\end{table*}

\begin{figure*}[!ht]
\centering
\includegraphics[width=1.0\columnwidth]{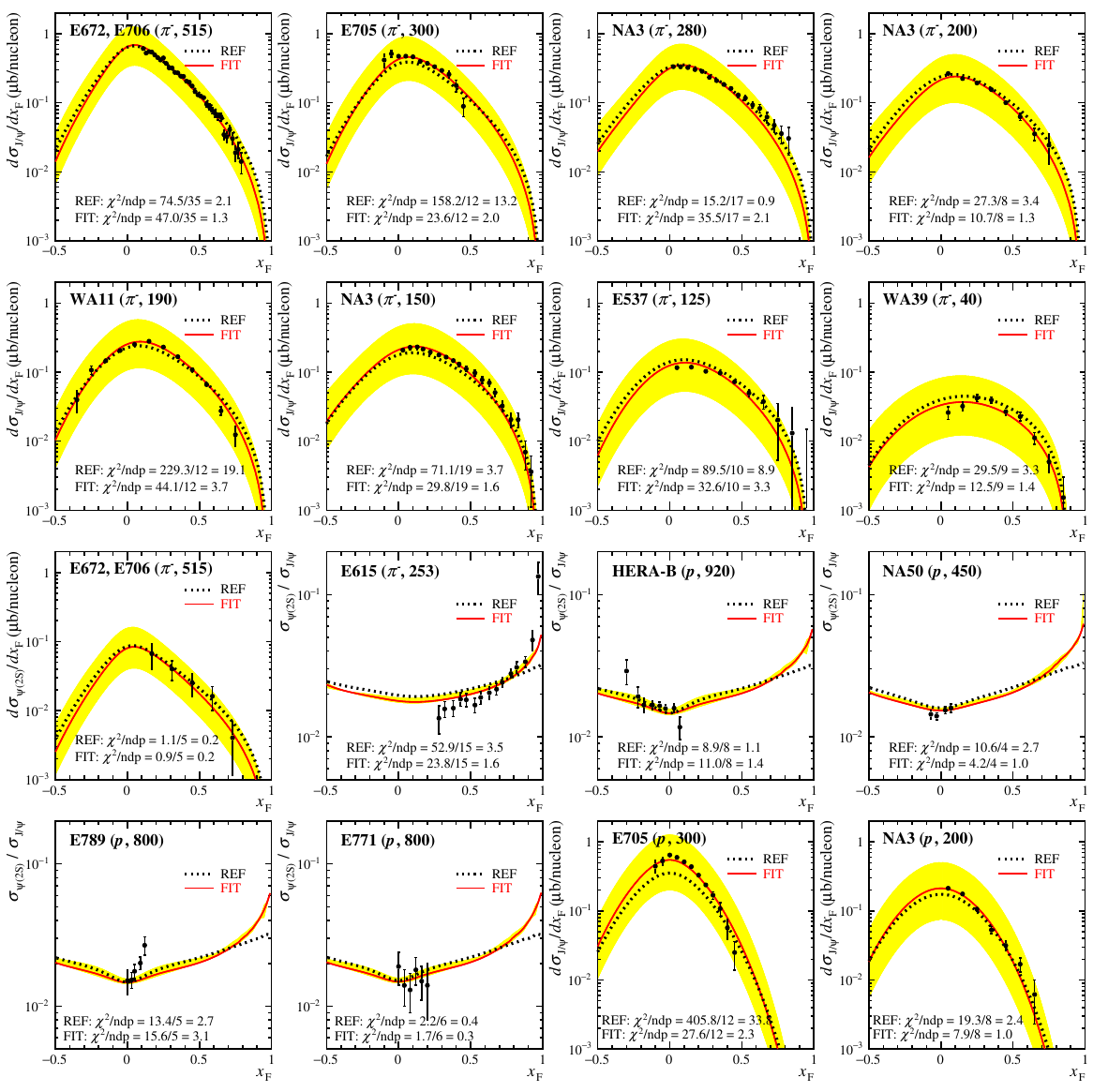}
\caption[\protect{}]{Same as Fig.~2 in the main text while the
    yellow bands represent the cross section uncertainties
    corresponding to the scale and charm quark mass systematic
    variations, with the fixed LDMEs from "Fit" in Table III.}
\label{fig_jpsi_PDF1_fixedLDME}
\end{figure*}

\begin{figure*}[!ht]
\centering
\includegraphics[width=1.0\columnwidth]{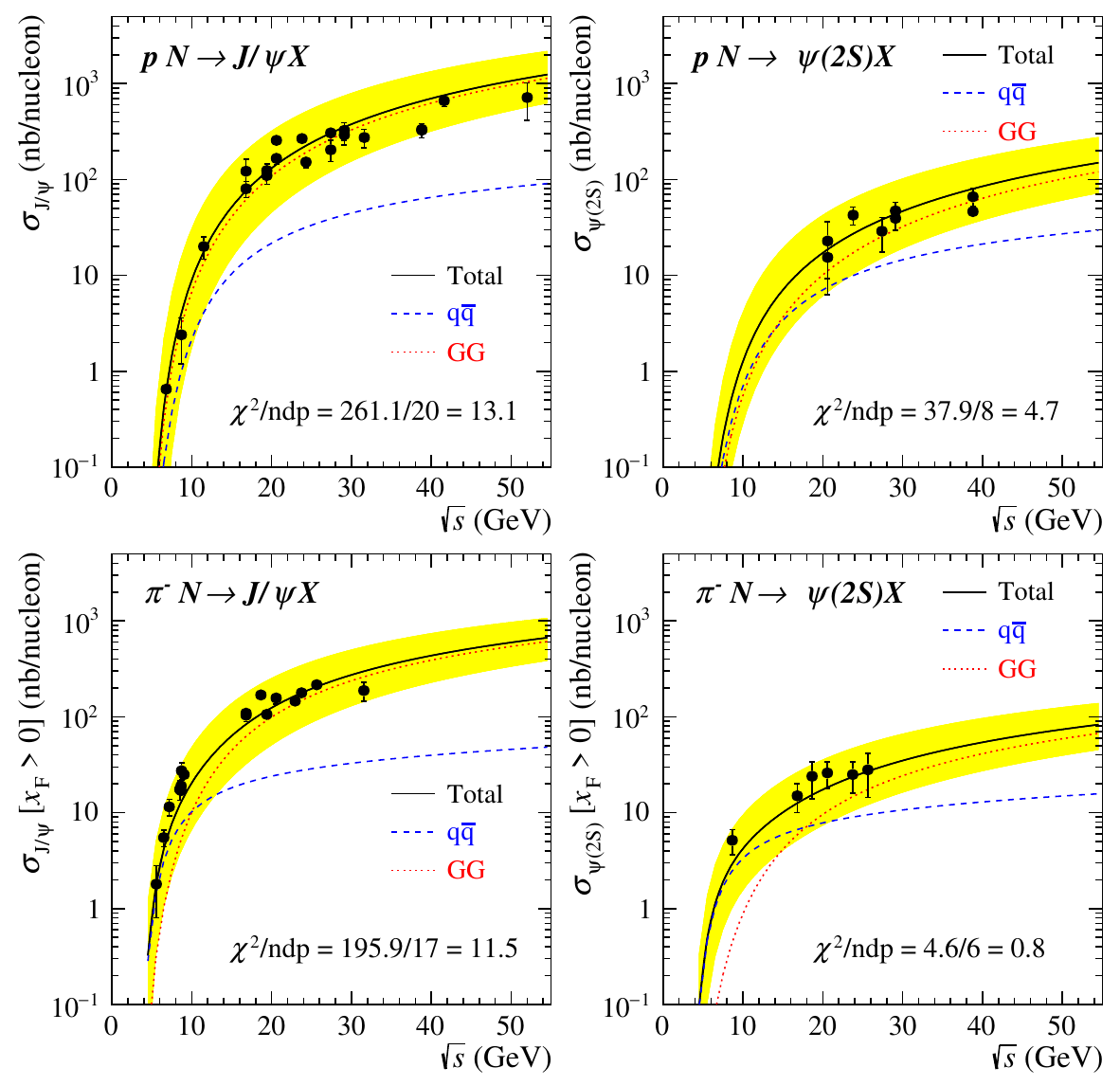}
\caption[\protect{}]{Same as Fig.~8 in the main text while the
    yellow bands represent the cross section uncertainties
    corresponding to the scale and charm quark mass systematic
    variations, with the fixed LDMEs from "Fit" in Table III.}
\label{fig_sdep_SMRS_fixedLDME}
\end{figure*}

\end{document}